\documentstyle[12pt,psfig]{article}
 1
\def\as{\alpha_{\rm S}}

\def\citenum#1{{\def\@cite##1##2{##1}\cite{#1}}}
\def\citea#1{\@cite{#1}{}}

\def\as{\alpha_{\rm S}}

\def\D{\Delta}

\def\g{\gamma}
\def\G{\Gamma}
\def\grad{\nabla}
\def\h{ \frac{1}{2}}

\def\hm{ {-{1\over 2}}  }

\def\o{\omega}

\def\pa{\partial}

\def\s{\sigma}

\def\({\left(}
\def\){\right)}

\def\citenum#1{{\def\@cite##1##2{##1}\cite{#1}}}
\def\citea#1{\@cite{#1}{}}

\def\l1vt{\vec{l_{1\perp}}}

\def\rt{r_{\perp}}
\def\bt{b_{\perp}}
\def\rt2{$r^2_{\perp}$}
\def\bt2{$b^2_t$}

\def\jol1{$J_0(\,l_{1\perp}\,r_{\perp}\,)$}

\def\citea#1{\@cite{#1}{}}

\def\beq{\begin{equation}}
\def\eeq{\end{equation}}
\def\bea{\begin{eqnarray}}
\def\eea{\end{eqnarray}}

\def\eq#1{{eq.~(\ref{#1})}}

\def\bbbz{{\mathchoice {\hbox{$\sf\textstyle Z\kern-0.4em Z$}}
{\hbox{$\sf\textstyle Z\kern-0.4em Z$}}
{\hbox{$\sf\scriptstyle Z\kern-0.3em Z$}}
{\hbox{$\sf\scriptscriptstyle Z\kern-0.2em Z$}}}}
\def\npb#1#2#3{    {\it Nucl. Phys. }{\bf B#1} (19#2) #3}
\def\plb#1#2#3{    {\it Phys. Lett. }{\bf B#1} (19#2) #3}
\def\prd#1#2#3{    {\it Phys. Rev. }{\bf D#1} (19#2) #3}

\def\prl#1#2#3{    {\it Phys. Rev. Lett. }{\bf #1} (19#2) #3}

\def\zpc#1#2#3{    {\it Z. Phys. }{\bf C#1} (19#2) #3}

\def\sjnp#1#2#3{   {\it Sov. J. Nucl. Phys. }{\bf #1} (19#2) #3}

\def\l{\lambda}
\relax
\begin{document}
\begin{titlepage}
\noindent
 20 April 1996   \hfill  CBPF-NF-020/96 \,\,\,\, {\bf hep - ph }\\[4ex]
\begin{center}

{\Large\bf{QCD EVOLUTION OF THE GLUON}}\\[1.4ex]
{\Large \bf { DENSITY IN A NUCLEUS}}\\[9ex]
{\large \bf { A. L.
Ayala  F$^{\underline{o}}$ ${}^{a)\,b)}$${}^*$\footnotetext{ ${}^*$ E-mail:
ayala@if.ufrgs.br}
,
 M. B. Gay  Ducati ${}^{a)}$$^{**}$\footnotetext{${}^{**}$
E-mail:gay@if.ufrgs.br}}}\\
 and\\
{ \large \bf{ E.M. Levin ${}^{\dagger}\,{}^{c)\,d)}$
\footnotetext{$^{\dagger}$ E-mail: levin@lafex.cbpf.br;levin@ccsg.tau.ac.il} 
}} \\[1.5ex]

{\it ${}^{a)}$Instituto de F\'{\i}sica, Univ. Federal do Rio Grande do Sul}\\
{\it Caixa Postal 15051, 91501-970 Porto Alegre, RS, BRAZIL}\\[1.5ex]
{\it ${}^{b)}$Instituto de F\'{\i}sica e Matem\'atica, Univ. 
Federal de Pelotas}\\
{\it Campus Universitario, Caixa Postal 354, 96010-900, Pelotas, RS,
BRAZIL}\\
{\it ${}^{c)}$  LAFEX, Centro Brasileiro de Pesquisas F\'\i sicas  (CNPq)}\\
{\it Rua Dr. Xavier Sigaud 150, 22290 - 180 Rio de Janeiro, RJ, BRAZIL}
\\
{\it$ {}^{d)}$ Theory Department, Petersburg Nuclear Physics Institute}\\
{\it 188350, Gatchina, St. Petersburg, RUSSIA}\\[6.5ex]
\end{center}
{\large \bf Abstract:}
The Glauber approach to the gluon density in a nucleus, 
suggested by A. Mueller, is  developed and studied in detail. Using
 the GRV parameterization for the gluon density in a nucleon, the value as well
 as energy and $Q^2$ dependence of the gluon density in a nucleus  is
calculated. It is shown that the shadowing corrections are under
 theoretical control and  are essential in the region of small $x$.
They change
crucially the value of the gluon density as well as the value of the
anomalous dimension of the nuclear structure function, unlike of the
nucleon one. The systematic theoretical way to treat the correction to the
Glauber approach is developed and a new evolution equation
 is derived and solved. It is shown that the solution of the new evolution
equation can provide a selfconsistent matching of  ``soft" high energy
phenomenology with  ``hard" QCD physics.
\end{titlepage}

\section{Introduction.}

 In this paper we  discuss the QCD evolution for
the gluon density in a nucleus. The gluon density is the most 
important physical observable that governs the physics at high energy
(low Bjorken $x$) in deep inelastic processes \cite{r1}. Dealing with nucleus 
we have to take into account the shadowing correction (SC) due to 
rescattering of the gluon inside the nucleus, which is the main 
point of interest in this paper. We show that SC can be treated theoretically
in the framework of perturbative QCD (pQCD) and can be calculated using the
information on the behavior of the gluon structure function for the nucleon.

Our calculations were  performed in the double log approximation
 ( DLA) of pQCD, or in other words in the GLAP evolution
 equations \cite{r4} for the region of low  $x$ ( high energy ). In the DLA 
we consider
the kinematic region where $\as \ln(1/x) \,\ln (Q^2/Q^2_0)\,\,\sim\,\,1$ while
$\as \ln(1/x)\,\,\ll\,\,1$ and $\as \ln (Q^2/Q^2_0) \,\,\ll\,\,1$ as well as 
$\as\,\,\ll\,\,1$, where $Q^2$ is the virtuality of the photon
and $\as$ is the QCD coupling constant.
 In terms of the anomalous dimension ( see the next section
for details) it means that we restrict ourselves by  considering   only the 
leading term in the anomalous dimension, namely $\gamma$ 
being $\as\,\ll\,\gamma\,\ll\,1$.

\begin{figure}[htbp]
\centerline{\psfig{figure=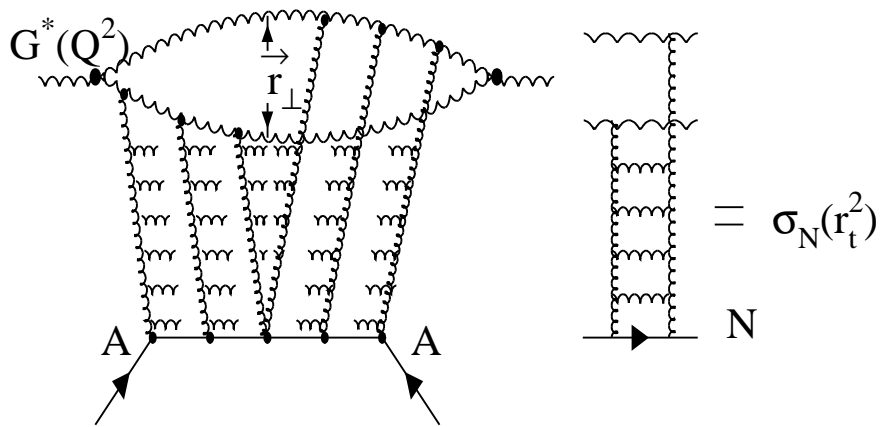,height=100mm,clip=t}}
\caption{\em The structure of the parton cascade in the Glauber formula.
 $A$ denotes
the nucleus, N - the nucleon,  $G^{*}(Q^2)$ - the virtual gluon and
$\s_N (r^2_t)$ is the nucleon cross section.}
\label{Fig.1}
\end{figure}

The advantages of the DLA are (see Fig.\ref{Fig.1} for notations): 

1) we can neglect the change in the
distance between quark (gluon) and antiquark (gluon) $r_t$ during the passage 
of the $q \bar q$ ( GG) pair through the nucleus. This simplifies the derivation
of all formulae for the SC and leads to an eikonal picture of classical
 propagation of the $q \bar q$ (GG) pair with high energy which receives
 independent  kicks due to rescattering in a nucleus. 

2) the cross section of the $q \bar q $ (GG) pair with transverse
 separation $r_t$ can be expressed through the gluon deep inelastic structure
 function for a nucleon.

We will discuss both points in the next section in  more detail. However,
it is worthwhile mentioning that the DLA allows us to obtain the simplest and
closest expression for the SC in DIS with a nucleus.

The main goal of the paper is
(i)  to present a study of the SC in the Glauber approach  using  
 available information on the gluon distribution in the nucleon based 
mainly on new experimental data from HERA \cite{r2} and on the solution
of the GLAP evolution equations  \cite{r4}; and (ii) to find the generalization
of the Glauber approach which will give the theoretical basis for the 
selfconsistent description of the gluon structure function for nuclei.

 Considering the deep inelastic scattering (DIS), we have to answer two
 principal questions: (i) why and how the cross section of DIS 
 with nucleus which is equal to $\s(\gamma^* A) \,=\,A \,\s( \gamma^* p)$ at
$x\,\approx\,1$ changes its A-dependence and becomes $\s(\gamma^* A)\,
\propto\, A^{\frac{2}{3}} $ at $x \,\rightarrow \,0$ ; and (ii) why 
$\s( \gamma^* A)$,  which   is proportional 
to $A^{\frac{2}{3}}$ at small values of $x$ and
 $ Q^2$,  approaches to $\s(\gamma^* A) = A \s(\gamma^* p)$ at
large values of $Q^2$ even in the region of small $x$. The last statement is
obvious  from the intuitive physical picture because at large values of $Q^2$
 $\s(\gamma^* p)$ is small and the virtual photon probes the number of
 nucleons in a nucleus.

The first question has been answered in the framework of the parton model
( see refs.\cite{r24} \cite{I1}\cite{r25}) and the answer is based on the
 space-time picture of DIS in the rest frame of the nucleus. Indeed, the
incident electron penetrates the nucleus and radiates the virtual photon
 whose lifetime $\tau_{\gamma^*}\,\propto\,\frac{1}{ m x}$ \cite{r24}.
We can recover three different kinematic regions:

1. $\tau_{\gamma^*}\,=\,\frac{1}{ m x}\,\,<\,\,R_{NN}$, where $R_{NN}$ is
the characteristic distances between the nucleons of the nucleus.
This virtual photon can be absorbed only by one nucleon and the total cross 
section is  $\s(\gamma^* A) \,=\,A \,\s( \gamma^* p)$.

2. $ R_A\,\,>\,\,\tau_{\gamma^*}\,=\,\frac{1}{ m x}\,\,>\,\,R_{NN}$, where
 $R_A$ is the nucleus radius.
In this kinematic region the virtual photon can interact with the group of
nucleons. However, $\s(\gamma^* A)$ is still proportional to $A$ since the
number of nucleons in a group is much less than $A$.

3. $\tau_{\gamma^*}\,=\,\frac{1}{ m x}\,\,>\,\,R_{A}$. Here, before reaching
the front surface of the nucleus, the virtual photon ``decomposes"
 in the developed parton cascade which then interacts with the nucleus.
It can be shown \cite{I1} that the absorption cross section of the virtual
 photon will now be proportional to the surface area of the nucleus
$\s(\gamma^* A)\,
\propto\, A^{\frac{2}{3}} $, because the wee partons of the parton cascade
are absorbed at the  surface and do not penetrate into the centre of
 the nucleus.

However, the above simple picture cannot help us to answer the second question.
Indeed, we can use it to explain the $A$ dependence of the initial partonic 
distributions but in the GLAP evolution equations the  $A$ dependence
is factorized out and do not affect the $Q^2$ evolution. Therefore, we have
 to change the evolution equations to incorporate the physical phenomenon 
which we have formulated as the second question. To recover the physical
origin of the new evolution equations in nucleus let us consider the
 oversimplified  structure of the parton cascade:  the 
virtual photon decays only in quark - antiquark pair. In this  case 
the cross section can be written in the form:
\beq \label{i1}
\s(\gamma^*, A)\,\,\propto\,\,
\int^1_0 d z \int d^2 r_t    \Psi(z,r_t) \,\sigma_A(zQ_0,r^2_t)\,\Psi^* (z,r_t)
\eeq
where $\s_A$ is the cross section for $\bar q q $ interaction with
  the nucleus,  $ z$ is the fraction of energy of the photon ($Q_0$)
 carried by  quark and $r_t$ is
 the transverse separation between quark and antiquark. $\Psi$ is the wave
function of the virtual photon, which is known and $ |\Psi|^2$ is equal to
$ a^2 K^2_1 ( a r_t) [ z^2 + ( 1 - z)^2]$, where $K_1$ is  the McDonald
function and $a^2 = Q^2 z ( 1 -z)$. The main contribution in the \eq{i1}
comes from the region where $a r_t\,<\,1$. Expanding $K_1$ and integrating 
over $z (1 - z)\,< \frac{1}{Q^2 r^2_t}\,<\,\frac{1}{4}$ we derive
\beq \label{i2}
\s(\gamma^*, A)\,\,\propto\,\,
 \frac{1}{Q^2} \int^{\infty}_{\frac{4}{Q^2}} \frac{d r^2_t}{r^4_t} 
    \,\sigma_A(\frac{Q_0}{Q^2 r^2_t},r^2_t)\,
\eeq
For very small $r_t$ the cross section in QCD is small and proportional to
 $r^2_t$. Such small cross section leads to $\s_A \,\propto\,A\,\frac{1}{Q^2}
\ln Q^2$, since the probe with small cross section interacts with all nucleons
in the nucleus. One can see that
 this is  the first term of the GLAP evolution equations. With more complicated
parton cascade we are able to reconstruct the GLAP evolution equations in a 
full. However, even at small $r^2_t$,  $\s_A = A \s_N$ only if the rescattering
of $\bar q q $ - pair in a nucleus is small. The parameter which controls the
value of the rescatterings is the number of collisions which is equal
$\nu\,=\,2 \,\s_N \,\rho\,R_A$, where $\rho$ is the nucleon density
 in the nucleus. If $\nu\,\ll\,1$ we can neglect the rescatterings,but
if $\nu \,\gg\, 1$ $ \s_A\,=\,2 \pi R^2_A$ and does not depend on nucleon
cross section. Therefore, we can trust the $\ln Q^2$ contribution in \eq{i2}
only for $\nu= r^2_t \rho R_A \,<\,1$ or for $Q^2 > \rho R_A$. Really, this
 condition depends on $x$ too, since the nucleon cross section is a function of
$x$, namely we will show that $\s_N\,\propto\,r^2_t xG(x,\frac{1}{r^2_t})$.

Therefore, the lesson  learned from this simple exercise is that we
 can  trust the GLAP evolution for the nucleus structure function only
for $Q^2 \,>\,Q^2_0(x)$. It means that we  have to solve the GLAP evolution
 equation with the initial condition on the line $Q^2 = Q^2_0(x)$ or
we have to change the evolution equations if we want to solve them in an
usual way, namely, starting from the structure functions at $ Q^2=Q^2_0$.

The previous attempts to attack this problem were related to the GLR equation
\cite{r1,r21}, in which the interaction ( recombination )
between two partons from different
parton cascades was taken into account ( see refs.\cite{I2,r21,I3,I4,I5,I6} ). 
 It was shown that the GLR equation is able to describe
the main features of the experimental data on
the deep inelastic scattering off nucleus \cite{I7,II7}.  The
  applications of the
 same ideas to description of other process such as J/$\Psi$ production
\cite{I8} and Drell-Yan process \cite{I9} was also with reasonable success
\cite{I10,I11}.

However, the GLR equation was derived in the limited kinematic region where
more complicated recombination processes have been proven to be negligible.
Roughly speaking, we can trust the GLR equation only for small $x$ and
 big values of $Q^2$.

In this paper we reanalyse the situation with the shadowing corrections in QCD 
for nucleus gluon structure function starting with the Glauber approach.
The Glauber formula for the gluon structure function in the nucleus
has been proven by A. Mueller in ref.\cite{r7} but this was remained unnoticed
by the majority of experts because his  paper was devoted to quite different
problem.  In section 2 we rederive the Mueller formula and discuss the main
properties of the Glauber approach in QCD.

In section \ref{numres}, we present the result of our calculations 
based on Glauber approach and point out which information 
on nuclear gluon distribution is needed in order to provide a 
reliable calculation. We adopt to following semiquantative way to study
the mechanism and value of the SC: we used the GRV parameterization
 for the gluon structure function in the nucleon, which describes quite well
the current
experimental data. However, we would like rather to study 
the main properties of the SC than to provide  reliable predictions
 for an experiment.

In section \ref{cgf} we consider the correction to the Glauber formula 
and discuss the generalization of the Glauber approach for deep inelastic 
scattering off nucleus. Here, we discuss the GLR evolution equation which
was the basis of all previous attempts to go beyond  the Glauber approach
and suggest and solve a  more general evolution equation which
is correct in the whole kinematic region without the  limitations of the GLR 
equation.

A summary and
final discussion are presented is section \ref{conclu}.

\section{ The Glauber approach in QCD .}
\label{gaqcd}
\begin{figure}[htbp]
\centerline{\psfig{figure=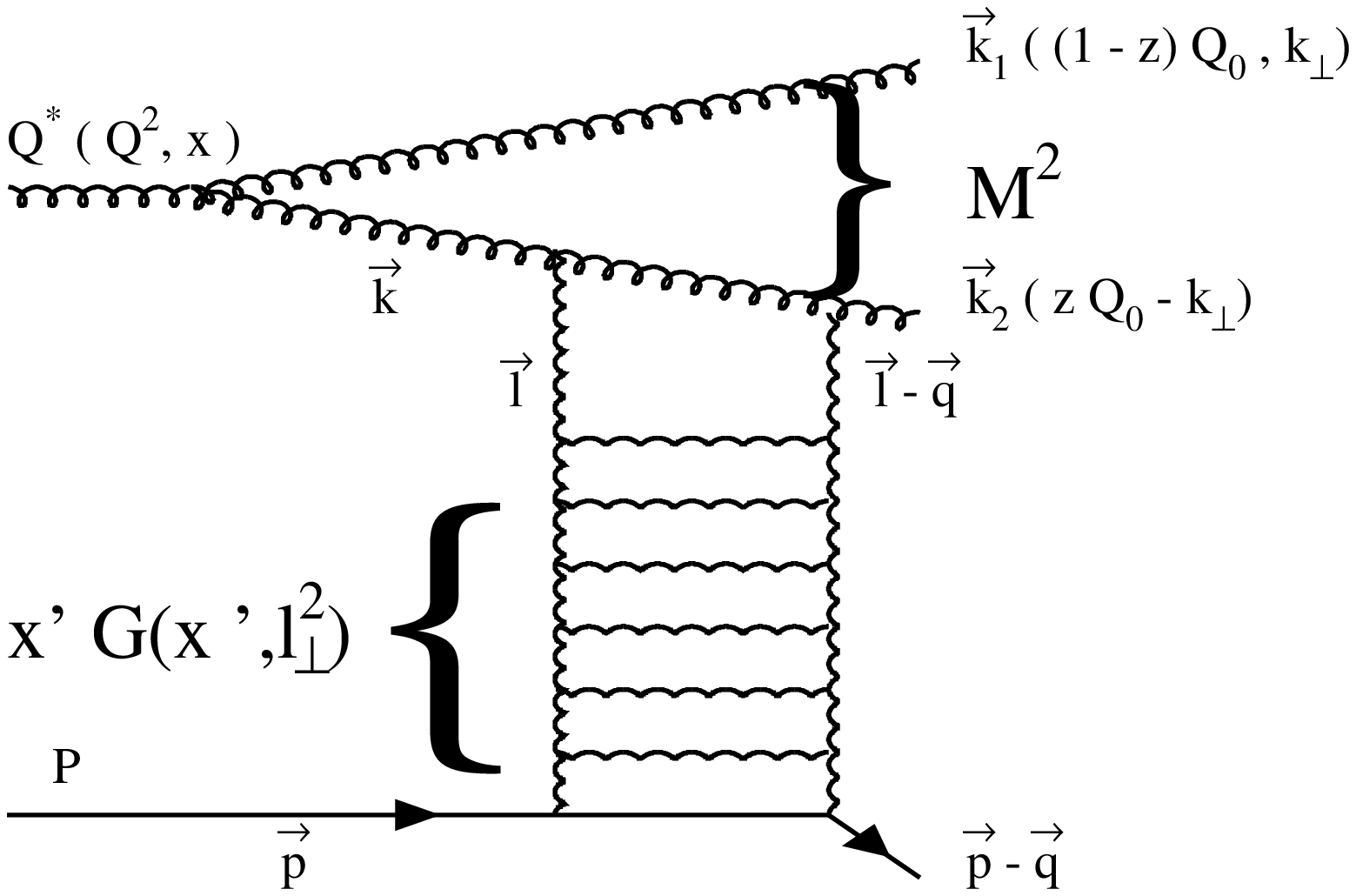,height=100mm}}
\caption{\em Kinematics of diffractive production of a gluon pair.
$Q^*$ is the virtual colorless probe of the gluon density. }
\label{Fig.2}
\end{figure}

In this paper we will use the following notation (see Fig.\ref{Fig.1} and
 Fig.\ref{Fig.2}):

$Q^{2}$ denotes the virtuality of the gluon in deep inelastic scattering (DIS);

$m$ is the mass of the proton nucleus specified otherwise;

Bjorken $x$ = $x_{Bj} \equiv \frac{Q^2}{s}$ where $\sqrt{s}\,=\,W$ is the c.m.
energy of the incoming gluon.

$\vec{k}_{t}$ denotes the transverse momentum of the quark or gluon;

$\vec{r}_{t}$ is the transverse separation between quark (gluon) and 
antiquark (gluon);

$\vec{b}_{t}$ is the impact parameter of the reaction which is the variable
conjugated to the momentum transfer ($\vec{q}_{t}$); 

$\vec{l}_{1 t}$ denotes the transverse momentum of the gluon attached to the
quark - antiquark (gluon-gluon) pair.

We will use the GLAP evolution equations \cite{r4} for the parton 
densities in momentum space. For any function $f(x)$, we define the moment 
$f(\omega)$ as
\beq
f(\omega)=\int_{0}^{1} d x \, x^{\o} f(x) \, .
\label{mom}
\eeq

Note that the moment variable $\o$ is chosen  such that the $\o = 0$ moment
measure the number of partons, and the $\o = 1$ measure their momentum.
An alternative moment variable $N=\o -1$ is often found in the literature. The
$x$-distribution can be reconstructed by considering the inverse Mellin
transform. For example, for the gluon distribution it reads
\beq
x G(x,Q^2) = \frac{1}{2 \pi i} \int_{C} d \o \,\, x^{-\o}\, g(\o,Q^2) \, ,
\label{gdef}\
\eeq
where $ g(\o,Q^2)$ is the moment of the gluon distribution. The contour of
integration $C$ is taken to the right of all singularities.

The GLAP evolution equations have the solutions of the form
\beq
g(\o,Q^2) = g(\o)e^{\g (\o) lnQ^2}\, ,
\label{gsol}
\eeq 
where $\g (\o)$ denotes the anomalous dimension, which in the leading 
$ln(1/x_{B})$ approximation (LLA) of pQCD, is a function of $\as/ \o$
and can be presented as the following series \cite{r5}:
\beq 
\g ( \omega ) \,\,=\,\,\frac{\as N_c }{\pi}\cdot \frac{1}{\omega} \,\,+\,\,
\frac{2  \as^4 N_{c}^4 \zeta ( 3 )}{\pi^4} \cdot\frac{1}{\omega^4}\,\,+\,\,O
( \frac{\as^5}{ \omega^5} )
\label{gexp1}
\eeq
where $\zeta (3)$ is the Riemman zeta function, and $N_c$ is the number of colors.
 In DLA, only the first term in the above series is taken into account.

 The amplitude is normalized such that
\beq
\frac{d \s}{d t}\,\,=\,\,\pi \,|f(s,t)|^2 \,\, ,
\eeq
\beq
\s_{tot}\,\,=\,\,4\pi\,\,Im\,f (s,0) \,\, ,
\eeq
and the scattering amplitude in $b_{t}$-space is defined as
\beq
a(s,b_{t})\,\,=\,\,\frac{1}{2 \pi}\,\,\int\,\,d^2 q_{t} \,\,
e^{- i {\vec{q}}_{t}\,\cdot\,{\vec{b}}_{t}}\,\,f(s,t = - q^2_{t}) \, .
\eeq
In this representation
\beq
\s_{tot}\,\,=\,\,2\,\,\int \,d^2 b_{t} \,\,Im\,\,a (s, b_{t} ) ,
\label{st}
\eeq
\beq
\s_{el}\,\,=\,\,\int \,d^2 b_{t} \,\,|\,a (s, b_{t} )|^2 .
\label{sel}
\eeq
 
The normalization of the nucleus wave function $\Psi_{A} ( r_1,...r_i,r_A)$ is 
\beq
\int \Psi_{A} ( r_1,...r_i,r_A) \Psi_{A}^{*} ( r_1,...r_i,r_A) 
\prod_{i=1}^{A}d^{3} r_{i} = A \,\, , 
\eeq
where $A$ is the number of nucleons in the nucleus, 
and the nucleon form factor in the nucleus is defined as
\beq    
F_{A}(q_{z},b_{t}) = \int \,d \,z_1\,
 e^{i q_{z} z_1} \Psi_{A} (z_1, b_{t}, r_2,...r_i,r_A) 
\Psi_{A}^{*} (z_1,b_{t}, r_2,...r_i,r_A) \prod_{i=2}^{A}d^{3} r_{i} \, .
\label{fadef}
\eeq

Throughout this paper we use the Gaussian parameterization for
$F_{A}(q_{z},b_{t}) $, namely
\beq
F_{A}(q_{z},b_{t}) = \frac{A}{\pi R_{A}^{2}} e^{-\frac{b_{t}^{2}}{R_{A}^{2}}
-\frac{R_{A}^{2}}{4} q_{z}^2} \, ,
\label{fapar}
\eeq
where the mean radius of the nucleus  $R_{A}^{2}$ is equal to
\beq
R_{A}^{2}=\frac{2}{5} R_{WS}^{2} \, \, ,
\eeq
and $R_{WS}^{2}$ is the size of the nucleus in the Wood-Saxon parameterization\cite{r33}, 
which we choose $R_{WS}^{2}= r_{0} A^{1/3}$ with $r_{0}=1.3 fm$. 

In nonrelativistic theory for the nucleus we can neglect the change of energy
for the recoil nucleon. Its energy is $E_{p'}= m + q^{2}/2m$ in the rest 
frame of the nucleus and $q^2 /2m \ll q_{z}$.

\subsection{Passage of the $\bar q q$ ($G G$) pair through the 
target}

The idea how to write the Glauber formula in QCD was originally formulated in 
Ref.\cite{r6} and it has been carefully developed in Ref.\cite{r7}. 
It is easier to explain
the main idea considering the penetration of quark anti-quark pair,
produced by the virtual photon, through the
target. While the boson projectile is traversing the target, the distance
$r_{t}$ between the quark and anti-quark can vary by amount $\D r_{t} 
\propto R_{A} \frac{k_{t}}{E}$ where $E$ denotes the energy of the pair in the
target rest frame and $R_{A}$ is the size of the target (see Fig.\ref{Fig.2}).
\begin{figure}[phtb]
\centerline{\psfig{figure=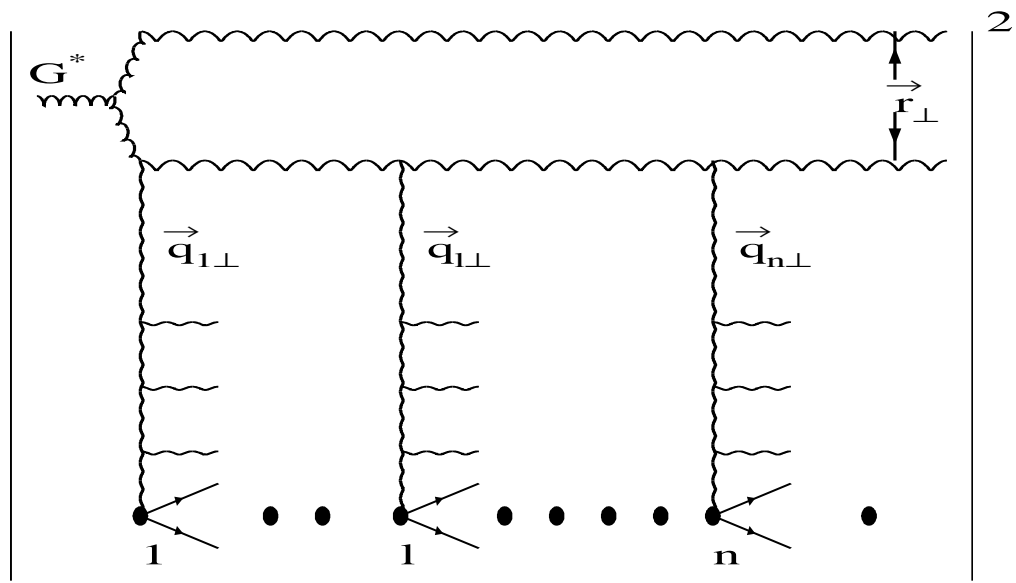,width=160mm,height=160mm}}
\caption{\em The structure of parton cascade in the Glauber ( Mueller )
 formula. }
\label{Fig.3}
\end{figure}

The quark transverse momentum is $k_{t} \propto 1/r_{t}$. 
Therefore
\begin{equation}
\Delta \,r_t\, \propto \,R\,\,\frac{k_t}{E}\,\,\ll\,\,r_t \, ,
\end{equation}
and is valid if
\begin{equation}
\,r^2_t \, s \,\,\gg\,\,\,2\,m\,R \, ,
\end{equation}
where $s=2mE$. In terms of Bjorken $x$, the above condition looks as 
follows
\begin{equation}
x\,\,\ll\,\,\frac{1}{2\,m\,R} \, .
\label{16}
\end{equation}
 Therefore the transverse distance between quark and antiquark is a  good 
degree of freedom \cite{r7}\cite{r8}\cite{r9}. As has been shown 
by A.Mueller, 
not only quark - antiquark pairs can be considered in such way. 
The propagation of a gluon through the target can be treated in a similar
way as the interaction of gluon - gluon pair with definite transverse
separation $r_t$ with the target. It is easy to understand if we remember
 that virtual colorless graviton or Higgs boson is a probe of the gluon
 density. 

The total cross section of the absorption
of gluon($G^*$) with virtuality $Q^2$ and Bjorken $x$ can be written 
in the form:
\beq \label{sig1}
\sigma^A_{tot}(\,G^*\,)\,\,=\,\,
\eeq
$$
\int^1_0 d z \,\,\int \,\frac{d^2 r_t}{2 \pi}\,\,
\int\frac{d^2 b_t}{2 \pi}
\Psi^{G^*}_{\perp} (Q^2, r_t,x,z)\,\,
 2 \{1\,\,-\,\,exp[ -\,\h \sigma(r^2_t)\,\,
S(b^2_t)\, ]\,\}\,\,[{\Psi^{G^*}_{\perp}} (Q^2, r_t,x,z)]^* \, ,
$$
where $z$ is the fraction of energy 
which is carried by the gluon, $\Psi^{G^*}_{\perp}$ is the wave 
function of the transverse polarized gluon and $\sigma(r^2_t)$ is the
cross section of the interaction of the pair with transverse separation
$r_{t}$ with the nucleon, and $S(b_{t})$ is the profile function of a nucleus 
which we will specify later.

The physical interpretation of Eq.$(17)$ is very simple, if one 
notices that the factor in curly brackets is the total cross section
for the $GG$ pair with transverse separation $r_{t}$ passing through the
nucleus 
\beq
\sigma^{SC}_{tot}(r_{t}) = 2 \int\frac{d^2 b_t}{ \pi}
\{1\,\,-\,\,e^{-\,\h \sigma_{N}(r^2_t)\,\, S(b^2_t)\, }\,\} \, .
\label{sigsc}
\eeq
 
Indeed, the above formula is a solution of the $s$-channel unitarity 
relation:
\beq
2\,\,Im\,a(s,b_{t})\,\,=\,\,|a(s,b_{t})|^2 \,\,+\,\,G_{in}(s,b_{t})\, ,
\label{ima}
\eeq
where $a$ denotes the elastic amplitude for the $GG $ pair with a 
transverse separation
 $r_{t}$, and $G_{in}$  is the contribution of all the 
inelastic processes. The inelastic cross section is equal to
\beq
\s_{in}\,\,=\,\,\int \,\,d^2 \,b_{t}\,\,G_{in} ( s, b_{t})
\,\,=\,\,\int d^2 b_{t} \,\,\( \,\,1\,\,-\,\,
e^{-\,\,\s_{N}(r_{t},q^2_{t} = 0 )\,\,S(b^2_{t})}\,\,\) \, .
\label{sin}
\eeq
We assume that
the form of the final state is a uniform parton distribution that
follows from the QCD evolution equations. Note that we neglect
the contribution of all
diffraction dissociation processes to the inelastic final state
(in particular to the ``fan" diagrams which give an  important
contribution),
as well as  diffraction dissociation in the region of small
masses, which cannot
 be presented as a decomposition of the $GG$ wave function.
We  evaluate this input  hypothesis in section \ref{cgf}.
 
In the language of Feynman diagrams, Eq.(\ref{sigsc}) 
sums all diagrams of Fig.\ref{Fig.2} in which
 the $GG $ pair rescatters with the target, and exchanges
 ``ladder" diagrams each of which corresponds to the
gluon density.
This sum has already been performed by Mueller\cite{r7},
and we will only comment
 how one can obtain the result, without going into details.
 We borrow the presentation of the end of this subsection as well as the next
two ones borrow  Ref.\cite{GLMV}.
 
The  simplest way is to consider the inelastic cross section
(see Refs.\cite{r7,r1,r11}), which has the direct
 interpretation through the parton wave function of the hadron (see
 Fig.\ref{Fig.3}), as all partons that are produced  are on the mass shell.
In leading $ln{1}/{x}$ approximation we have two
orderings in time:
\newline
\newline 
1. The time of emission of each ``ladder" by the fast $GG $ pair,
which should obey the obvious ordering for $n $ produced ``ladders"
(see Fig.\ref{Fig.3}) 
\beq
t_1\,\,>\,\,t_2\,\,...\,\,>\,\,t_{i}\,\,>\,\,t_{i + 1}\,\,...>\,\,
t_{n} \, ;
\eeq
\newline
\newline 
2. Each additional ``ladder"  which should live for a shorter time than the
previous one. This gives a second ordering (see Fig.\ref{Fig.3}):
\beq
t_1 - t'_1\,\,>\,\,t_2 - t'_2\,\,>\,\,...\,\,> \,\,t_i - t'_i\,\,>\,\,...
\,\,>\,\,t_n - t'_n \, ;
\eeq
Each ``ladder" in the leading log approximation is the same
 function of $t_i - t'_i$ which we denote 
as $\s(t - t',r^2_{t})$. This
fact allows us to carry out the integration over $t_i$ and $t_i - t'_i$,
which gives
for $n$ emitted cascades:
\beq
\s_n \,\,=\,\,\frac{1}{(n!)^2} \int^t d t_1 \int^{t - t'} d ( t_1 - t'_1) \,
\,\s^n(t_1 - t'_1,r^2_{t}) \, .
\eeq
For the case of the $GG $ pair,  both last integrations are not
logarithmic, and in the leading log approximation
we can safely replace the above
integral by
\beq
\s_n\,\,=\,\,\,\frac{1}{(n!)^2}  \,
\,\s^n(t - t',r^2_{t}) \, ,
\eeq
where $t - t'$ is of the order of $\frac{1}{q_{\parallel}}$ due to uncertainty
 principle, where
 $q_{\parallel}\,\,=\,\,\frac{Q^2}{s}$.
One $n!$ is compensated by the number of possible
diagrams, since the order of the "ladders" are not fixed.
 Therefore, the contribution of
the $n$th ``ladder" exchange gives
\beq
\s_n\,\,=\,\,\,\frac{1}{n!}  \,
\,\s^n( x,r^2_{t}) \, .
\eeq
Applying the AGK cutting rules\cite{r11}  we reconstruct the total cross
section which results in Eq.(\ref{sigsc}).

\subsection{$\s(r_{t},q^2_{t})$ at $q_{t}$ = 0.}

The expression for
$\s(r_{t},q^2_{t})$ at $q_{t}=0.$ was first written
down in Ref.\cite{r6} 
( see Eq.(8) of this paper). It turns out that $\s$ can be
expressed through  the unintegrated parton function 
 $\phi$ first introduced  in the BFKL papers\cite{r12} 
and widely used in Ref.\cite{r1}. The relation of this
function to the Feynman diagrams and the gluon density can
be calculated using the following equation:
\beq
 \as (Q^2) x G(x, Q^2)\,\,=\,\,\int^{Q^2} \,\,d l^2_{t}\,\,
\as(l^2_{t})\,\,
\phi(x,l^2_{t})
\eeq
Using the above equation we reproduce the results of Ref.\cite{r6}, 
which reads
\beq
\s(r_{t},q^2_{t})\,\,=\,\,\frac{16 C_F  }{N^2_c - 1}\,\,\pi^2
\int \phi(x,l^2_{t})\,\,\{\,\,1\,\,-\,\,e^{i\,\vec{l}_{t}
\vec{r}_{t}}\,\,\}\,\,\frac{\as(l^2_{t})}{2 \pi}\,
\frac{d^2 l_{t}}{l^2_{t}}
\eeq
where $\phi \,\,=\,\,\frac{\pa x G (x,Q^2)}{\pa  Q^2}$.
 
We evaluate this integral using Eqs.(\ref{gdef}),(\ref{st}) and (\ref{sel})
and integrate over the azimuthal angle.
  Introducing a new variable $\xi = r_{t} l_{t}$, 
the integral can be written in  the form:
\beq
\s(r_{t},q^2_{t})\,\,=\,\,\frac{16 C_F \as }{N^2_c - 1}\,\,\pi^2
\int_C \frac{d \o}{ 2 \pi i} \,x^{-\o} \,g(\o)\,\,\g(\o)\,\,(r^2_{t})^{1 - \g(\o)}
\int^{\infty}_{0} d \,\,\xi\,\, \frac{1 - J_0(\xi)}{ (\xi)^{3 - 2 \g(\o)}}
\eeq
 Evaluating the integral over $\xi$ (see Ref.\cite{r20} {\bf 11.4.18})
we have:
\beq
\s(r_{t},q^2_{t})\,\,=\,\,\frac{8 C_F \as }{N^2_c - 1}\,\,\pi^2
\int_C \frac{d \o}{ 2 \pi i} \,x^{-\o}\,g(\o)\,\,\g(\o) \,\,
(\frac{r^2_{t}}{4})^{1 - \g(\o)}\,\,\,\frac{\G(\g(\o)) \,\,
\G(1 - \g(\o))}{(\,\G(2 - \g(\o)\,))^2}
\eeq
In the double log approximation of pQCD where $\g(\o)\,\,\ll\,\,1$
 the cross section  for $N_c$ = 3 reads:
\beq
\s(r_{t},q^2_{t})\,\,=\,\,\frac{\as(\frac{4}{r^2_{t}})}{3}
\,\,\pi^2\,\,r^2_{t}\,\,
\(\,\, x G^{GLAP}(x, \frac{4}{r^2_{t}})\,\,\)
\label{srt}
\eeq
This result coincides with the value of the
cross section given in Refs.\cite{r14,r15} 
(if we neglect the factor 4 in the argument of the gluon
density). We checked that Eq.(2.16) of Ref.\cite{r16} also
leads to the same answer, unlike the value for $\s$ quoted
 in Ref.[16] (see Eq.(2.20)).

In the case of the passage of the $GG$ pair, Eq.(\ref{srt})
reads (for $N_{c}=3$)
\beq
\s (r_{t},q^{2}_{t}=0)=\frac{3 \as (\frac{4}{r^{2}_{t}})}{4} \pi^{2}
r^{2}_{t} \left( x G_{N}^{GLAP} (x, \frac{4}{r^{2}_{t}})\right)
\label{srt0}
\eeq
Eqs (\ref{srt}) and (\ref{srt0}) are valid only on DLA of pQCD because
we obtained these expressions considering $l_{i t} \ll k_{t}$ and 
neglecting the longitudinal part of the momentum $l_{i}$ 
($l_{i \parallel} \ll l_{i t}$). It is very essential to 
realize this limitation  of our approach for all further applications.

\subsection{The $b_{t}$ dependence of the scattering amplitude with a nucleon}

To deal with the SC we need to know the amplitude not only at $q_{t}=0$,
but at all values of momentum transfer, so that we can calculate the
profile function of the amplitude in impact parameter space. The gluon 
density in a nucleon depends only weakly on $q_{t}$ in the GLAP evolution
(see Ref.\cite{r1} for details).  Therefore, all the $q_{t}$-dependence
 comes from the form factor of the $G_{1} G_{2}$ ($\bar q q$) pair with the
 transverse
separation $r_{t}$ and the form factor of the target nucleus.

The form factor for $G_{1} G_{2}$ ($\bar q q$) pair with 
the transverse separation $r_t$ is equal to
\beq
F_{G_{1} G_{2}} (q^{2}_{t}) = \Psi^{i}_{GG} \left(\frac{(\vec{k_1}_t 
-\vec{k_2}_t)}{2} \cdot \vec{r}_t \right)  
{\Psi^{f}_{GG}}^{*} \left(\frac{({\vec{k_1}_t}^{\prime} 
-{\vec{k_2}_t}^{\prime})}{2} \cdot \vec{r}_t \right) 
\label{fgg}
\eeq
where $k_{i}$ ($k_{i}^{\prime}$) denotes the momentum of the gluon `$i$'
before and after collision.
Each of the wave functions is exponential, and a simple sum of 
different attachment  of gluons lines gives
\beq
F_{G_{1} G_{2}} (q^{2}_{t}) = e^{i \frac{\vec{q}_{t} \cdot \vec{r}{t}}{2}}
\left(1 - e^{i \vec{l}_t \cdot \vec{r}_t  }\right)
\label{fgg2}
\eeq

We have absorbed the last factor in the expression for the cross section, 
while the transform gives $q_t$ dependence of $F_{G_{1} G_{2}}$. After
integration over the azimuthal angle it has the form
\beq
F_{G_{1} G_{2}} (q^{2}_{t}) = J_{0}\left( \frac{q_t r_t}{2} \right)
\label{fgg3}
\eeq

The target form factor cannot be treated theoretically in pQCD. In our
problem it consists of the nuclear form factor and the nucleon distribution
in the nucleus. For our purpose the phenomenological exponential 
parameterization for the nucleon form factor will suffice, namely
\beq
F_N (q_{t}^2) = e^{-\frac{B}{4} q_{t}^{2}}
\label{fn}
\eeq
with  the slope $B=10$ Gev$^{-2}$\cite{r17} , extracted from experimental data on
hadron - hadron collisions, if we put the Pomeron slope 
$\alpha^{\prime}= 0$ \cite{r17}. For nucleon distribution in a nucleus we use 
Eq.(\ref{fapar}). Finally, the resulting $b_t$ distribution looks as follows
\beq
S(b_{t}^{2}) = \int S_N \left( (b_{t}-b_{t}^{\prime})^2 \right)
F_A \left( q_z , b_{t}^{\prime} \right) \frac{d^{2} b_{t}^{\prime}}{\pi}
\label{sbt1}
\eeq
where
\bea
S_N (b_{t}^{2}) = \frac{1}{4 \pi^2} \int d^2 q_{t} e^{i \vec{b}_{t} \cdot
\vec{q}_{t}}   F_{G_{1} G_{2}} (q_{t}^{2}) F_{N} (q_{t}^{2})   \nonumber \\
= \frac{1}{\pi B} I_{0} \left( \frac{b_t r_t}{B} \right) 
e^{- \frac{b_{t}^{2} + \frac{r_{t}^{2}}{4}}{B}}
\label{sn1}
\eea

Considering $R^{2}_{A} \gg B \gg r^2_t$ one notice that it can be safely 
neglected the  $b_t$-dependence in gluon nucleon interaction. Indeed, $S_N (b_t)$
is a steep function of $b_t$ in comparison with $F_A (q_z ,b_t)$. It means 
that we can make the integral over $b_{t}^{\prime}$ in Eq. (\ref{sbt1}) putting
$b_{t}^{\prime}= b_t$ in $F_A (q_z ,b_{t}^{\prime})$. The result is
\beq
S_{A}(b_t) = F_N (q_z ,b_t)
\label{san}
\eeq

\subsection{The $q_z$ dependence of nucleon density (gluon lifetime cutoff).}

To calculate the value of $q_z$ we have to consider the process of diffractive
dissociation pictured in Fig.(\ref{fig5}). This process is the AGK cut with
small multiplicity of produced hadron of the first diagram 
for the SC \cite{r11}.
We can find $q_z$ from the obvious equation
\beq
( Q + q)^2 = M^2
\eeq
which gives
\beq 
q_z = \frac{M^2 + Q^2}{2 Q_0} = \frac{M^2}{2 Q_0} + m x
\label{qz1}
\eeq
since $Q_z = Q_0 - \frac{Q^2}{2Q_z} \,\, \longrightarrow Q_0$
 
Now, we have to calculate $M^2$ through $k_t$ and the fraction of energy $z$ 
(see Fig.\ref{fig5}). Using the technique of ref.\cite{r17}, one obtains
\beq
M^2 = (k_1 + k_2)^2 = \frac{k_t^2}{ z} + \frac{k_t^2}{(1 - z)}
= \frac{k_t^2}{z (1-z)}
\label{m2}
\eeq

To calculate $z$ we have to consider the gluon structure function that
enters in the value of $\s_N (r_t^2)$ (Eq.(\ref{srt})). Indeed, using
$( k_2^{\prime} - l)^2 =0$ (see Fig.(\ref{fig5})) we have
\beq
(k_2^{\prime} - l )^2 = k_2^{\prime 2} - 2(k_{\mu}^{\prime} l^{\mu}) + l^2=0
\label{k2}
\eeq
since $l^2 \ll k_2^{\prime 2}$, we have from Eq.(\ref{k2})
\bea
k_2^{\prime 2}= - k_t^2 = 2 \,(k_{\mu} l^{\mu}) = 2 \,\left(k_0 l_0 -
( k_0 + \frac{k_t^2}{2 k_0})(l_0 + \frac{l^2}{2 l_0}) \right) \nonumber
\\ = -k_0 \frac{l^2}{2 l_0} =- 2 z Q_0 x^{\prime} m
\label{k22}
\eea
which gives
\beq
z=\frac{k_t^2}{2 x^{\prime} Q_0 m}
\eeq
Substituting $z$ in the expression for $M^2$ we have $M^2= 2x^{\prime}
m Q_0$. Finally 
\beq
q_z^2= (x^{\prime} + x)^2 m^2
\label{qz2}
\eeq 
where $x^{\prime} > x$.

Eq.(\ref{qz2}) has to be  substituted in Eq.(\ref{fapar}) to obtain the dependence
of the nucleon distribution in a nucleus on $q_z$. The physical meaning of this 
$q_z$ dependence is very simple. The life time of the virtual gluon is equal
to
\beq  \label{tau}
\tau = \frac{Q_0}{Q^2} = \frac{1}{2 m x}
\eeq

If $\tau$ is smaller then the size of the nucleus $\tau < R_A$ the gluon cannot
interact with all nucleons in a nucleus. The number of possible collisions
is of the order of
\beq \label{nucol}
\rho \, \tau = \frac{\rho}{2 m x} 
\eeq
where $\rho$ is the density of nucleons in a nucleus. Eq.(\ref{fapar})
and Eq.(\ref{qz2}) give us a practical way to introduce the finite life time of
the virtual gluon in our calculation.

\subsection{The wave function of virtual gluon.}

\begin{figure}[hptb]
\centerline{\psfig{figure=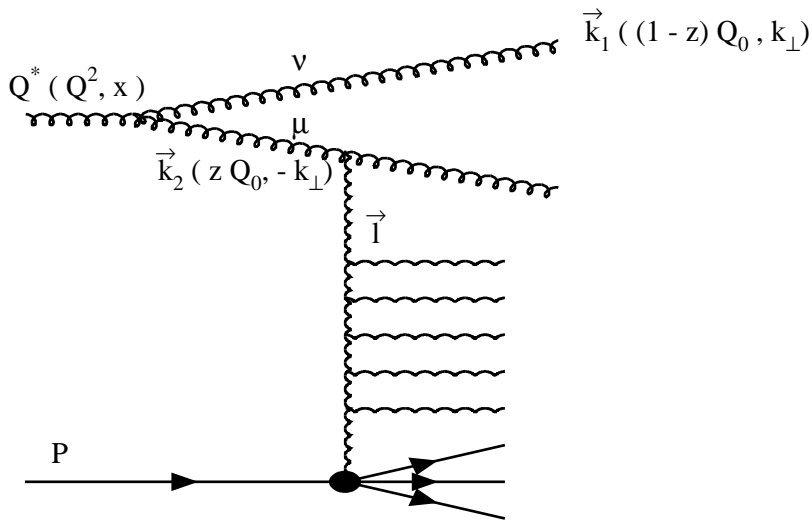,height=100mm}}
\caption{\em The vertex of a virtual colorless probe
 ( $Q^*$ ) to two gluons. }
\label{fig4}
\end{figure}

Here, we are going to discuss the last ingredient of the master formula
of Eq.(\ref{sigsc}), namely, the wave function of initial gluon with 
virtuality  $Q^2$. Basically, it was done by A. Mueller in Ref.\cite{r7},
 however, we will discuss it in this section to 
clarify the approximations that we have
to do to get Mueller's answer. It is easier to discuss the
colorless probe with virtuality $Q^2$ that interacts with gluons 
than the virtual gluon.  There is a
number of such probes, for example, the graviton or Higgs meson. We 
do not need to specify what particle we use as a probe. We only need to write
down the momentum structure for the meson-two gluon vertex. For a scalar 
particle such vertex has the following structure
\beq
\G_{\mu \nu} = g \left( (k_1 k_2) g_{\mu \nu} - k_{1\mu} k_{2\nu}   \right) \, ,
\label{fver1}
\eeq
where all notation is clear from Fig.(\ref{fig4}) (see, for example Ref.\cite{r19} for the vertex of Higgs meson to two gluons).

We intent to use a frame in which the nucleon with momentum $p$ is essentially
at rest ($p_{+} \ll Q_{+}$) and where 
\beq
Q=(Q_{+},Q_{-}, Q_t) =( Q_{+}, -\frac{Q^2}{Q_{-}},\mbox{\boldmath{0}})\, .
\eeq

To find the wave function we can use the technique developed in Ref.\cite{r18}. 
however, before doing so we need to specify the polarization of gluons that
work in our case. The problem is that we know that the gluon which interacts
with the target in Fig.(\ref{fig4}) ($k_2$ in our case) has the longitudinal
polarization at high energy (small $x$ ), while
 the second gluon indeed has been produced on mass shell and has only
transversal polarization (see, for example, Ref.\cite{r1} for a
detailed discussion). It means that we cannot treat the first gluon as a real
particle on the mass shell, where gluons have only transverse polarization.
Therefore, before introducing the wave function of the probe we have to make
use of the polarization vectors and the Weisszaker-Williams transform from 
longitudinal polarized gluon to transverse one. Indeed 
$k_{(i)}^{\mu} . e^{(i)}_{\mu}=0$,
where  $ e^{(i)}_{\mu}$ is the polarization vector of the `$i-th$' gluon.
Therefore
\beq 
k_{(i)}^{\parallel}\cdot e^{(i)}_{\parallel} 
= -\vec{k}_{(i)}^{t} \cdot \vec{e}^{(i)}_{t} 
\label{ke}
\eeq
Since only components along $Q_+$ of vectors $k_{(1)}^{\parallel}$  and
$k_{(2)}^{\parallel}$ are big, we obtain
\bea
z_i Q_+ e^{(i)}_{-}=  - \vec{k}_{(i)}^{t} \cdot \vec{e}^{(i)}_{t} \nonumber
\eea
or
\beq
Q_+ e^{(i)}_{-} =  -\frac{\vec{k}_{(i)}^{t} \cdot  \vec{e}^{(i)}_{t}}{z_i}  
\label{qe}
\eeq 
Finally using Eq.(\ref{qe}) we can rewrite the vertex $\G_{\mu \nu}$ in the
form
\beq
\G_{\mu \nu} = g \frac{k^{t}_{\mu} k^{t}_{\nu}}{z (1-z)}
\label{fver2}
\eeq
where we sum over two possibilities: gluon $2$ interacts with the target and
gluon $1$  interacts with the target.

We anticipate that sufficiently small transverse distances($r_t$) will 
contribute to the processes ($ r_{soft} \gg r_t \gg \frac{1}{Q}$). In this
case all interactions between gluon 1 and 2 before the interaction with the
target are small since they are of the order of $\as (r^2_t)$ and
$\as (r_t) \ll 1$. Therefore we have to calculate only the contribution of
two gluon state to the wave function and using the technique of Ref.\cite{r18}
and Eq.(\ref{fver2}) for $\G_{\mu \nu}$ we have
\bea
\Psi_{\mu \nu} (z, k_t) &=& \frac{g  k^{t}_{\mu} k^{t}_{\nu}}{\sqrt{k_{+ 1} k_{+ 2}}
\left( Q_{-} - \frac{k_{t}^{2}}{k_{1 +}} - \frac{k_{t}^{2}}{k_{2 -}}\right)}
\nonumber \\
& \cdot &\frac{1}{z (1-z)}  \label{psi1}  \\
&=& \frac{g  k^{t}_{\mu} k^{t}_{\nu}}{\sqrt{z (1-z)} 
\left(Q^2 z(1-z) + k^2_t \right)}  \nonumber
\eea
Going to $r_t$- representation one obtains
\beq
\Psi_{\mu \nu} ( r_t, z) = - \grad_{\mu} \grad_{\nu} K_{0} (a r) 
\frac{1}{\sqrt{z (1-z)}}  
\label{psi2}
\eeq
where $a^2 = Q^2 z (1-z)$ and $K_{0} (a r )$ is the McDonald function. Making
use of the properties of the McDonald functions we derive
\beq
\Psi_{\mu \nu} = \frac{ 1}{\sqrt{z (1-z)}} 
\left\{ a^2 K_2 ( a r) \frac{ r_{t \mu}
r_{t \nu}}{r^2} - \frac{a K_{1} ( a r ) \delta_{\mu \nu}}{r} \right\}
\label{psi3}
\eeq
where $\mu, \, \nu \, = \, 1,\,  2$.

\subsection{The Mueller formula.}

Now we have all ingredients that we need to derive the Glauber formula for
the percolation of $GG$ pair through the nucleus. Using the well known 
relationship between cross section and the deep inelastic structure function
we can derive the following formula for $x G_{A} (x, Q^2)$, for $N_c = 3$ 
( see Ref.\cite{r7})
\beq
x G_{A}(x,Q^2) = \frac{2}{\pi^2} \int_{0}^{1} d z \int \frac{d^2 r_t}{\pi}
\int \frac{d^2 b_t}{\pi} | \Psi(r_t,z)|^2 2 
\left\{ 1 - e^{\hm \s_{N}^{GG} ( x^{\prime}, r^2_t ) S(b^2_t) } \right\}
\label{xga1}
\eeq
where $ x^{\prime} = \frac{1}{r^2_t z Q_0}$.
To specify the region of integration we have to put in Eq.(\ref{xga1}) the
wave function of Eq.(\ref{psi3})  
\bea
|\Psi (r_t,z)|^2 = \sum_{\mu \nu} \Psi (r_t,z) \Psi^* (r_t,z) =
\nonumber \\ 
 \frac{ 1}{z (1-z)} \cdot
\left\{ \left( a^2 K_2 (a r )  - \frac{a K_1 (a r )}{r} \right)^2 +
\frac{1}{r^2} ( a K_1 ( ar ) )^2 \right\}
\label{psisq}
\eea
where $\sum_{\mu \nu}$ represents the sum over polarization of the gluon.
The main contribution in Eq.(\ref{xga1}) comes from the region of small
$z$ ($ a \,r_t \ll 1$), where 
\beq
|\Psi (r_t,z)|^2 = \frac{2}{z r_{t}^{4}}
\label{psiap}
\eeq
which can be easily derived from the expansion of McDonald function at 
$ a r_t \ll 1$. In all terms of the expansion, except the first one, the
integral over $z$ is convergent.

The condition of $ a \, r_t \ll 1$ means that 
\beq
z (1\, - \,z ) \,< \,\frac{1}{Q^2 r_{t}^{2}} \,<\,\frac{1}{4}
\label{z1}
\eeq
Introducing the new variable
\beq
x'= \frac{1}{2 z Q_0 r^{2}_{t} m }
\label{xp1}
\eeq
instead of $z$, one sees that Eq.(\ref{z1}) can be rewritten in the form
\beq
x' > \frac{Q^2}{2 m Q_0} = x_{Bj} = x
\label{xp2}
\eeq
Substituting Eqs.(\ref{psiap}) and (\ref{z1}) in Eq.(\ref{xga1}), we derive
the Mueller's formula
\bea
x G_A(x,Q^2) = \frac{4}{\pi^2} \int_{x}^{1} \frac{d x'}{x'} 
\int_{\frac{4}{Q^2}}^{\infty} \frac{d^2 r_t}{\pi r_{t}^{4}} 
\int_{0}^{\infty} \frac{d^2 b_t}{\pi}  2
\left\{ 1 - e^{\hm \s_{N}^{GG} ( x^{\prime},r^2_t ) S(b^2_t) } \right\}
\label{xga2}
\eea
The lower limit in $r_t$ integration comes from 
\eq{z1}.
 
It is easy to see that the first term in the expansion of Eq.(\ref{xga2})
with respect to $\s$ gives the GLAP equation in the region of small $x$.
Using Gaussian parameterization for $S(b_t)$ ( see Eq.(\ref{fapar})) we can
take the integral over $b_t$ and obtain the answer ($N_c = N_f = 3$)
\bea
x G_A(x,Q^2) = \frac{2 R_{A}^{2}}{\pi^2} \int_{x}^{1} \frac{d x'}{x'} 
\int^{\frac{1}{Q^2_0}}_{\frac{1}{Q^2}}  \frac{d r_t^2}{ r_{t}^{4}} 
\left\{ C + ln(\kappa_{G} ( x', r_{t}^{2})) + 
E_1 (\kappa_{G} ( x', r_{t}^{2}))  \right\}
\label{xga3}
\eea
where $C$ is the Euler constant and $E_1$ is the exponential integral
(see Ref.\cite{r20} Eq. {\bf 5.7.11}) and
\beq
\kappa_{G} ( x', r_{t}^{2}) = \frac{3 \as A \pi r^2_t}{2 R_{A}^{2}}
x' G_{N}^{GLAP} (x', \frac{1}{r^2_t} )
\label{kapa}
\eeq

To understand the physical meaning of this equation it is  instructive
to write down the evolution equation for the gluon density.
Indeed,
\beq
\frac{\pa^2 x G(x,Q^2)}{\pa \ln(1/x) \,\pa \ln Q^2}\,\,=
\label{dxga1}
\eeq
\bea
\frac{N_C \as}{\pi}\,\,x \,G^{GLAP}(x,Q^2)\,\,+\,\,\frac{2}{\pi^2 }\,
\sum_{k=1} \frac{(- 1)^k}{k k!}\,\,
\frac{1}{( R^2_A Q^2)^{ k}}\,\,
\( \frac{\pi N_c \,A \as x G^{GLAP} (x,Q^2)}{2}\)^{k + 1}
\nonumber
\eea
The first term corresponds to the usual GLAP equations, while the second
one takes into account the SC.
 It should be stressed that the term with $k$ = 1 (if
treated as an equation \cite{r1}),
is the same term that appears
 in the nonlinear GLR equation which sums the ``fan" diagrams.
 This term has been calculated using quite
a different technique \cite{r21,r22}.
The coefficient in front of the other terms reflects the fact that all
 correlations between the gluons have been neglected, despite the fact
that gluons are uniformly distributed in the disc of radius $R_A$.
 
 Mueller's formula is not a nonlinear equation, it is the analogue
of the Glauber formula for the interaction with a nucleus,
which gives us the possibility
 to calculate the shadowing corrections using the solution of the
 GLAP evolution equation. Hence, this formula should be used as an
input to obtain the complete
  effect of the SC,
 for the more complicated evolution equation, such as 
GLR\cite{r1}, or the generalized evolution equations (see Ref.\cite{r23}).
 
\begin{figure}[htbp]
\centerline{\psfig{figure=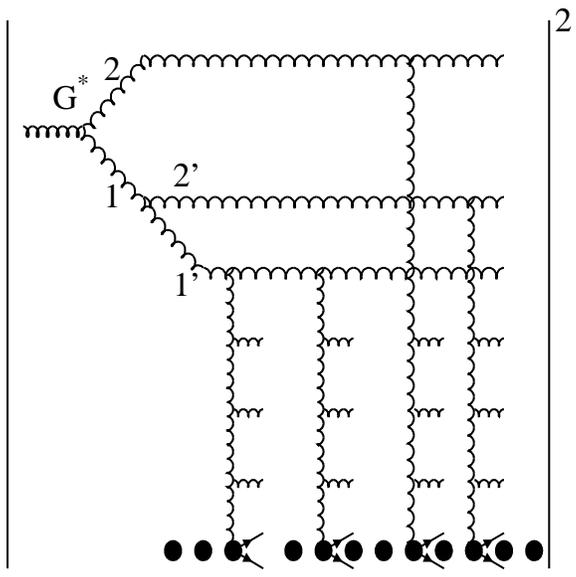,height=100mm}}
\caption{\em  The interaction with nucleons that is not taken into
account in the Glauber ( Mueller) formula. }
\label{fig5}
\end{figure}

Calculating the subsequent iterations of the Mueller's
 formula
gives us a practical way to estimate the value of the SC from more
complicated Feynman diagrams, that have to be taken into account (diagrams 
such as in Fig.\ref{fig5}). Certainly, the iteration of Mueller's 
formula is not the most efficient way
 to calculate the SC correction in the region  of extremely
  small $x$, but  it 
 could give sufficiently reliable results for the HERA kinematic region.

At first sight \eq{dxga1} looks like the Operator Product Expansion.  However,
it should be stressed that the Mueller formula itself  takes into account not
 only the high twist contributions to the deep inelastic structure function but
also changes crucially the leading twist  term since it  contains the
 full  integral over $r_t$.
 
\subsection{An instructive example.}

Before discussing the numerical results of integration of the master 
equation (\ref{xga3}) we will consider one example which gives the
physical picture there originated. 

The first observation is that Mueller formula itself gives a natural cutoff
for the large distance contribution and can be used as a model for the behavior
of the gluon distribution in a target even at very small values of $Q^2$.

Doing all calculations in double log approximation of pQCD ( or in other 
words, using the GLAP evolution equation in the region of small $x$ ), we have
to assume that the anomalous dimension $\g (\o) \ll 1$. It means that we can
confidently take the gluon density independent of the integral at the low 
limit $r^2_t= 4/Q^2$ and only integrate over the factor $r_t^2$ in the 
expression of $\s^{GG}$.
Straightforward integration leads to the answer
\beq
xG_A (x,Q^2) = \frac{N_c^2 -1}{4 \pi^2} \int_{x}^{1} \frac{d x'}{ x'}
R_{A}^{2} Q^2 \left\{ C \,+\, ln(\kappa_G) + (1 + \kappa_G) E_1 (\kappa_G) +1
- e^{-\kappa_G} \right\}
\label{xgaap1}
\eeq
where
\beq
\kappa_G (x')= \frac{3 \as \pi A}{2 Q^2 R^2_A} x' G^{GLAP} (x',Q^2)
\label{kapaap}
\eeq
The last integration over $x'$ has to be done numerically.

One can see from Eq.(\ref{xgaap1}) that the answer mostly depends on $\kappa_G$.
For $\kappa_G \ll 1$ Eq.(\ref{xgaap1}) gives:
\beq
xG_A (x,Q^2) = \frac{3 \as A}{\pi} \int_{x}^{1} \frac{d x'}{ x'}
 x' G_{N}^{GLAP} (x',Q^2) ln\(\frac{Q^2}{Q_{0}^{2}(x')}\)
\label{xgaap2}
\eeq
where $ Q_{0}^{2}(x')$ is the solution of the equation $\kappa_G = 1$.

\begin{figure}[htbp]
\centerline{\psfig{figure=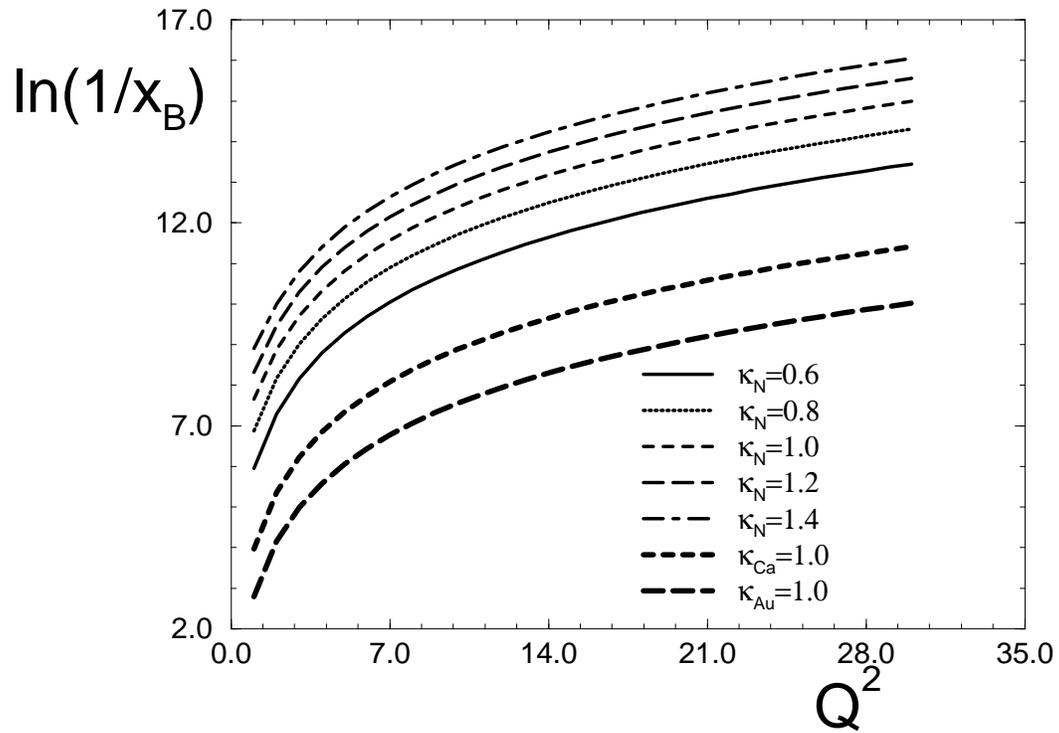,width=140mm}}
\caption{\em The contours of $\kappa$ for Nucleon, Ca and Au. }
\label{kap}
\end{figure}

This result is very close to GLAP equations. The difference is due to the fact
that one cannot substitute the gluon structure function at low limit of 
integration over $r_t$ for the first term of expansion for 
Eq.(\ref{xga3}). This particular
contribution is of the order of $1/\g (\o)$. It is necessary to improve Eq.
(\ref{xgaap1}) by adding a term
\bea
\D xG_A (x,Q^2) & = & \frac{N_c \as }{\pi}  \int_{x}^{1} 
\frac{d x'}{ x'} 
\int^{Q^2}_{max \{ Q_{0}^{2}, Q_{0}^{2}(x')\} } \frac{ d Q'^{ 2}}{Q'^{2}} 
xG^{GLAP} (x',Q^2)  \nonumber \\
 & - &  \frac{N_c \as }{\pi} \int_{x}^{1} 
\frac{d x'}{ x'} x' G^{GLAP} (x',Q^2) 
ln(\frac{Q^2}{max \{ Q_{0}^{2}, Q_{0}^{2}(x')\} })  
\label{xgaap3}
\eea
where $Q_0^{2}$ is the initial virtuality from which we start the GLAP
 evolution. 
Eq.(\ref{xgaap2}) illustrates the important point that the SC  provides us a
new scale in the evolution which crucially depends in $x'$. It means that the
first correction for the evolution equations can be found, just introducing
this scale, namely
\beq
xG_A (x,Q^2) = \frac{3 \as A}{\pi} \int_{x}^{1} \frac{d x'}{ x'}
\int^{Q^2}_{Q_{0}^{2}(x')} \frac{ d Q'^{ 2}}{Q'^{2}} x' G_{N}^{GLAP} (x',Q^2) 
\label{xgaap4}
\eeq
 In Fig.\ref{kap} are plotted the contours of $\kappa$ for a nucleon target
 ( A = 1 in  \eq{kapaap} )  that give 
an idea in which kinematic region we expect big SC.

The large SC corresponds to $\kappa_G \gg 1$ where Eq.(\ref{xgaap1}) gives
\beq
xG_A (x,Q^2) = \frac{2}{\pi^2} R^{2}_{A} Q^2 \int_{x}^{1} \frac{d x'}{x'}
\left( C + ln(\kappa_{G}) \right)
\label{xgaap5}
\eeq

Sum of Eqs.(\ref{xgaap4}) and (\ref{xgaap5}) gives the answer
\bea
xG_A (x,Q^2) & = & \theta\left( Q^2-Q_{0}^{2}(x)\right) 
\frac{3 \as A}{\pi} \int_{x}^{1} \frac{d x'}{ x'}
\int^{Q^2}_{Q_{0}^{2}(x')} \frac{ d Q'^{ 2}}{Q'^{2}}
x' G_{N}^{GLAP} (x',Q^2)  \nonumber \\
&+& \theta\left(Q_{0}^{2}(x) -Q^2 \right) 
\frac{2}{\pi^2} R^{2}_{A} Q^2 \int_{x}^{1} \frac{d x'}{x'}
\left( C + ln(\kappa_{G}) \right)
\label{xgaap6}
\eea 
\begin{figure}[htbp]
\centerline{\psfig{figure=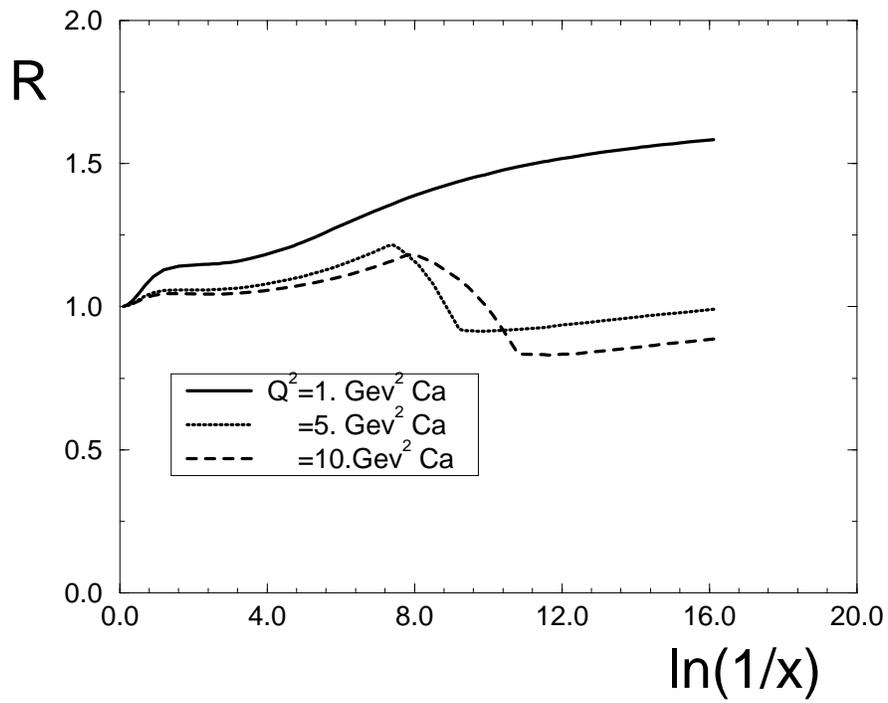,height=100mm}}
\caption{\em Ratio $xG_{A}^{APPROX}$/ $xG_{A}^{FULL}$. }
\label{rap}
\end{figure}

Eq.(\ref{xgaap6}) is very approximate but Fig.(\ref{rap}) shows that this
equations describes the full formula quite well except the region where
$Q^2 \approx Q_{0}^{2} (x)$. The nice feature of this equation is that it 
illustrates in explicit way how the SC  work. First, they provide
the new scale for the transverse momentum inside the parton cascade 
($Q_{0}(x)$). It means that we expect the GLAP evolution only for distances
$r_{t} < 1/Q_{0} (x)$. Second, the SC generate the surface term ( the second
in Eq.(\ref{xgaap6}) ) which gives $\s \propto A^{2/3} $ as for normal hadron
nucleus interaction. Of course, all these properties have been anticipated
(see Refs.\cite{r24,r25,r26}) and Eq.(\ref{xgaap6}) gives the simplest
 example how
they manifest themselves in the case of the deep inelastic scattering. 

Now, let us try to understand $x'$ dependence of our integrand. In our formula
it is
implicitly assumed that each gluon can interact with all nucleons in a nucleus.
It is not the case because each gluon lives a certain time which is
$\tau_{G} = \l /m x'$ where coefficient $\l$ is not quite well known,
but $\l > 1/2$. It means that the gluon can interact only with $T=R^{2}_{A}
\rho \tau_{G}$ nucleons unlike $T$ becomes smaller than $A$. 
Substituting $T$ instead of $A$ we have the following answer for
Eq.(\ref{xgaap4}) 
\beq
xG_A (x,Q^2) = \frac{3 \as}{\pi} \int_{x}^{1} \frac{d x'}{ x'}
T(x') \int^{Q^2}_{Q_{0}^{2}(x')} \frac{ d Q'^{ 2}}{Q'^{2}} x'
 G_{N}^{GLAP} (x',Q^2) 
\label{xgaapt1}
\eeq
at $x' > x_{A} = \l / m R_{A}$ we have no log integration over $x'$. Therefore,
basically, the GLAP equation can be reduced to the form
\beq
xG_A (x,Q^2) = \frac{3 \as A}{\pi} \int_{x}^{x_{A}} \frac{d x'}{ x'}
\int^{Q^2}_{Q_{0}^{2}(x')} \frac{ d Q'^{ 2}}{Q'^{2}} x' G_{N}^{GLAP} (x',Q^2) 
\label{xgaapt2}
\eeq
One can see in Eq.(\ref{xgaapt2}) that a new cutoff in $x$ appears
in the GLAP equation which has been discussed
 two decades ago in Ref.\cite{r27,r28}.

\subsection{ Theory status of the Mueller formula.}

In this section we shall recall the main assumptions that have been made 
to obtain
the Mueller formula.

1. The gluon energy ($x$) should be  high (small) enough to satisfy 
Eqs.(\ref{16}) and $\as ln(1/x) \leq 1$. The last condition means that we 
have to assume
the leading $ln (1/x)$ approximation of perturbative QCD for the nucleon gluon
structure function.

2. The GLAP evolution equations hold in the region of small $x$ or, in other
words, $\as ln(1/r^2_t) \leq 1$. One of the lessons from HERA data is the
fact that the GLAP evolution can describe the experimental data.

3. Only the fastest partons ($GG$ pairs) interact with the target and there are
no correlations (interaction) between partons from the different parton cascades
(see Fig.\ref{Fig.3}).

4. There are no correlations between different nucleons in a nucleus.

5. The average $b_t$ for $GG$ pair-nucleon interaction is much smaller than
$R_A$.

The first assumption allows us to treat successive rescatterings as independent
 and simplifies all formulas reducing the problem to an eikonal picture of the
classical propagation of a relativistic particle with high energy ($ E \gg 
\mu^{-1}$, where $\mu$ is the scattering radius in the nuclear matter) through 
the nucleus. The second one simplifies calculations but we can consider the BFKL evolution \cite{r12} instead of GLAP one. The
third assumption is artifact of the eikonal approach and we shall discuss it
in the following sections. The last two are usual assumptions to treat nucleus scattering. We
have used the specific Gaussian parameterization for $b_t$ dependence.
Also, one can easily generalize our 
formula in more general case, as Wood-Saxon parameterization \cite{r33}.

\section{pQCD calculations from  the Mueller formula.}
\label{numres}
We use the GRV parameterization \cite{r29} for the nucleon gluon distribution,
which describes all available experimental data quite well, including recent
HERA data at low $x_{B_j}$.  Moreover, GRV is suited for our purpose
 because
(i) the initial virtuality for the GLAP evolution is small
 ($Q_0^2 \approx 0.25GeV^2$) and we can discuss the contribution of the 
large distances in MF
having some support from experimental data;
 (ii) in this parameterization the most
essential contribution comes from the region where $\alpha _s ln Q^2 \approx 1$
and $\alpha _s ln 1/ x_{B_j} \approx 1$. This allows the use of the double 
leading log
approximation of pQCD, where the MF is proven \cite{r18}. It should be also
stressed here, that we look at the GRV parameterization as a solution of the GLAP
evolution equations, disregarding how much of the SC has been taken into
 account in this parameterization in the form of the initial gluon distribution.

 However, in spite of the fact that the GLAP evolution in the GRV
 parameterization starts from very low virtuality ( $Q^2_0 \,\sim\,0.25 GeV^2$)
it turns out that the DLA still does not work quite well in the accessible
kinematic region ($Q^2\,> 1 GeV^2, x \,>\,10^{-4}$). On the other hand, our
master equation (see Eq.\ref{xga2}) is proven in  DLA. Willing to develop a
realistic approach in the region of not ultra small $x$ ($x \,>\,10^{-4})$
we have to change our master equation (\ref{xga2}). We suggest to integrate
Eq.(\ref{dxga1}) and substitute the small $x$ kernel for the full GLAP kernel
in the first term of the r.h.s. This procedure gives
\bea
x G_A(x,Q^2)\,\, & = &\,\, x G_A(x, Q^2)(\,\eq{xga2}\,) \,\,+\,\,
A x G^{GRV}_N (x,Q^2)\,\,
\nonumber \\
& - &\,\,A\,\frac{\as N_c}{\pi} \,\int^1_x \,\int^{Q^2}_{Q^2_0} \,\,
\frac{d x'}{x'}\,
\frac{ d Q'^2}{Q'^2} \,x' G^{GRV}_N (x',Q'^2)\,\,.
 \label{FINANS}
\eea
 The above equation includes also $A x G^{GRV}_N (x, Q^2_0)$ as the initial
condition for the gluon distribution and gives $A x G^{GRV}_N (x, Q^2)$ as the
 first term of the expansion with respect to $\kappa_G$. Therefore, this
equation is an attempt to include the full expression for the anomalous
 dimension for the scattering off each nucleon, while we use the DLA to 
take into account all SC. Our hope, which we will confirm by numerical
calculation, is that the SC  are small enough for $x \,>\,10^{-3}$ and
we can be not so careful in the accuracy of their calculation in this kinematic
 region. Going to smaller $x$, the DLA becomes better and \eq{FINANS} tends
 to our master equation (\ref{xga2}).

\subsection{The gluon structure function for nucleon.}

\begin{figure}[htbp]
\centerline{\psfig{figure=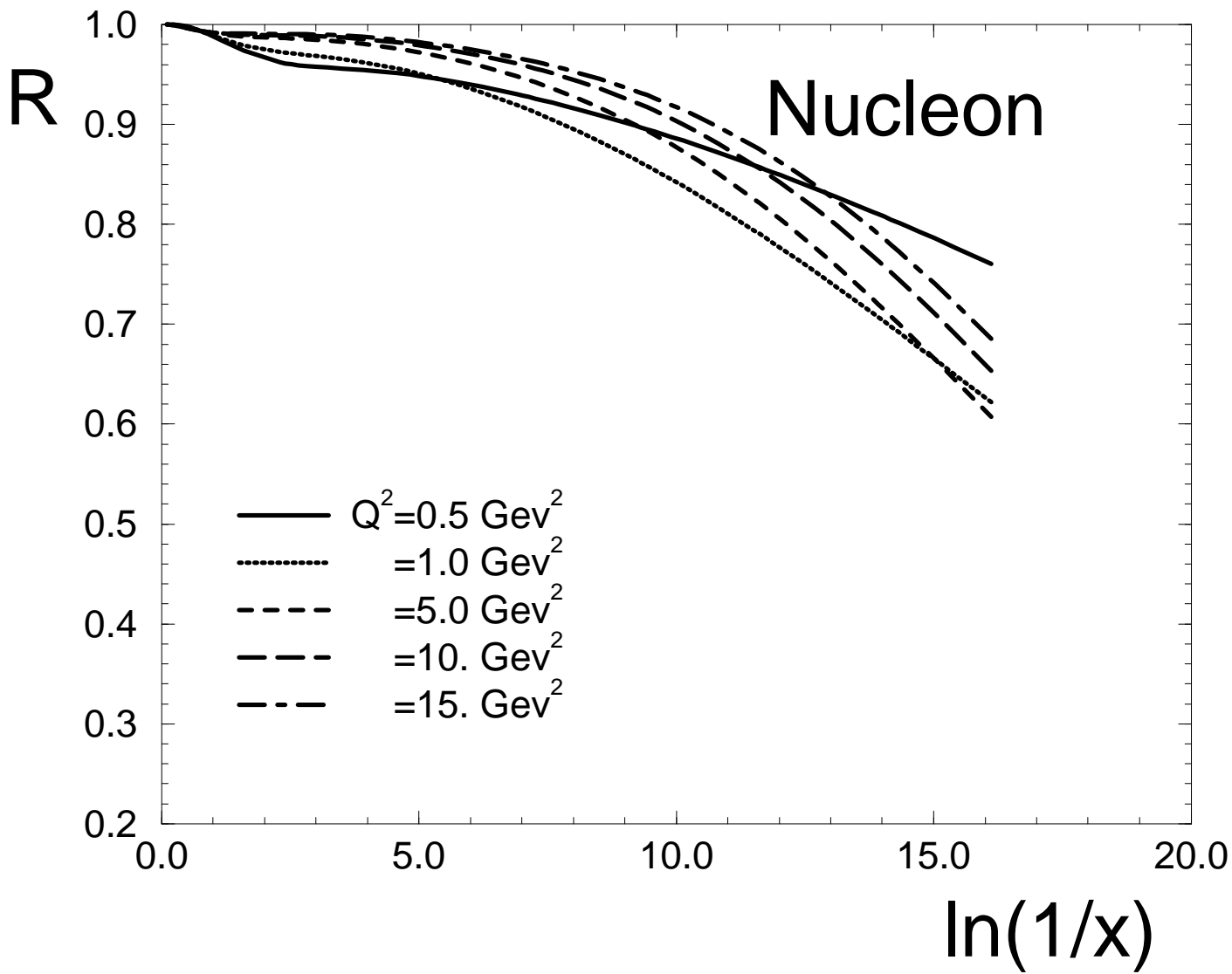,height=160mm}}
\caption{\em The SC  for nucleon (A=1) as a function of $ln(1/x)$ and $Q^2$,
 where ratio $R_1$ compares $xG^A $ with $xG\,\, (GRV)$ distribution. }
\label{r1n}
\end{figure}

In this subsection we are going to check how \eq{FINANS} describes the
gluon structure function for a nucleon, which is our main ingredient in the 
Mueller formula.  We calculate first  the ratio
\beq \label{RN}
R^N_1\,\,=\,\,\frac{xG^A(x,Q^2) (\eq{FINANS})}{x G^{GRV}_N (x,Q^2)}\,\,,
\eeq
for $A=1$, which is shown in Fig.\ref{r1n}. From this ratio we can see the
general behavior of the SC as a function of $ln(1/x)$ and $Q^2$. When compared
to the GRV gluon distribution, the $xG^A$ distribution presents a suppression
 which increases with $ln(1/x)$ and decreases with the virtuality $Q^2$. In the 
region of the HERA data, $3 \, < \, ln(1/x) \, < 10$, and $Q^2 > 2 \, 
GeV^2$\cite{r2}, the
SC are not bigger than $15 \, \%$. The SC give 
a contribution bigger than $20 \, \%$
 only at very small value of $x$, where we have no experimental data.

In the semiclassical approach (see \cite{r1}), the nucleon 
structure function is supposed to have  $Q^2$ and $x$ dependence as 
\beq \label{SEMICLAS}
x G^N(x,Q^2)\,\,\propto\,\,\{Q^2\}^{< \gamma >}\,\,
\{\frac{1}{x}\}^{ < \omega >}\,\,.
\eeq

We can calculate both exponents using the definitions
\beq
< \omega >\,\,=\,\,\frac{\partial \ln (x G^N (x,Q^2))}{\partial \ln(1/x)}\,\,.
\label{omega}
\eeq

\beq 
< \gamma >\,\,=
\,\,\frac{\partial \ln (x G^N(x,Q^2))}{\partial \ln (Q^2/Q^2_0)}\,\,;
\label{gama}
\eeq 

\begin{figure}[htbp]
\centerline{\psfig{figure=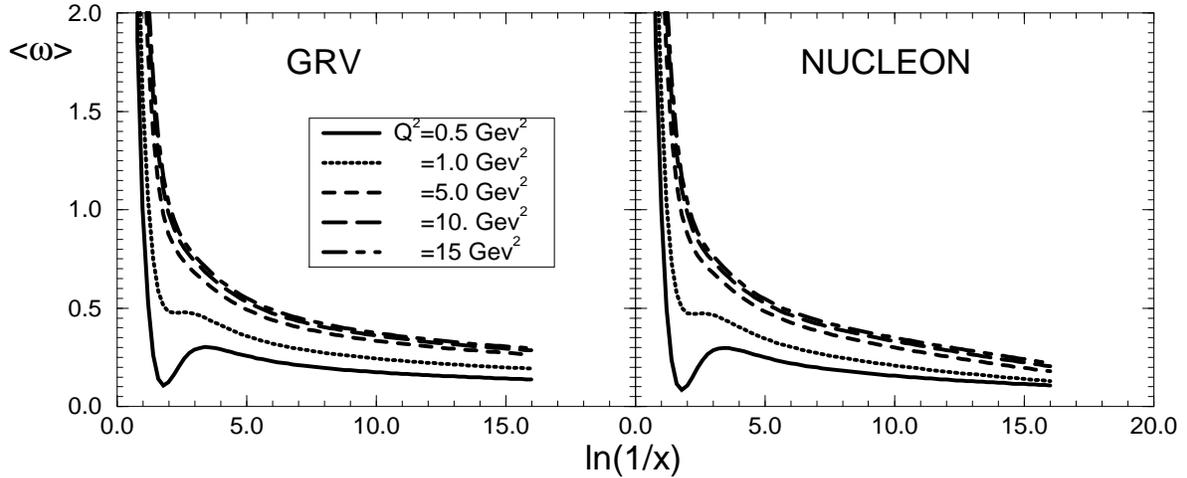,width=160mm,clip=}}
\caption{\em The effective power $<\omega >$
 calculated for $xG^A(A = 1)$ and the GRV 
distribution. }
\label{omn}
\end{figure}

\begin{figure}[htbp]
\centerline{\psfig{figure=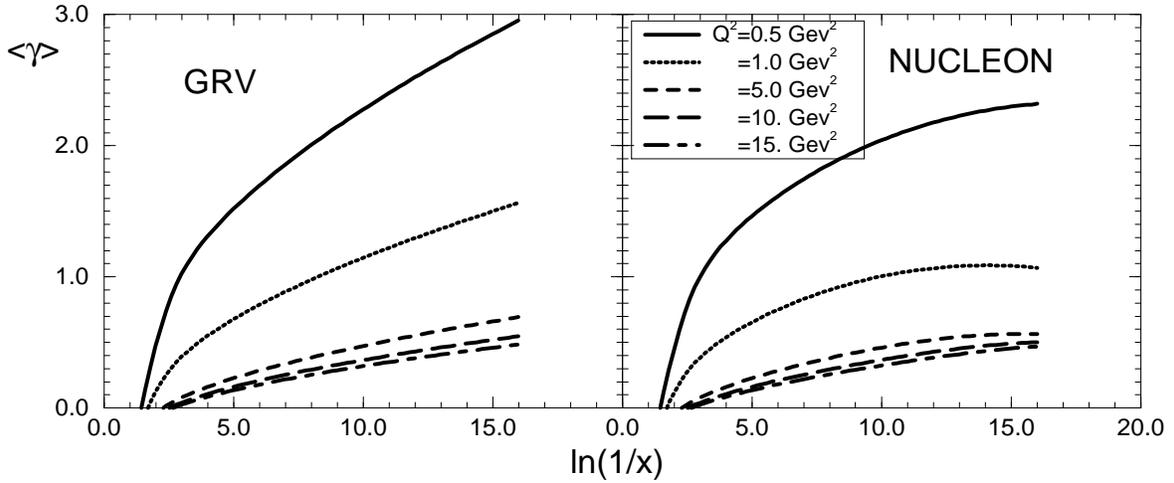,width=160mm}}
\caption{\em The effective power of $Q^2$ dependence calculated for
 $xG^A(A = 1)$ 
and the GRV  distribution. }
\label{gmn}
\end{figure}

The eq.(\ref{omega}) gives the average value of the effective power 
$<\omega > $  of the  gluon distribution, 
$xG(x,Q^2) \propto x^{-<\omega>} $, which is
suitable to study the small $x$ behavior of the gluon distributions. 
Fig.\ref{omn} shows the calculation of $<\omega>$ the nucleon distribution
calculated using  eq.(\ref{FINANS})
and for  GRV gluon distribution, both as  functions of $ln(1/x)$ for different 
values of $Q^2$. From the figure, we can see that the effective powers 
of $xG^A(A=1)$ and $xG(GRV)$ have the same general behavior in the small $x$ 
limit but the nucleon distribution is slightly suppressed.
We calculate also, in the same kinematical region, the exponent $<\gamma>$, 
given by eq (\ref{gama}). This  is the average value of the anomalous 
dimension, which describes the effective dependence of the distribution in 
$Q^2$ variable. Figs.\ref{gmn} shows $<\gamma>$ for  the 
nucleon and GRV distributions, indicating that
 the $Q^2$ dependence is slightly soften by the SC.

Comparing figures \ref{omn} and \ref{gmn}, we can conclude that even these 
more detailed characteristic of the gluon structure function have not been
seriously affected by the SC in the nucleon case. 

\begin{figure}[htbp]
\centerline{\psfig{figure=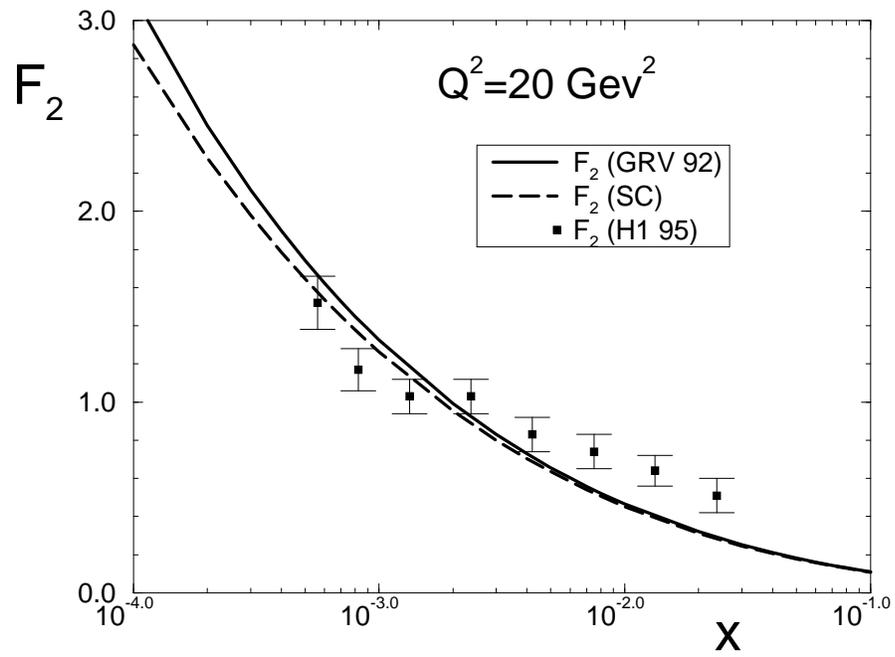,height=100mm}}
\caption{\em $F_2$  from $xG^A$  and  the  GRV 
distribution, compared with experimental data \protect\cite{r2}. }
\label{f21}
\end{figure}

We also use the GLAP evolution equations to predict the value of the deep
inelastic structure function $F_2$ from the $xG^A$ gluon distribution.
 Summing the GLAP
evolution equations for each quark flavor, 
the function $F_2$ may be written \cite{r30}
\bea
F_2 =\frac{\as (Q^2)}{\pi} \sum_{q} e_{q}^{2} \int^{{Q}^2}_{{Q}^{2}_{0}}
 \frac{d Q'^2}{Q'^2} 
\int^{1-x}_{0} [ z^2 + (1-z)^2 ] \frac{x}{1-z} G^{N}(\frac{x}{1-z}, {Q'}^2)
\label{eqf21}
\eea
where the sea quark distributions   have been  neglected in comparison with 
the gluon distribution.
Fig.\ref{f21} shows the prediction for $F_2$ from $xG^A$  and from the  GRV 
distribution, compared with experimental data. As we can see, the magnitude of 
the suppression due to the  SC is less than $10 \%$ in the region of the 
HERA data and this suppression 
is smaller than the experimental error.

From the above results we can conclude that 
\eq{FINANS} gives a good description for the gluon structure 
function for nucleon and describes the available experimental data.
 The MF provides a good description of the SC and can be taken as 
a correct first approximation
in the approach to the nucleus case.

\subsection{The gluon structure function for nucleus.}

In the framework of perturbative approach
it is only possible to calculate 
the behavior of the gluon distribution at small distances. The initial gluon 
distribution should be taken from the experiment. Actually the initial
virtuality $Q_0^2$ should be big enough to guarantee that we are dealing
with the leading twist contribution. Our main assumption is that we start
the QCD evolution with a small value of $Q_0^2$ considering that the 
 MF is a good
model for high twist contributions in   DIS off nucleus.

The scale of the SC governs by the value of $\kappa_A$, namely
they are big for $\kappa_A\,>\,$1 and  small for $\kappa_A\,<\,$ 1.
Fig.6 shows the plot of $\kappa_A$ = 1 for different nuclei. One can see
that the SC should be essential for heavy nuclei starting from Ca at the
accessible experimentally kinematic region.

\begin{figure}[p]
\centerline{\psfig{figure=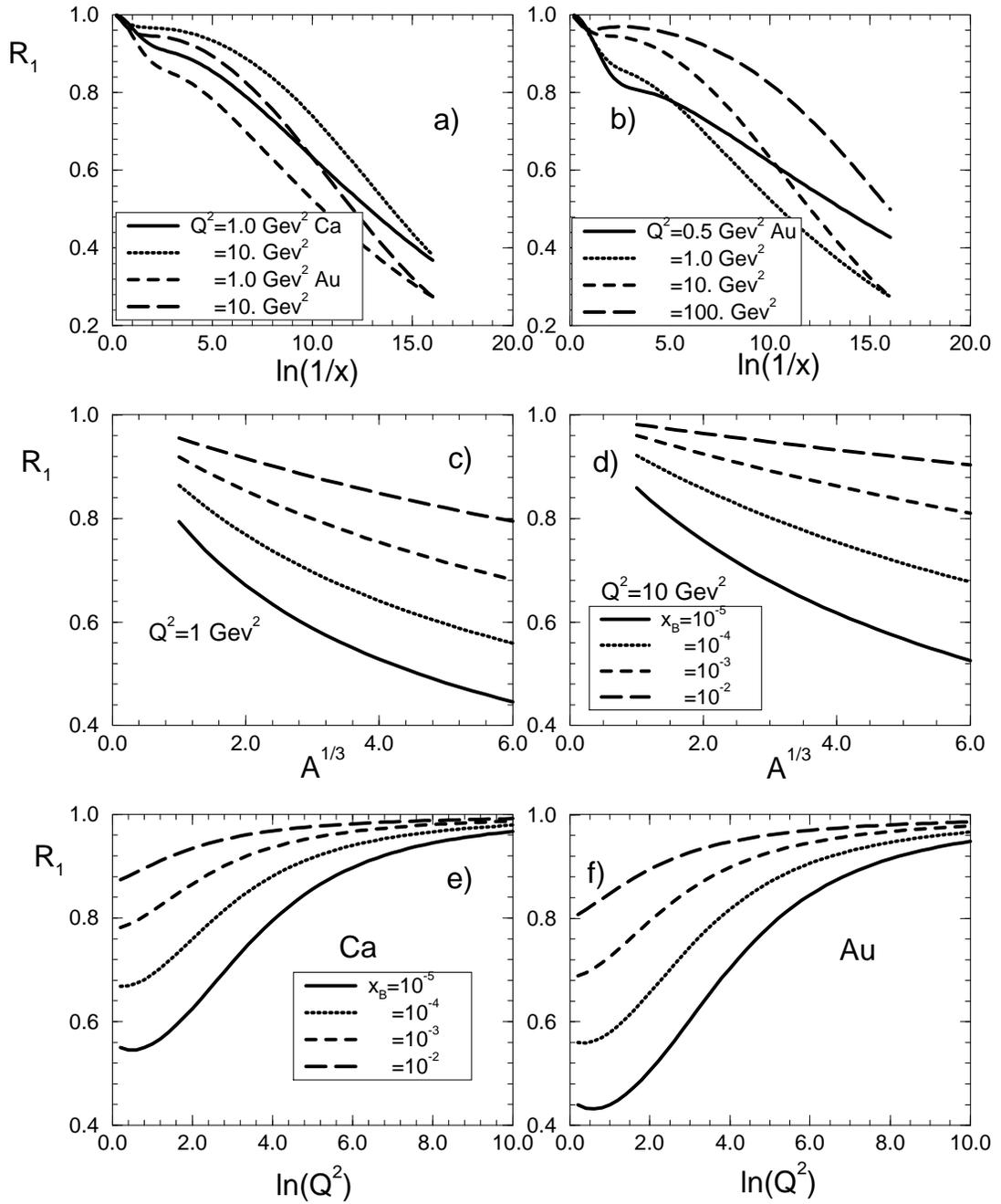,width=160mm,height=180mm}}
\caption{\em $R_1$ as a function of $ln(1/x)$,
 $lnQ^2$ and $A^{1/3}$: a) $R_1$ as a function of $ln(1/x)$ 
for different nucleus and different values of A; b)
$R_1$ as a function of
$ln Q^2$ for different values of $x_B$ for Au;
 c)  and  d) $R_1$ as a function of $A^{1/3}$ for different $Q^2$; e) and f)
 $R_1$ dependence on $Q^2$ for Ca and Au.}
\label{r1a}
\end{figure}

Now we extend the definition of $R_1$ for the nucleus case
\beq
R_1\,\,=\,\,\frac{xG^A(x,Q^2)}{A x G^{GRV}_N (x,Q^2)}\,\,,
\label{ratr1}
\eeq
where the numerator is calculated using eq.(\ref{FINANS}). Figure \ref{r1a}
shows the results for the calculations of $R_1$ as a function of
the variables $ln(1/x)$, $lnQ^2$ and $A^{1/3}$. Fig.\ref{r1a}a presents the 
ratio $R_1$ for two different values of $Q^2$ and for different nuclei.
The suppression due to the SC increases with $ln(1/x)$ and is much bigger than
 for
the nucleon case. For $A=40$ (Ca) and $Q^2=10 \, GeV^2$, the suppression varies
from $4 \, \%$ for $ln(1/x)= 3$ to $25 \, \%$ for $ln(1/x)= 10$. For $A=197$ 
(Au) the suppression is still bigger, going from $6 \%$ to $35 \%$ in the same
kinematic region. Fig. \ref{r1a}b shows the same ratio for different values of 
$Q^2$ for the gold. The suppression decreases with $Q^2$.
Figs. \ref{r1a}c and d show  the $R_1$
ratio as a function of $A^{1/3}$ and $x$ for a fixed value of $Q^2$.
As 
expected, the SC increases with $A$. An interesting feature of this figure 
is the
fact that the curves tend to straight lines as $x$ increases. It occurs 
because, as $x$ grows, the structure function $xG (GRV)$ becomes smaller, 
and the correction term of (\ref{FINANS}) proportional to $\kappa$ dominates. 
Since $\kappa$
is proportional to $A^{1/3}$, the curves behave as straight lines.
The decrease of suppression with $Q^2$ is illustrated in more detail in Figs.
\ref{r1a}e and f which presents $R_1$ as a function of $\ln Q^2$ for different
values of $x$ for Ca and Au, respectively. 
The effect is pronounced for small $Q^2$ and 
$x$ and diminishes as $ln Q^2$ increases.

\begin{table}
{\bf Table 1:} Values of $R_{1 N}$ and $\alpha$ for parameterization 
$R_1 \,=\,R_{1N}\,A^{-\,\alpha}$.\\
\vspace{0.5cm}
\begin{center}
\begin{tabular} {||l|l|l|l|l||}
\hline
\hline
\multicolumn{1}{||c|}{} & \multicolumn{2}{c|}{$Q^2=1 GeV^2$}
 & \multicolumn{2}{c||} {$Q^2 = 10 GeV^2$}
\\
\cline{2-5}
$x$ & $R_{1N}$ & $\alpha$ & $R_{1N}$ & $\alpha$ \\
\hline
$10^{-2}$ & 0.94   & 0.0416 & 0.98 & 0.014\\
\hline
$10^{-3}$ & 0.92 & 0.0616 & 0.94 & 0.034\\
\hline
$10^{-4}$ & 0.88 & 0.094 & 0.92 & 0.0563\\
\hline
$10^{-5}$ & 0.8 & 0.145 & 0.86 & 0.093\\ 
\hline
\hline
\end{tabular}
\end{center}
\end{table}
This picture ( Fig.\ref{r1a} ) shows also that the gluon structure function is
far away from the asymptotic one. The asymptotic behavior $R_1\,\rightarrow 1$
 ( see Figs.\ref{r1a}e and f )
occurs only at very high value of $Q^2$ as well as in the GLR approach
( see ref. \cite{I3} ). The asymptotic $A$-dependence (  $R_1\,\propto\,
A^{-\frac{1}{3}}) $ ) has not  been seen in the accessible kinematic range of
 $Q^2$ and $x$ ( see Figs. \ref{r1a}c and d and Table 1 ). This result also
 has been predicted in the GLR approach \cite{I2}. We want also to mention that
parameterization $R_1\,=\,R_{1N}\,A^{- \alpha}$ does not fit the result
 of calculations   quite well for
$1Gev^2\,\leq \,Q^2\,\leq\, 20 GeV^2$ and $ 10^{-2}\,\leq\,x\,\leq\,10^{-5}$.
For $x\,\sim\,10^{-2}$ the parameterization $R_{1}\,=\,R_{1N}
 \,-\,R' \,A^{\frac{1}{3}}$ with 
parameters $R_{1N}$
 and $R'$ for each value of $Q^2$,  works much better reflecting that
 only the first correction to the Born term is essential in the Mueller formula.

\begin{figure}[hptb]
\centerline{\psfig{figure=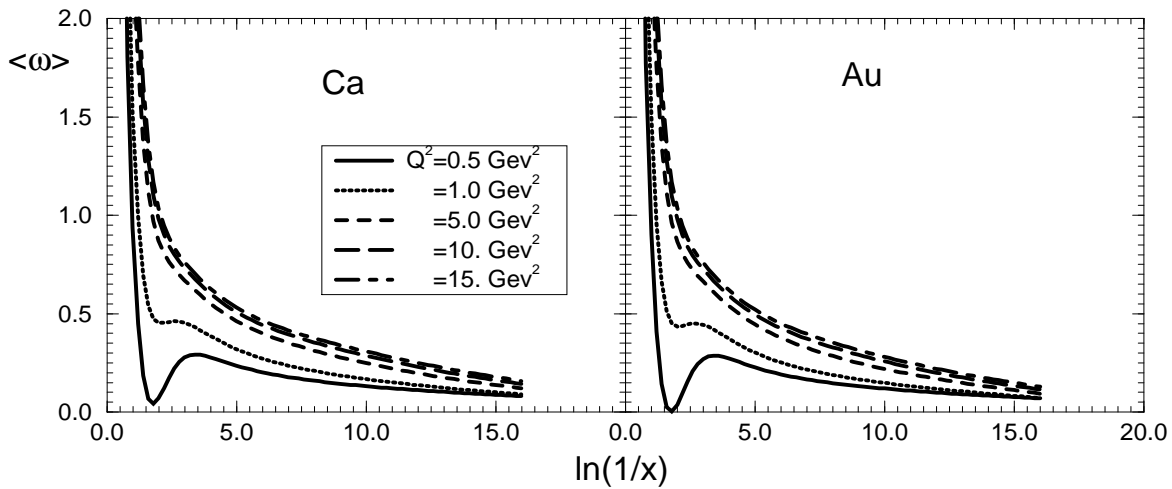,width=160mm}}
\caption{\em $<\omega>$ for different values of $Q^2$ and $A$. }
\label{oma}
\end{figure}

We extend also the calculation of the exponents $<\omega>$ and $<\gamma>$
of the semiclassical approach
for the nuclear case. We calculate the effective power of the nuclear gluon
distribution $<\omega>$ using the expression
\beq
< \omega >\,\,=\,\,\frac{\partial \ln (x G^A (x,Q^2))}{\partial \ln(1/x)}\,\,.
\label{omegaa}
\eeq
Fig.\ref{oma} shows the results as functions of $ln(1/x)$ for different values 
of $Q^2$ and different nucleus. The SC decreases the effective power of
 the nuclear distribution, giving rise to a flattening of 
the distribution in the 
small $x$ region.

 It is also interesting to notice that at small values of $Q^2$, the
effective power tends to be rather small, even in the nucleon case, at
very small $x$. However it should be stressed that the effective power
remains bigger than the intercept of the so called "soft" Pomeron \cite{SOFTPO},
even in the case of a sufficiently heavy nucleus (Au), for $Q^2\,>\,1GeV^2$.
Nowadays, many parameterizations \cite{CAPELA} with matching of "soft" and
"hard" Pomeron have appeared triggered by new HERA data on diffraction
dissociation \cite{DDHERA}. These parameterization used Pomeron-like behavior
namely, $xG(x,Q^2) \propto x^{- \omega(Q^2)}$. However, if the Pomeron is a 
Regge pole, $\omega$ cannot depend on $Q^2$, and the only reasonable
explanation is to describe $\omega(Q^2)$ as the result of the SC. Looking at
Fig.\ref{oma} we can claim the SC from the MF cannot provide sufficiently 
strong SC  to reduce the value of $\omega$ to $0.08$, a typical value 
for the soft Pomeron \cite{SOFTPO}, at least for $Q^2 \geq 1 GeV^2$.

The calculation of the effective value of the anomalous
dimension $\gamma$ may help us to estimate  what distances work in the
SC corrections. This effective exponent is given by
\beq 
< \gamma >\,\,=
\,\,\frac{\partial \ln (x G^A(x,Q^2))}{\partial \ln (Q^2/Q^2_0)}\,\, .
\label{gma1}
\eeq 
\begin{figure}[htbp]
\centerline{\psfig{figure=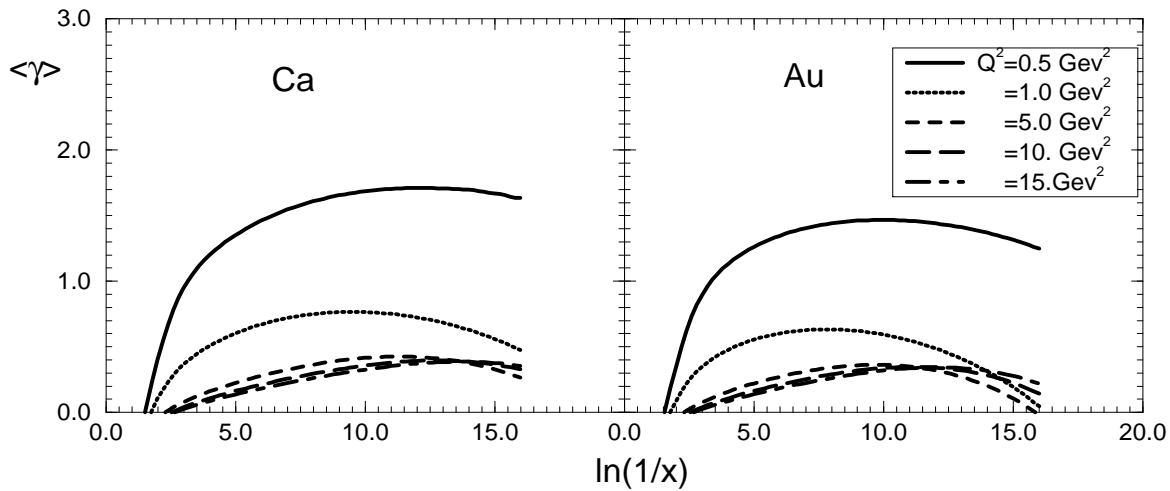,width=160mm}}
\caption{\em $< \gamma > $ for different $Q^2$ and $A$. }
\label{gma}
\end{figure}

Fig.\ref{gma} shows the results as functions of $ln(1/x)$ for different
values of $Q^2$ and for two nuclei.  We see that the values of $\gamma$ at 
$ln(1/x)\, \leq \,5$, for both Ca and Au, is very close to the results for 
GRV  and for nucleon case.  At smaller values of $x$, the anomalous
dimension presents a sizeable reduction, which increases with A. For
$ln(1/x)\, > \,15$, $<\gamma>$ tends to zero unlike in the GLAP evolution
equations ( see Fig.10 for the GRV parameterization). Analysing the $Q^2$ 
dependence, we see that $<\gamma>$ is bigger than $1$ only for 
$Q^2= 0.5 \, GeV^2$. For $Q^2= 1.0 \, GeV^2$, the anomalous
dimension is close to $1/2$, and for  $Q^2 > 5.0 \, GeV^2$ it is always
smaller than $1/2$. 

 Using semiclassical approach, we see that

\beq
\kappa \propto \frac{1}{Q^2}( Q^2)^{\g}\, ,
\label{kappa}
\eeq
and if $\gamma \geq 1$, the integral over $r_t$ in  the master equation
(\ref{FINANS}) becomes divergent, concentrating at small distances.

If  $ 1\,>\,\gamma \geq 1/2$, only the first SC term, namely, the second term in expansion
of the master equation, is concentrated at small distances, while higher
order SC are  still sensitive to small $r_t$ behavior. 
Fig.\ref{gma} shows
that this situation occurs for $Q^2 \, > \, 1 \, GeV^2$, and even for 
$Q^2 = 1 GeV^2$ at very small values of $x$. 
We will return to discussion of these properties of the anomalous 
dimension behavior in the next section.

In subsection 2.4 we have discussed that the virtual gluon  can interact with
 the target only during  the finite time $\tau$ (see \eq{tau} ) undergoing
$\rho \,\tau \,<\,\rho\,R_A$ collisions. In the framework of the Glauber
approach the easiest way to take into account the finite life time of
the gluon is to include in our calculation the longitudinal  part of the
transferred momentum ( $q_z$ ) to a nucleon during the collision
 ( see, for example,
 Ref. \cite{cut1} ).  Using \eq{san} and \eq{qz2}, we can obtain:
\beq \label{cut}
x \,G_A( x, Q^2)\,\,=\,\, A\,x\,G_N(x, Q^2) \,\,-\,\,A\,\frac{\as\,N_c}{\pi}
\int^1_x\,\,\int\,\frac{d x'}{x'}\,\frac{d Q'^2}{Q'^2}\,\,
L(q_z)\,\,x'\,G_N(x',Q'^2)
\eeq
$$
+\,\,
\frac{ 2 R^2_A}{\pi^2}\,\int^1_x\,\frac{ d x'}{x'}\,\int^{\frac{1}{Q^2_0}}_{
\frac{1}{Q^2}}\,\frac{d r^2_t}{r^4_t}\,\,\{\,C\,+\,
\ln(L(q_z)\,\kappa_G(x',r^2_t))
+\,\,E_1(L(q_z)\, \kappa_G(x',r^2_t)\,)\}
\,\,.
$$
where  $$L(q_z)\,\,=\,\,e^{-\,\frac{R^2_A}{4}\,m^2\,( x \,+\,x')^2}\,\,.$$

\begin{figure}[hptb]
\centerline{\psfig{figure=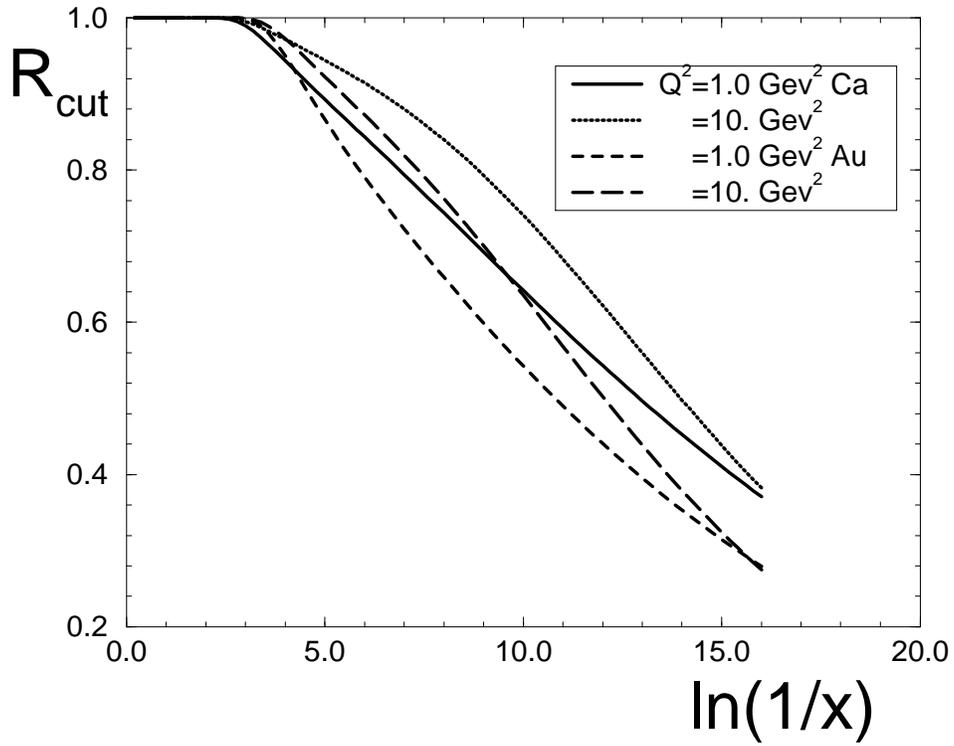,width=140mm}}
\caption{\em $ R_1$  for CA and Au with gluon life time cutoff.}
\label{r1cut}
\end{figure}

Fig. \ref{r1cut} shows the result of our calculations.
Comparing Fig.\ref{r1a} with this picture, one can see that the finite life
 time of the virtual gluon affects  the behavior of the gluon structure
function only at sufficiently large $x$ ( $x\,\geq\,10^{-2}$ ) diminishing
the value of the SC in this kinematic region. More interested in the small
$x$ behavior of the gluon structure function in nuclei we neglect  
the finite life time of a gluon through all calculations below.
 
\subsection{The behavior of $\alpha_s$.}
We calculated the ratio
\beq
\frac{<\as^A>}{<\as^N>} =   \frac{ x G^{GLAP}_N(x,Q^2)}{ xG_A(x, Q^2) } \frac{
 \left[ 
\frac{\pa }{\pa \ln\frac{1}{x}\pa \ln Q^2} (xG_{A}(x,Q^{2})) \right]}{
 \left[ 
\frac{\pa }{\pa \ln\frac{1}{x}\pa \ln Q^2} (xG^{GLAP}
_{N}(x,Q^{2})) \right]}
 \, ,
\label{alphas}
\eeq
\begin{figure}[htbp]
\begin{center}
\begin{tabular}{c c}
\psfig{file=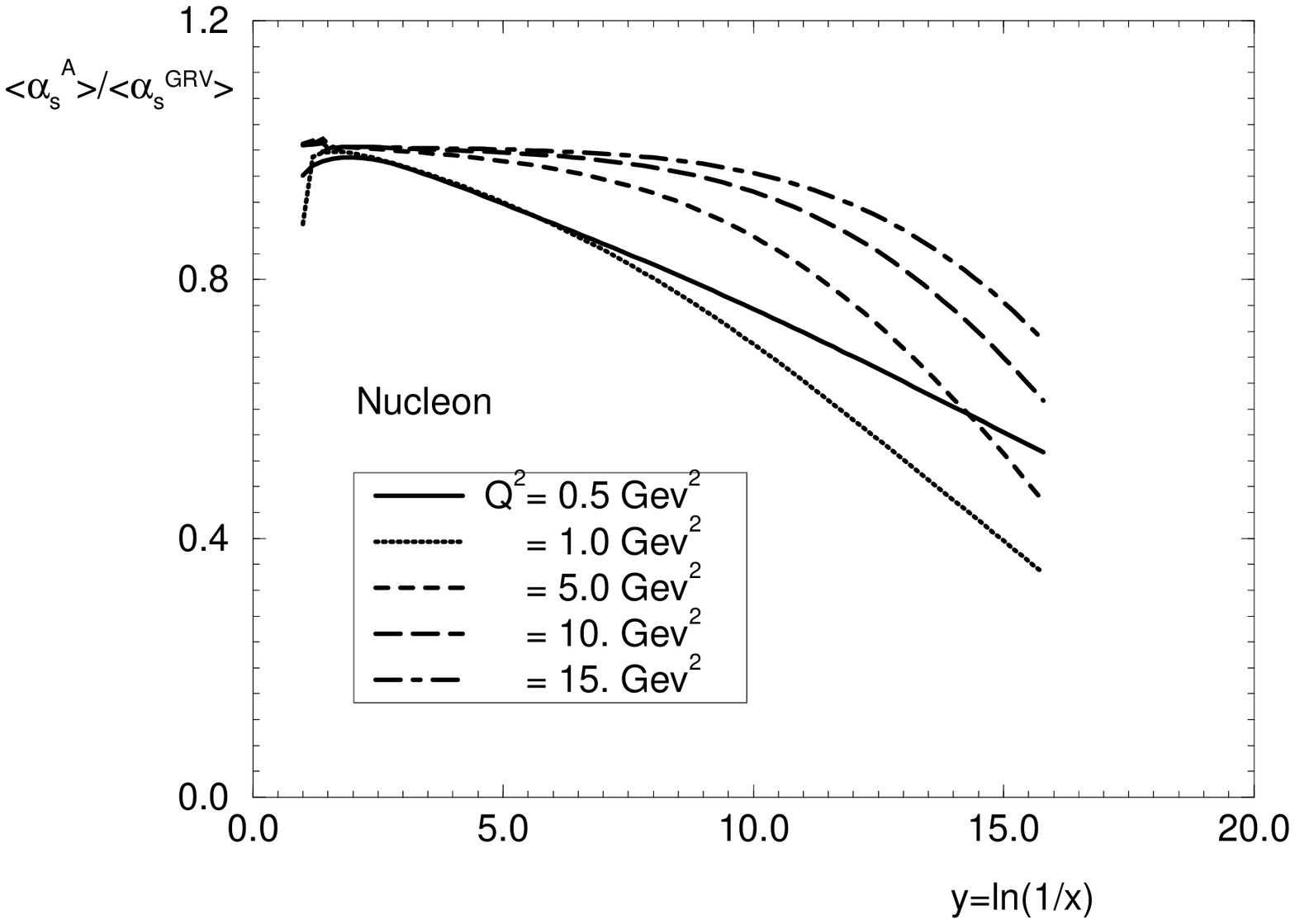,width=90mm} & \psfig{file=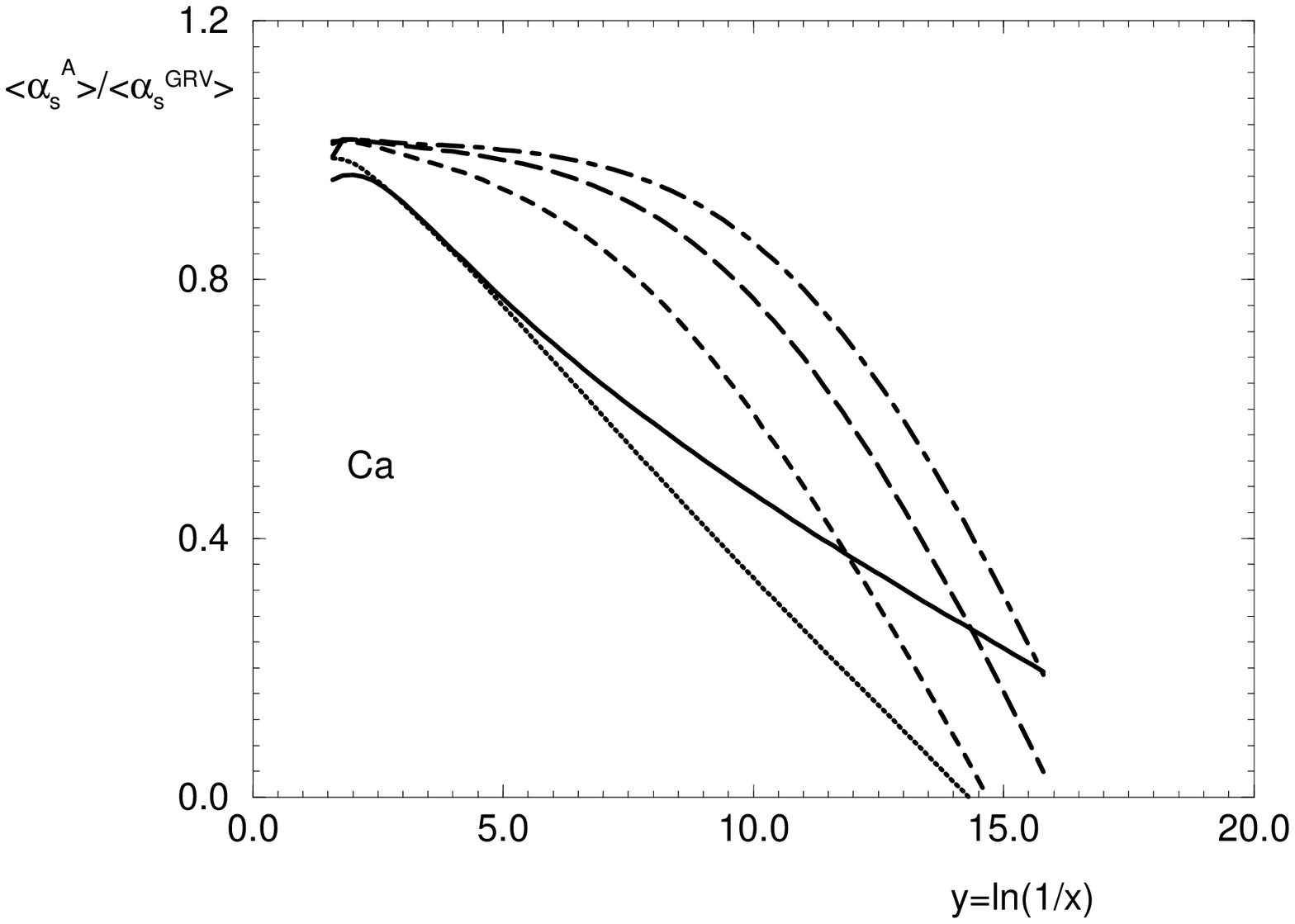,width=90mm}
\\
\psfig{file=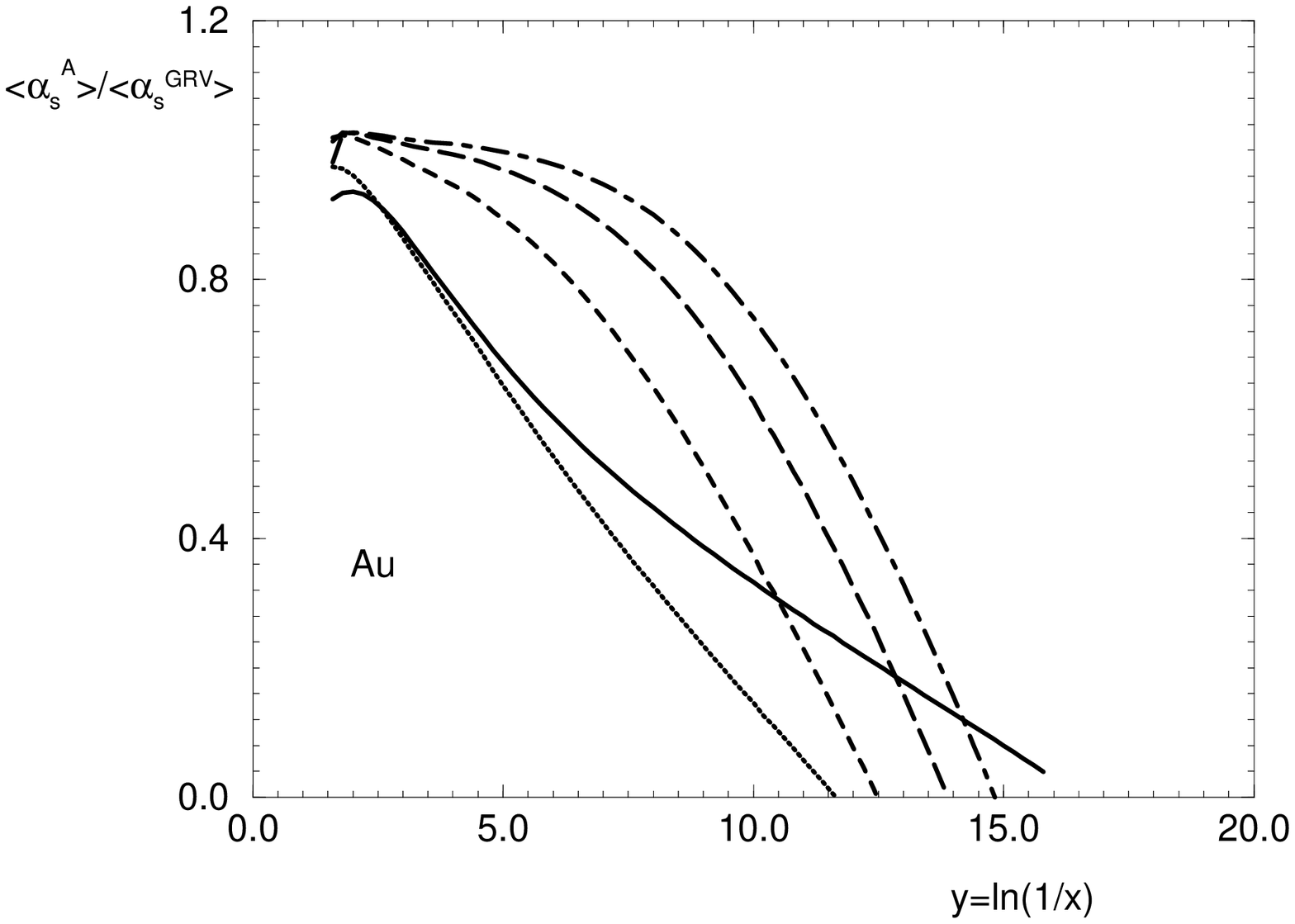,width=90mm} & \\
\end{tabular}
\end{center}
\caption{\it The ratio $<\as^A >/<\as^{GRV}>$.} 
\label{alfs}
\end{figure}
which is equal to $1$ in the case of the GLAP equation. In some sense
it characterizes the value of the effective QCD coupling constant in a nucleus.
 This
constant becomes smaller due to SC and in a restricted kinematic
region we can treat the parton cascade as a parton system evolving
according to the linear GLAP equations but with smaller coupling constant.
The result is shown in Fig.\ref{alfs}. 
 The physical meaning of
the weakening of the coupling constant is very simple. At low-$x$, the gluon
density increases and in a typical parton cascade cell there are many partons
with different colors.  The color average gives smaller mean color charge
at small gluon densities. One manifestation of the parton 
interaction weakening is the bigger value of
$\Lambda$ QCD
 for nuclei at small $x$. From Fig.\ref{alfs} one can see that
the GLAP evolution equation can be used to describe the evolution of the
gluon structure function for the nucleon in HERA kinematic region
 ( $x\,> \,10^{-4}$ ) but only for sufficiently large photon virtualities 
 ( $Q^2\,\geq \,5\,GeV^2$ ). At smaller values of $Q^2$ we have to develop
more sophisticated approach for the adequate treatment. In the case of nuclei,
the GLAP evolution equations with screened ( reduced ) value of $\as$ 
could be considered only as the first rough approximation even for large $Q^2$
( $Q^2\,\geq\,5\,GeV^2$ ).

\section{Beyond  the Glauber Approach.}
\label{cgf}
In this section we discuss  the corrections to the Glauber approach
 (the  Mueller formula of Eq.(66) ) as well as the way to construct a complete
 theory for deep inelastic scattering off a nucleus.

\subsection{The second iteration of the Mueller formula.}
To understand how big could be the corrections to the Glauber approach
 we calculate the second iteration of the Mueller formula of Eq.(66). As
has been discussed, Eq.(66) describes the rescattering of the fastest gluon
( gluon - gluon pair ) during the passage through a nucleus
 ( see Figs. 1 and 2 ). In the second iteration we take into account also
the rescattering of the next to the fastest gluon. This is a well defined task
 due to the strong ordering in the parton fractions of energy in the
parton cascade in leading $ln (1/x)$ approximation of pQCD that we are 
dealing with. Namely:
\beq \label{95}
x_B\,\,<\,\,x_n\,\,<\,\,...\,\,<x_1\,\,<\,\,1\,\,;
\eeq
where 1 corresponds to the fastest parton in the cascade.

Therefore, in the second interaction we include the rescatterings of the 
gluons with the energy fraction $1$ and $x_1$ ( see Fig.5 ). Doing the first
iteration we insert in Eq.(66) $G_N(x,Q^2)\,=\,G^{GRV}_{N}(x,Q^2)$. For the
second iteration we calculate the gluon structure function using Eq.(66)
 substituting
\beq \label{96}
x \,G_N\,\,=
\,\,\frac{x G^1_{A} (x, Q^2)}{A}\,\,-\,\,x\, G^{GRV}_{N} ( x,Q^2)\,\,;
\eeq
where $xG^1_A$ is the result of the first iteration of Eq.(66) that has been
discussed in details in section 3.

\begin{figure}[hptb]
\centerline{\psfig{figure=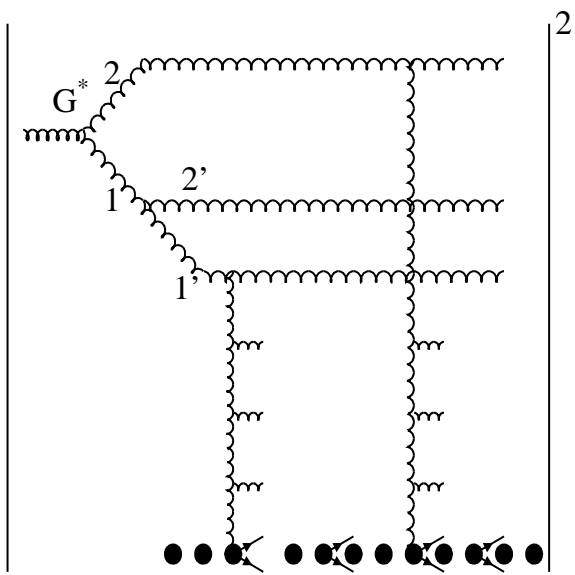,width=100mm}}
\caption{ \em The first term of the  second iteration 
 of \protect\eq{96}.}
\label{fig.16}
\end{figure}

Fig.\ref{fig.16}
  shows the need to subtract  $ xG^{GRV}_N$ in \eq{96} making the
second iteration. Indeed, in the second iteration we take into account the
rescattering of gluon $1'$ - gluon $2'$ pair off a nucleus. We picture
 in Fig.16  the
 first term of such iteration in which $G_{1'} G_{2'} $ pair has no
 rescatterings. It is obvious that it has been taken
into account in our first iteration, so we have to subtract it to avoid a
 double counting.

\begin{figure}[hptb]
\begin{center}
\begin{tabular}{ c c}
\psfig{file=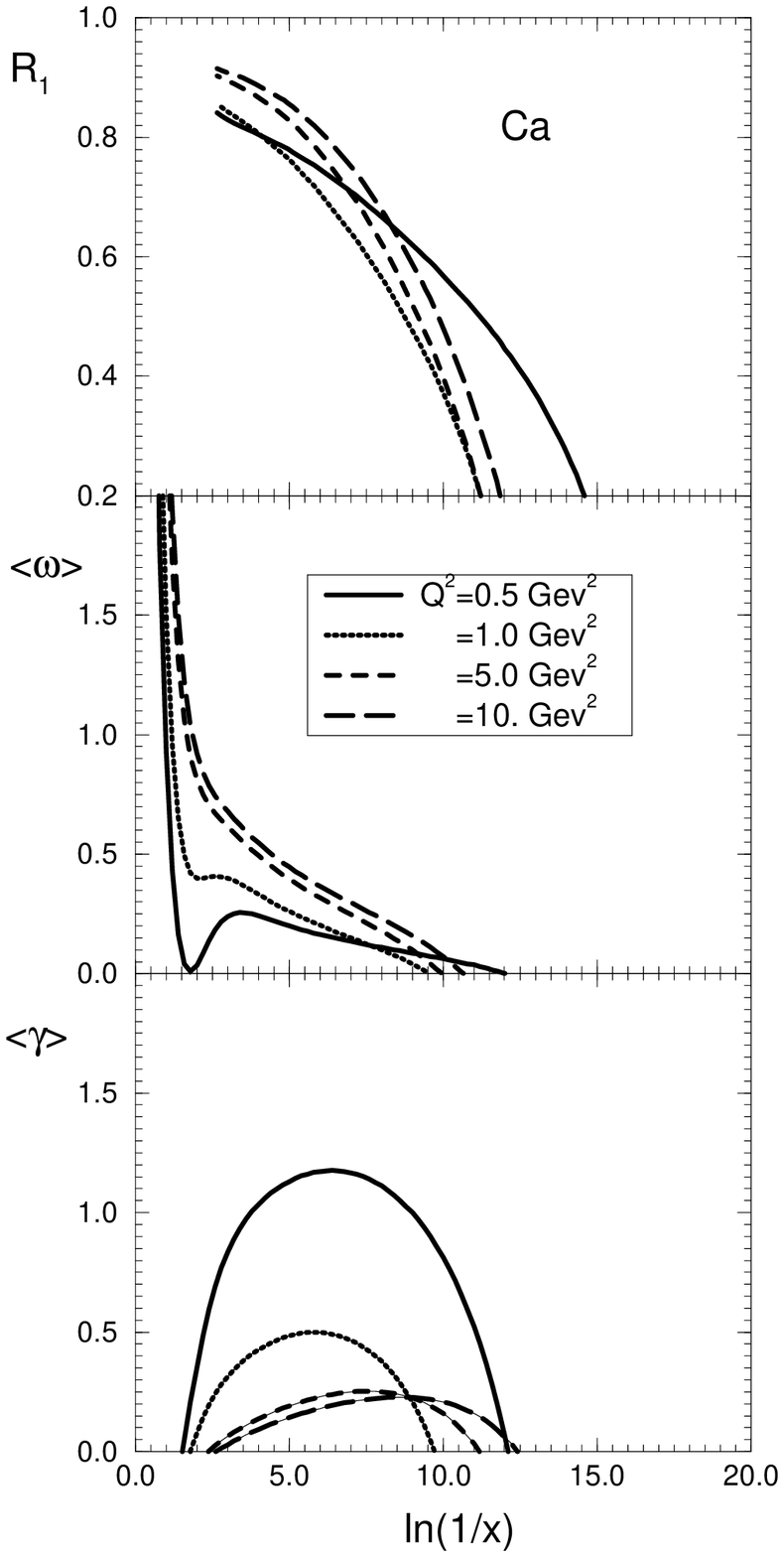,width=80mm} & 
\psfig{file=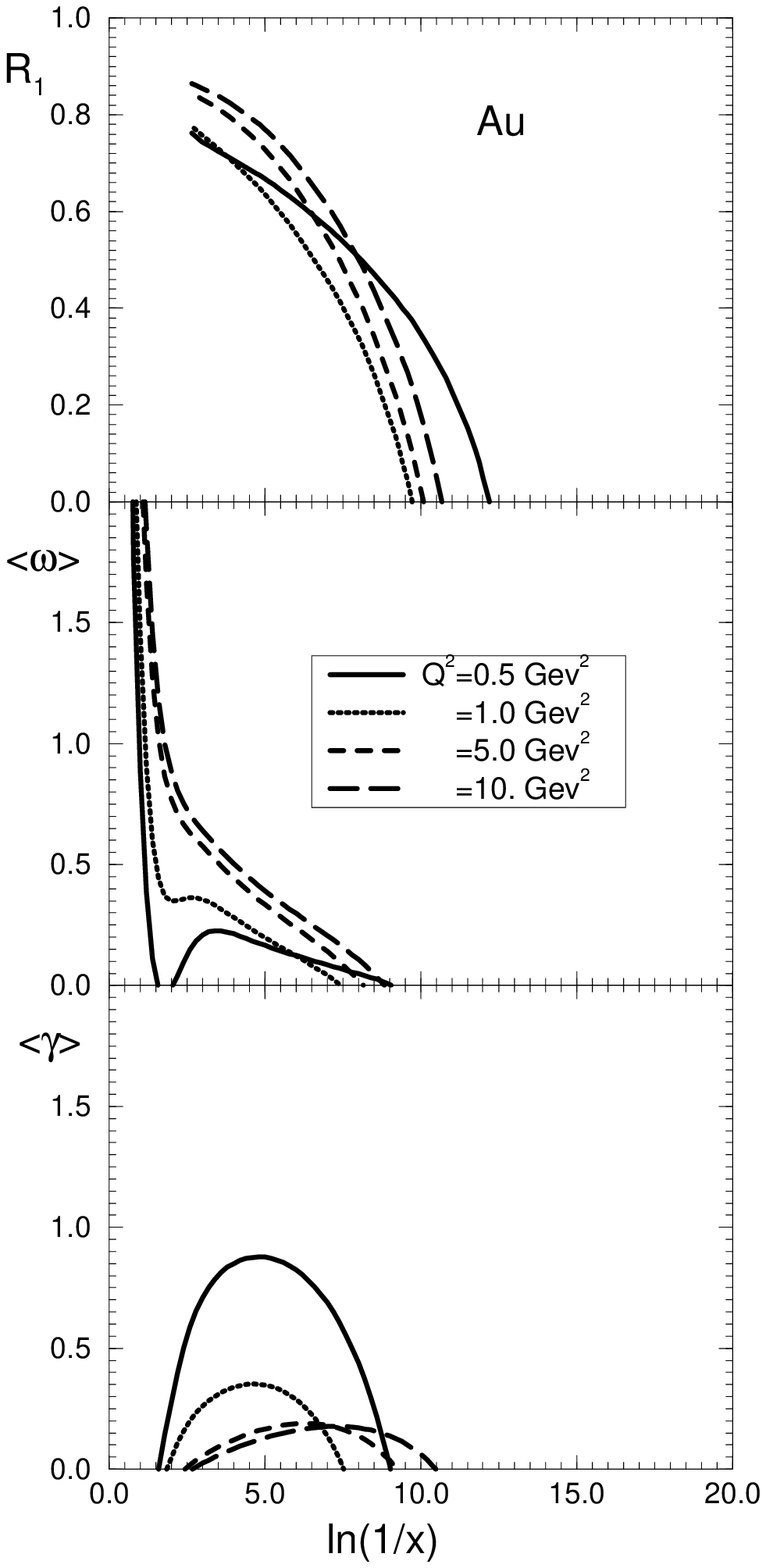,width=80mm}\\
\end{tabular}
\end{center}
\caption{ \em Second iteration calculations for $R_1$, $<\o>$, and $<\gamma >$
 for Ca and Au.}
\label{2ndit}
\end{figure}

One can see in Figs.\ref{2ndit} that the second iteration gives a big effect 
and changes crucially $R_1$, $< \gamma >$, and $<\omega >$.
The most remarkable feature
 is the crucial change of the value of the effective power $\omega(Q^2)
$ for the ``Pomeron" intercept which tends to zero at HERA kinematic region,
making possible the matching with ``soft" high energy phenomenology.
It is also very instructive to see how the second iteration makes
more pronounced all properties of the behavior of the anomalous dimension ( $<
\gamma>$) that we have discussed.
The main conclusions which we can make from Figs.\ref{2ndit} are:
(i) the second iteration gives a sizable contribution in the region $x\,\,<\,
\,10^{- 2}$ and for $x\,\,\leq \,\,10^{-3} $ it becomes of the order of the
first iteration; (ii) for $x\,\,<\,\,10^{-3}$ we have to calculate the
 next iteration. It means  that for such small $x$ we have to develop a 
different technique  to take into account rescatterings of  all the partons
 in the parton cascade which will be more efficient than the simple iteration
procedure for Eq.(66). However, let us first understand why the second iteration
becomes essential to establish small parameters that enter to our problem.

\subsection{Parameters of the pQCD approach.}

As has been discussed, we use the GLAP evolution equations for gluon 
structure function in the region of small $x$. It means, that we sum the Feynman
diagrams in pQCD using the following set of parameters:
\beq \label{97}
\as\,\,\ll\,\,1\,\,;\,\,\,\,\,\,\,\,\,\,\as\,\ln\frac{1}{x}\,\,<\,\,1\,\,;
\,\,\,\,\,\,\,\,\,\,\as\,\ln\frac{Q^2}{Q^2_0}\,\,<\,\,1\,\,;
\,\,\,\,\,\,\,\,\,\,
\as\,\,\ln\frac{Q^2}{Q^2_0}\,\,\ln\frac{1}{x}\,\,\approx\,\,1\,\,.
\eeq
The idea of the theoretical approach of rescattering that
 has been formulated in
 the GLR paper \cite{r1} is to introduce a new parameter \footnote{
In the GLR paper the notation for $\kappa$ was W, but in this paper 
we use $\kappa$ to avoid a misunderstanding since, in DIS, W is the energy
of interaction.}:
\beq \label{98}
\kappa\,\,=\,\,\frac{N_c \,\as\,\pi \,A}{ 2\,Q^2\,R^2_A}\,x G (x,Q^2)
\eeq
and sum all Feynman diagrams using the set of \eq{97} and $\kappa$ as parameters
of the problem, neglecting all contributions of the order of: $\as$,
 $\as \,\kappa$,
$\as \ln(1/x)$, $\as \ln(1/x)\,\kappa$, $\as \ln (Q^2/Q^2_0)$ and 
 $\as \ln (Q^2/Q^2_0)\,\kappa$. It should be stressed that Mueller formula
gives a solution  for such  approach. Indeed, Eq.(66) depends only on
$\kappa$ absorbing all \\
$(\,\as\,\ln (Q^2/Q^2_0)\,\ln(1/x)\,)^n$
contributions in $x G(x,Q^2)$. However, it is not a complete solution.
To illustrate this point let us compare the value of the second term of the
expansion of Eq.(62) with respect to $\sigma (r^2_t) $ with the first
correction due to the second iteration in the first term of such an expansion.
In other words we wish to compare the values of the diagrams in Fig.18a
and Fig.18b.
The contribution of the diagram of Fig.18a is equal:
\beq \label{99}
\Delta x G(x, Q^2)\,(\, Fig. 18a\, ) \,\,=\,\,\frac{R^2_A}{2\,\pi^2 }\,\,
\int \frac{d x'}{x'}\,\int\,d Q'^2 \,\,\kappa^2 ( x', \frac{Q'^2}{4})\,\,,
\eeq
where $x'$ and $Q'^2$ are the fraction of energy and the virtuality of 
 gluon $1$ in Fig.18a.

The diagram of Fig.18b contains one more gluon and its contribution is:
$$
\Delta x G(x, Q^2)\,(\, Fig. 18b\, ) \,\,=\,\,\frac{R^2_A}{\pi^2 2}\,\,
\frac{N_c \as}{\pi} \int\,\,\frac{d x'}{x'} \,\,\frac{ d Q'^2}{Q'^2}\,\,
\int \frac{d x"}{x"}\,\int\,d Q"^2 \,\,\kappa^2 (x",\frac{ Q"^2}{4})
$$
\beq \label{100}
\propto\,\,\frac{\as N_c}{\pi}\,\,\ln(1/x)\,\,\ln(Q^2/Q^2_0)\,
\,\Delta x G(x,Q^2)
(\,Fig.18a\,)\,\,,
\eeq
where $x'$ ($x''$) and $Q'^2$ ( $Q''^2$) are the fraction of energy and the
virtuality of gluon  $1$ ($1'$) respectively in Fig.18b.
\begin{figure}[p]
\begin{center}
\begin{tabular}{c c}
\psfig{file=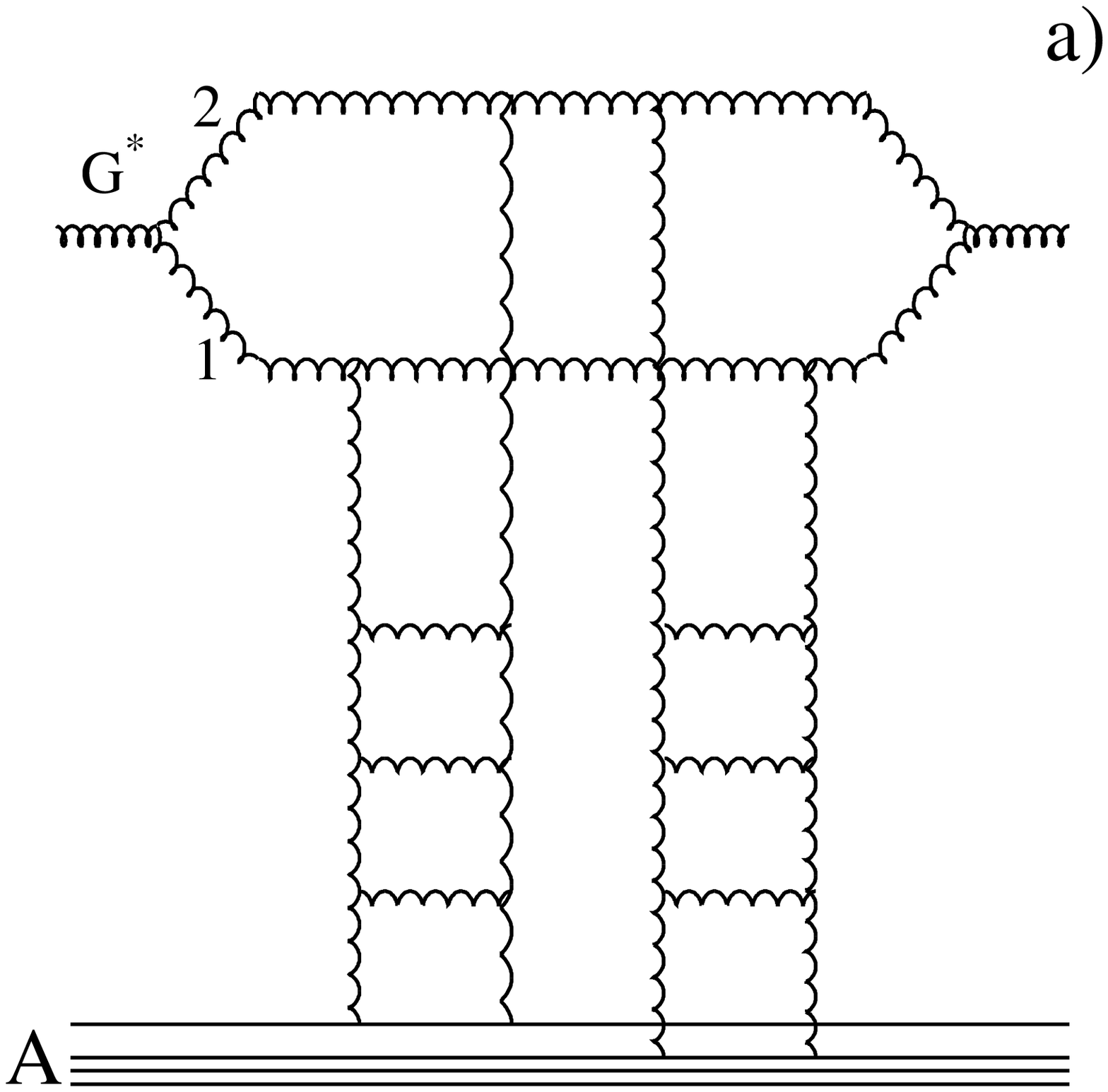,width=6.5cm} &
\psfig{file=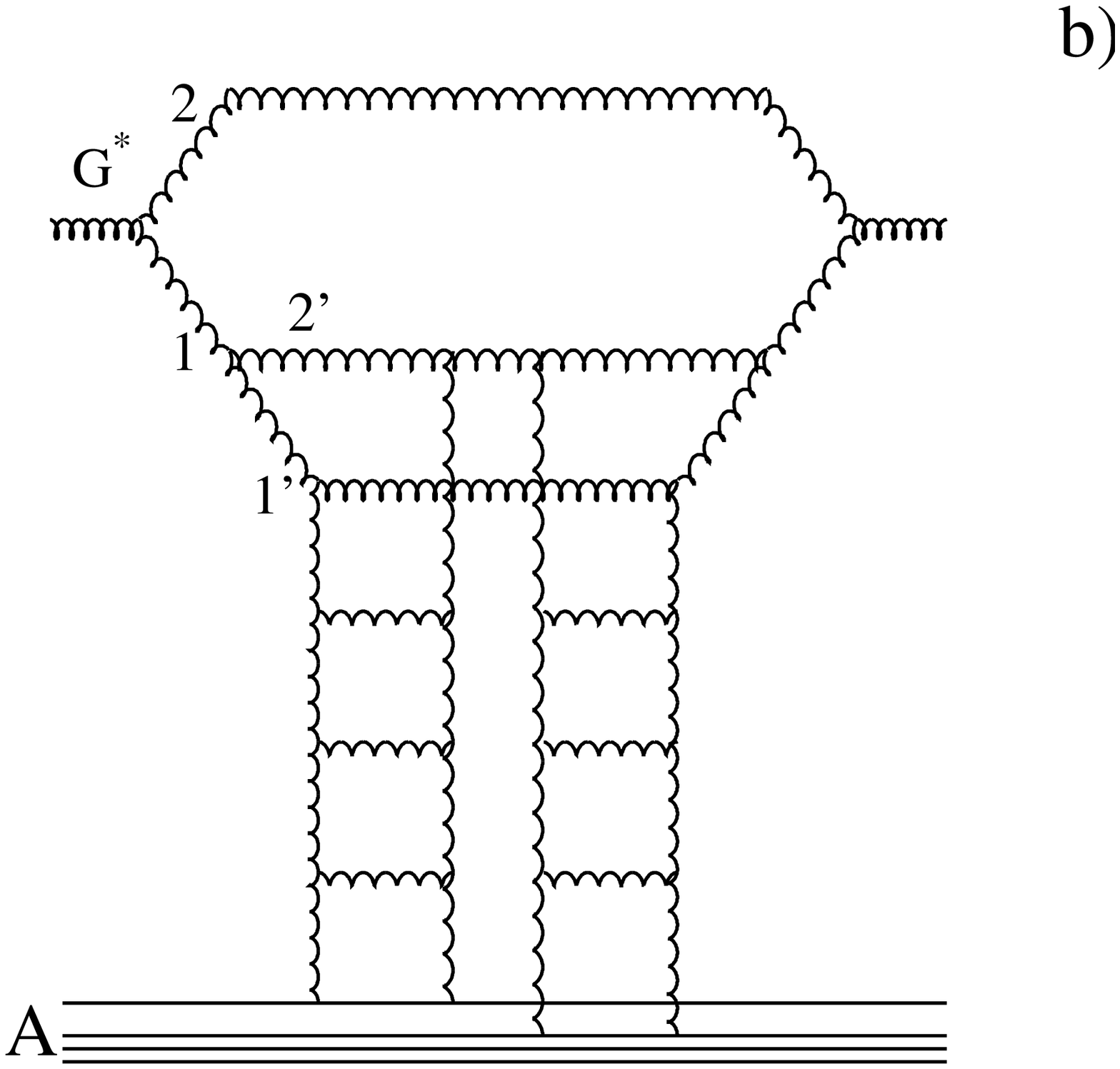,width=6.5cm}\\ 
\psfig{file=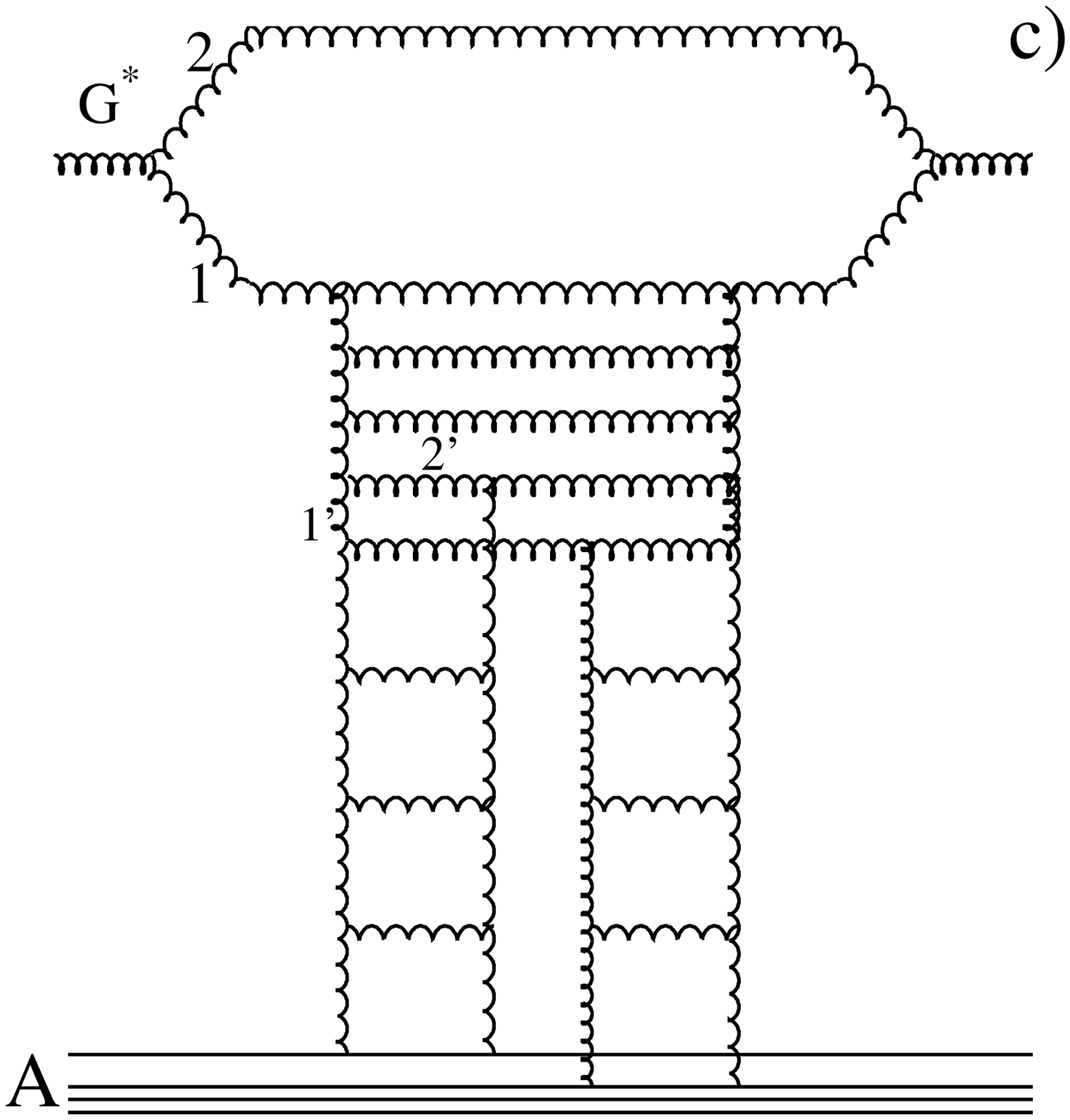,width=6.5cm} &
\psfig{file=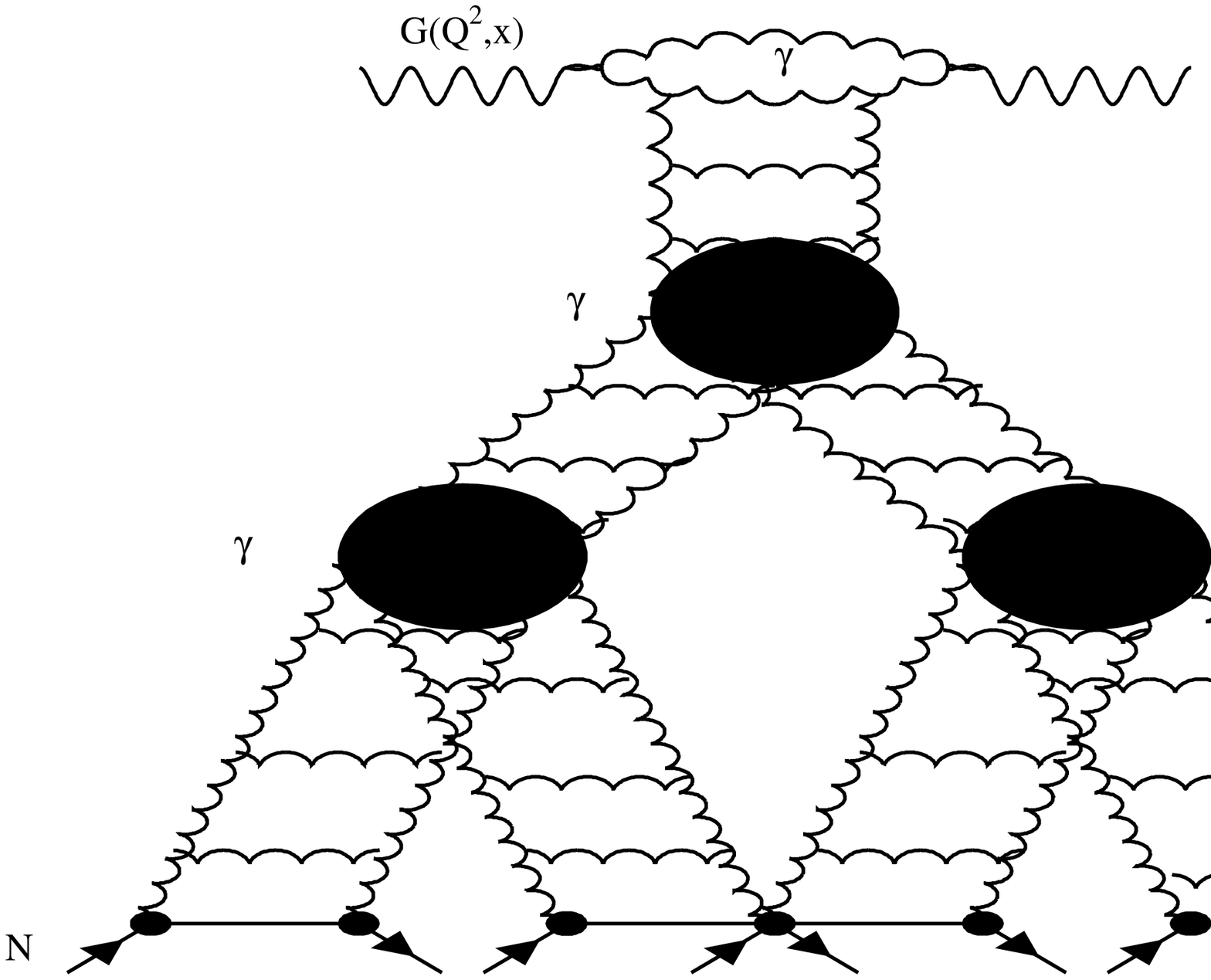,width=6.5cm}\\ 
\end{tabular}
\end{center}
\caption{\em Corrections to the Glauber approach.}
\end{figure}
Therefore, \eq{100} gives the contribution which is of the order of
\eq{99} in the kinematic region where the set of parameters of \eq{97} holds.
It means also that we need to sum all diagrams of Fig.18b type to obtain 
the full answer. In the diagram of Fig. 18b not only one but many gluons
 can be emitted. Such emission leads to so called ``triple ladder" interaction,
pictured in Fig.18c ( see ref.\cite{r1} ). This diagram is the first
from so called ``fan" diagrams of Fig. 18d. To sum them all we can neglect
 the third term in Eq.(65) and treat the remained terms as an equation
 for $x G(x,Q^2)$. It is easy to recognize that we obtain the GLR equation
  \cite{r1}\cite{r21}. Generally speaking the GLR equation sums the most
important diagrams in the kinematic region where $\as\,\ln(1/x)\,\ln(Q^2/Q^2_0)
\,\,\gg\,\,1$ and $\kappa\,<\,1$. Using the Mueller formula we can give
more precise estimates for the kinematic region where we can trust the
 GLR equation. Indeed, in Fig.19  we plot the ratio
\beq \label{101} 
R_C\,\,=\,\,\frac{\frac{R^2_A}{\pi^2 2}\,\,
\int \frac{d x'}{x'}\,\int\,d Q'^2 \,\,\kappa^2 (x', \frac{Q'^2}{4})}{
x G_A (x, Q^2) (\,\,Eq.(75)\,\,)\,\,-\,\,A \,x \,G^{GRV}_N (x, Q^2)}\,\,.
\eeq
If $R_C$ = 1, all the SC can be evaluated within good accuracy by the
 second term in Eq.(65).
\begin{figure}[htbp]
\centerline{\psfig{figure=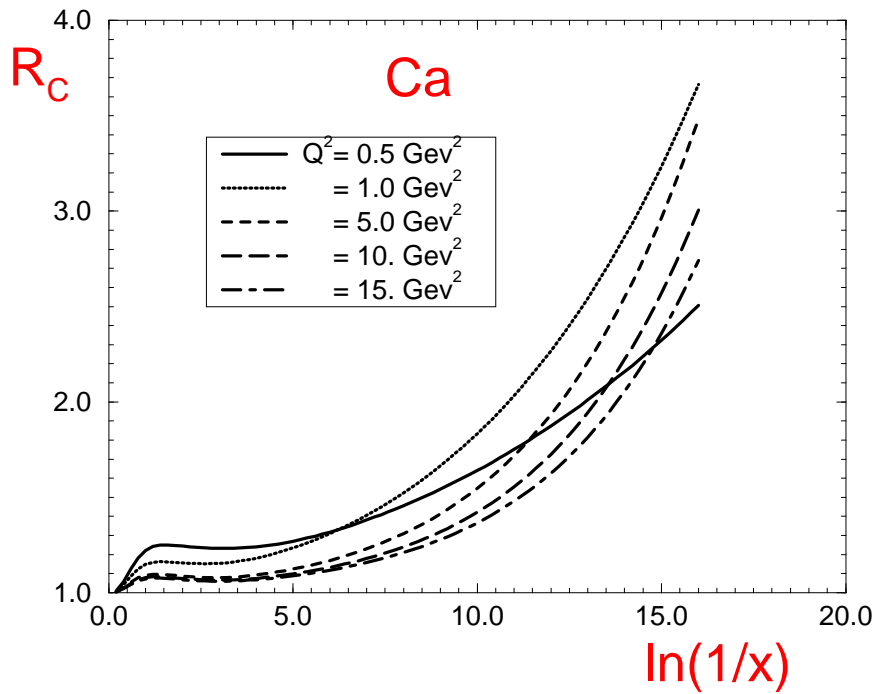,height=100mm}}
\caption{\em The ratio $R_C$ as a function of $ln(1/x)$ and $Q^2$. }
\label{gmac}
\end{figure}

From Fig.\ref{gmac} one can see that for Ca we can safely restrict
 by  the second term in the master equation and use the GLR equation
 to take into account the interaction of all parton in the parton cascade
 even for low values of
 $Q^2$ in the HERA kinematic region ($ x \,>\,10^{-5}$).
However, for the nuclei we have to develop more
general procedure for the interaction of the Mueller formula
 than the GLR equation for $x\,<\,10^{-2}$.

We need to make some very important remarks, concerning the whole
approach based on the Glauber - type shadowing corrections. It has been proven
\cite{Bartels}\cite{LRS} that keeping all parameters of \eq{96} and \eq{97}
and summing all Glauber - type interactions, as the Mueller formula does,
is not enough. 
It turns out that the interaction between partons from different parton cascades
that interact with the different nucleons are important. Fig.20a shows the
 first interaction of such a type that has to be taken into account.
 This diagram should be
compared with Fig.20b which shows the interaction included in the Mueller
formula as well as in the GLR equation. Fortunately, these new contributions
 are  proportional to $\frac{1}{N^2_c}$ and we will neglect them, considering
$N_c$ is large enough. The general procedure how to sum  all corrections
 of the order of
$\frac{1}{N^2_c}$  at least for the GLR equation has been developed in ref. 
\cite{r23}. In spite of the small parameter $\frac{1}{N^2_c}$ the 
parton selfinteraction can be essential for the case of a nuclear target
 but we postpone the detailed discussion of this problem to future publication.
\begin{figure}[hpbt]
\begin{center}
\begin{tabular}{c c}
\psfig{file=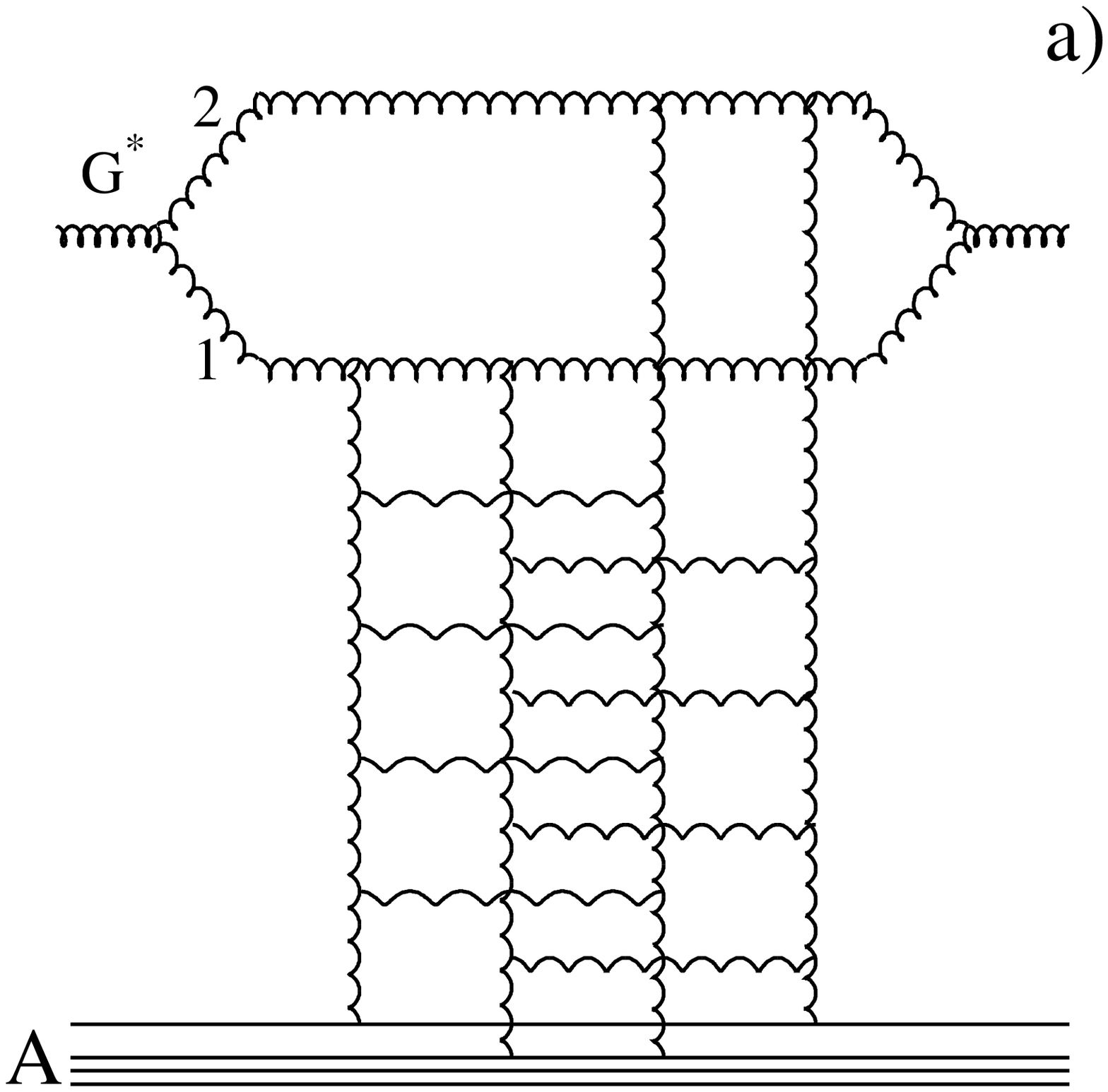,width=6.5cm} &
\psfig{file=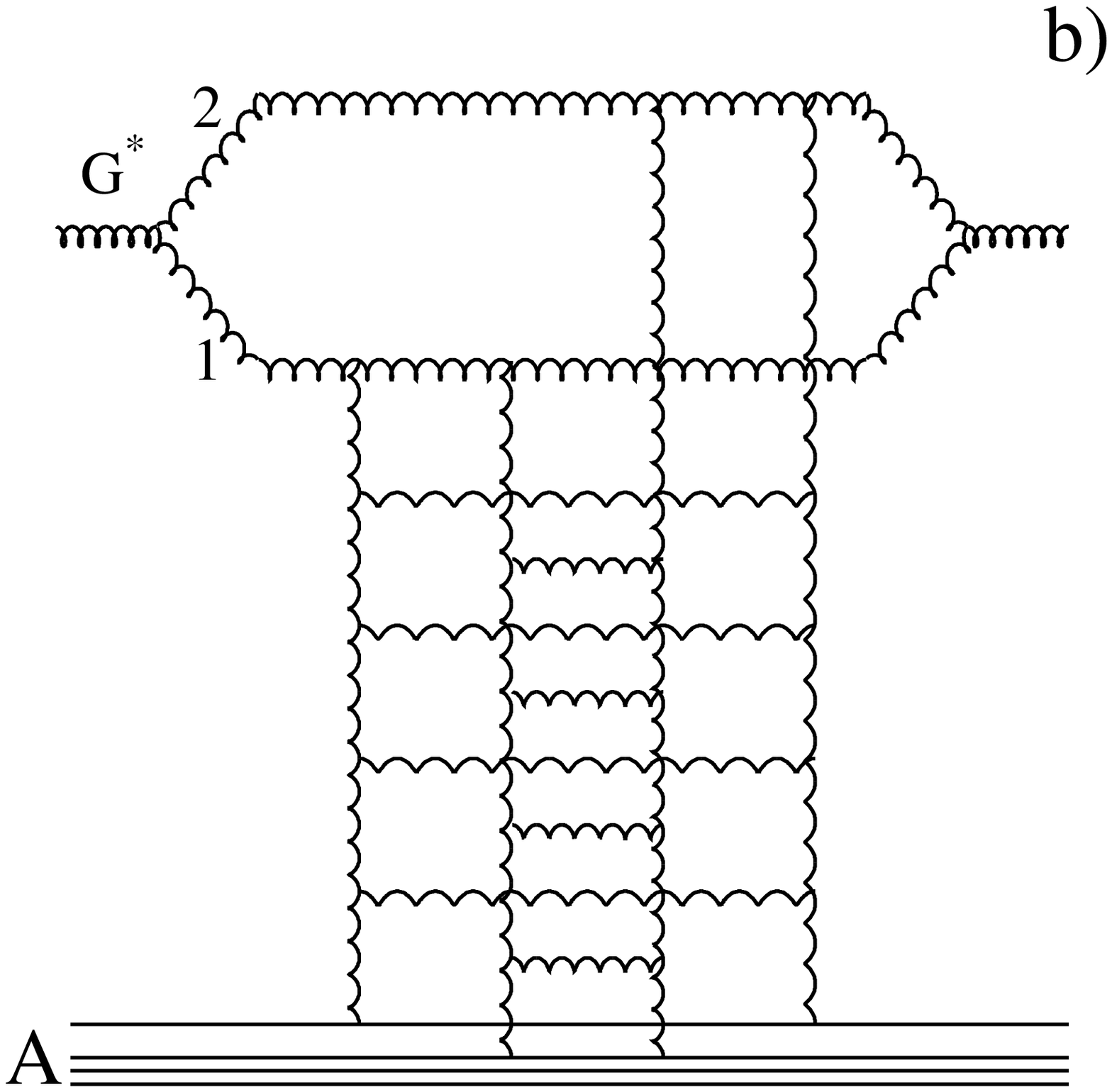,width=6.5cm}
\\ 
\end{tabular}
\end{center}
\caption{\em The Glauber approach ( Fig.20b) and corrections to the Glauber
 approach due to interaction between partons from different parton cascades.}
\end{figure}

\subsection{The GLR equation for nucleus.}
In this subsection we are going to discuss the GLR equation as a way  of
taking into account the interaction of all partons in the parton cascade
 with the target in spite of the criticism of the previous section.
Indeed, let us go back to the discussion of the behavior of
 the average anomalous dimension in the Mueller formula and in the first
 iteration of the MF  ( see Figs.14 and 19). The general feature of the both
 iterations is the fact that the resulting value of the anomalous dimension
 ( $< \gamma >$ )
 turns out to $ < \gamma > \,\leq\,\frac{1}{2}$ for $Q^2\,\geq\, 1 GeV^2$.
In this case the high order terms in the expansion of the MF are concentrated
 at large distances while the second one can be divergent at small distances
 if $< \gamma >\,\rightarrow\, \frac{1}{2}$. Therefore we can rewrite the
 Mueller formula in the form:
\beq \label{GLR1}
x G_A(x, Q^2)\,=\,\frac{4}{\pi^2}\,\int^1_x\,\frac{ d x'}{x'}\,\int^{\infty}_0
d \,b^2_t 
\eeq
$$
\{ \int^{\frac{4}{Q^2_0}}_{\frac{4}{Q^2}} \,\,\frac{d^2 r_t}{\pi \,r^4_t}
\,2\,[\,1\,-\,e^{ - \frac{1}{2}\sigma^{GG}_N (x',r^2_t)S(b^2_t)}\,]\,\,+
\,\,\int^{\infty}_{\frac{4}{Q^2_0}} \,\,\frac{d^2 r_t}{\pi \,r^4_t}
\,2\,[\,1\,-\,e^{ - \frac{1}{2}\sigma^{GG}_N (x',r^2_t)S(b^2_t)}\,]
\,\}\,\,,
$$
where $Q_0$ we choose of the order of 1 $GeV^2$. Recalling that
 $< \gamma >\,<\,1$
we can expand the first integral  in \eq{GLR1} and neglect all high 
contribution except the second term which can be important in the region where
$ <\gamma>\,\rightarrow\,\frac{1}{2}$. Differentiating with respect to
$\ln Q^2$ and $\ln(1/x)$ one obtain the GLR equation, namely
\beq \label{GLR2}
\frac{\partial^2 x G_A (x,Q^2)}{\partial \ln(1/x)\,\partial \ln Q^2}\,\,=\,\,
\frac{N_c \as}{\pi}\,x G_A(x, Q^2)\,\,-\,\,\frac{\as^2 A^2}{2 R^2_A}
\,\left(\, x G_A(x, Q^2)\,\right)^2\,\,.
\eeq
The second integral in \eq{GLR1} gives the initial condition for the GRL
equation. We want to draw your attention to two important outcomes from this
simple consideration. First, the initial condition  should be set only at 
sufficiently large value of $Q^2$, e.g., at $Q^2=Q^2_0\,\geq\,1 GeV^2$.
Second, we cannot use the MF to calculate this initial condition for
 the nucleus using the nucleon gluon structure function, since the
 corrections to MF for $Q^2 \,\leq 1 GeV^2$ is large. Therefore, 
 the solution of the GLR equation is only reliable
 in a kinematic region where it does not depend on any initial distribution.
The way out of this shortcoming is only to use the direct information
on the gluon structure  function in a nucleus. Therefore, in such  approach
 the main advantage of the Glauber formula is lost: the possibility to
calculate the nucleus structure function from the nucleon one. This is the
 reason to develop  a more  general approach in the next subsection.
\subsection{The generalization of the Glauber Approach.}
We suggest the following way to take into account the interaction of all
partons in a parton cascade with the target. Let us differentiate the Mueller
formula over $y \, = \,\ln (1/x)$ and $ \xi = \ln(Q^2/Q^2_0)$. It gives:
\beq \label{102}
\frac{\partial^2 x G_A( y,\xi)}{\partial y \partial \xi}\,\,=\,\,
\frac{2\,R^2_A \,\,Q^2}{ \pi^2}\,\,\{\,\,C\,\,+\,\,\ln \kappa\,\,+
\,\,E_1 ( \kappa )\,\,\}\,\,.
\eeq
Rewriting \eq{102} in terms of $\kappa$  given by
\beq \label{KAPPA}
\kappa \,\,=\,\,\frac{N_c \as }{2 Q^2 R^2_A}\,x G_A(x,Q^2)
\eeq
we obtain:
\beq \label{103}
\frac{\partial^2 \kappa( y,\xi)}{\partial y \partial \xi}\,\,+\,\,
\frac{\partial \kappa(y, \xi)}{\partial y}\,\,=\,\,
\frac{ N_c\, \as\,}{\pi}\,\,\{\,\,C\,\,+\,\,\ln \kappa(y, \xi)\,\,+
\,\,E_1 ( \kappa(y, \xi) )\,\,\}\,\,\equiv\,\,F(\kappa)\,\,.
 \eeq
Now, let us consider the expression of \eq{103} as the equation for  $\kappa$.
This equation sums all contributions of the order $ (\,\as\,y\,\xi\,)^n$ 
absorbing them in $x G_A (y,\xi)$, as well as all contributions of the order
of $\kappa^n$. For $N_c\,\rightarrow \,\infty $ \eq{103} gives the complete
solution to our problem. The great advantage of this equation in comparison
 with the GLR one is the fact that it describes the region of large $\kappa$
and provides the correct matching both with the GLR equation and with the
Glauber ( Mueller ) formula in the kinematic region where
 $\as y \xi\, \leq\, 1$.

Eq. (\ref{103}) is the second order differential equation in partial derivatives
and we need two initial ( boundary ) conditions to specify the solution.
The first one is obvious, namely, at fixed  $y$ and $Q^2 \,\rightarrow \,\infty$
$$
\kappa\,\,\rightarrow\,\,\frac{N_c \,\as\,\pi \,A}{ 2\,Q^2\,R^2_A}
\,x G^{GLAP}_N (x,Q^2)\,\,.
$$
The second one we can fix in the following way: at $ x = x_0 \,\,(y = y_0)$
which is small, namely, in the kinematic region where  $\as y \xi\, \leq\, 1$
\beq \label{104}
\kappa\,\,\rightarrow\,\,\kappa_{in}\,\,=\,\,
\frac{N_c \,\as\,\pi }{ 2\,Q^2\,R^2_A}
\,x G_A (x,Q^2)\,\,,
\eeq
where $x G_A$ is given by the Mueller formula ( see Eq.(66) ).
Practically, we can take $x_0 \,=\,10^{-2}$, because  corrections to the MF
are  small at this value of $x=x_0$.

\subsection{The solution to the generalized evolution equation.}
\subsubsection{The asymptotic solution}
First observation is the fact that \eq{103} has a solution which depends only
 on $y$. Indeed, one can check that $\kappa \,=\,\kappa_{asymp}(y)$ is
the solution of the following equation:
\beq \label{105}
\frac{d \kappa_{asymp}}{d y}\,\,=\,\,F(\,\kappa_{asymp}\,)\,\,.
\eeq
The solution to the above equation is: 
\beq \label{106}
\int^{\kappa_{asymp}(y)}_{\kappa_{asymp}(y=y_0)}\,\,
\frac{ d \kappa'}{F(\kappa')}
\,\,=\,\,y - y_0\,\,.
\eeq

\begin{figure}[hptb]
\begin{tabular} {c c}
\psfig{file=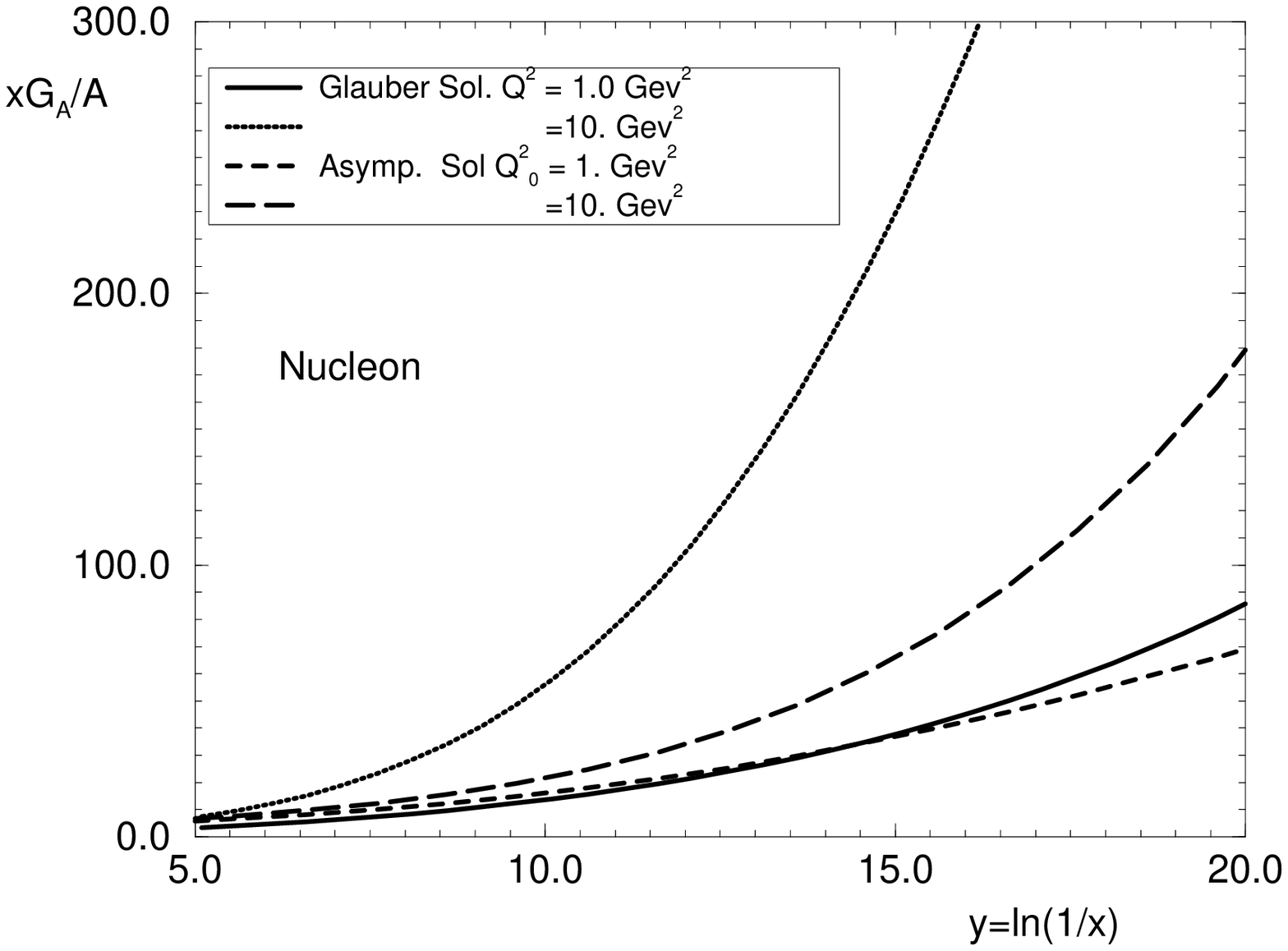,width=90mm} & \psfig{file=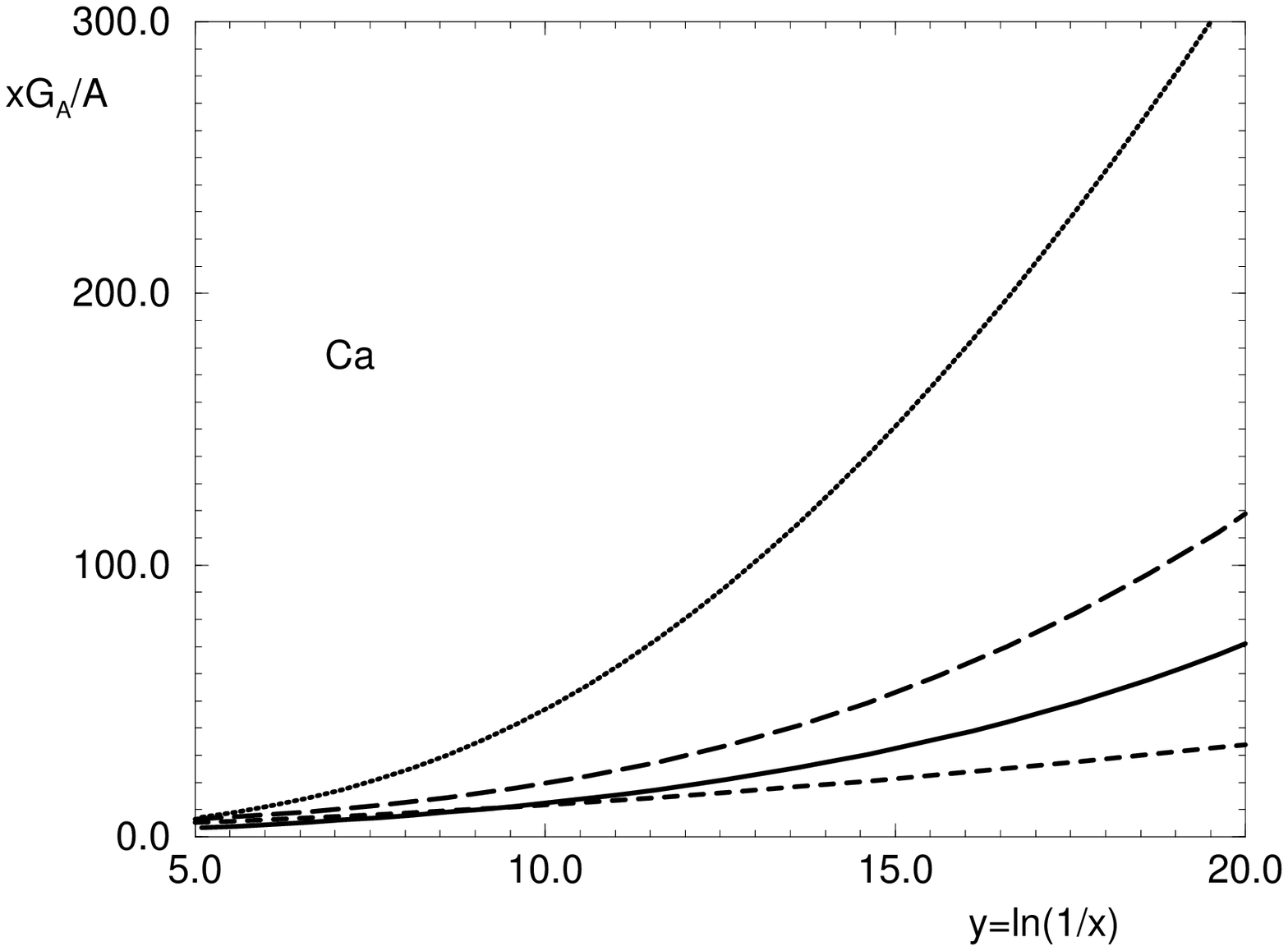,width=90mm}\\
\psfig{file=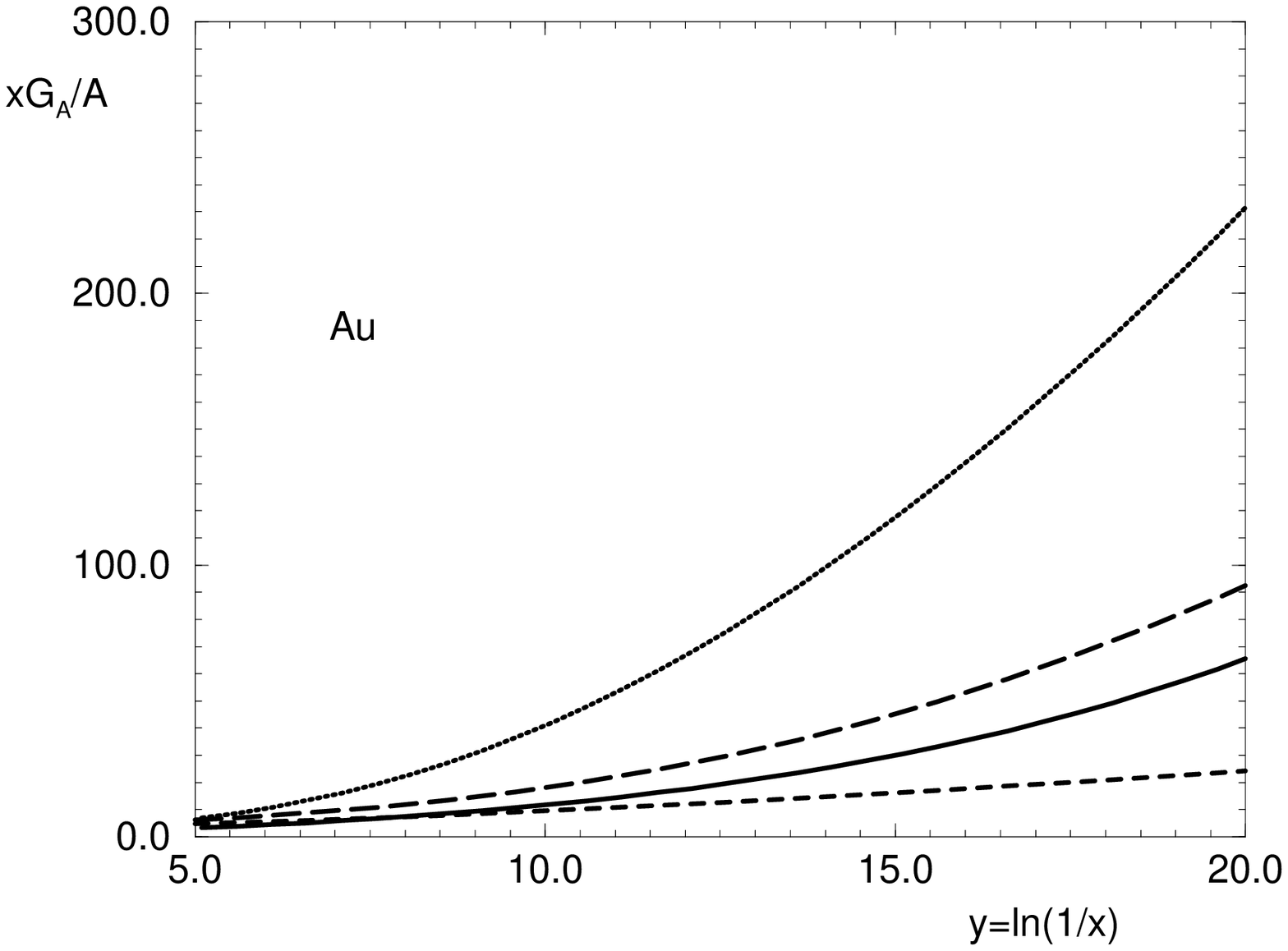,width=90mm} &  \\
\end{tabular}
\caption{ \em The Glauber approach  and asymptotic solution for
different nuclei.}
\label{asy}
\end{figure}

It is easy to find the behavior of the solution to \eq{106} at large value
of $ y$ since $F(\kappa)\,\,\rightarrow\,\,\bar \as \ln \kappa $ at
large $\kappa$ ( $\bar \as = \frac{N_c}{\pi}\,\as$ ). It gives
\beq \label{107}
\kappa_{asymp}\,\,\rightarrow \,\,\bar \as y \,\ln(\bar  \as y)\,\,\,\,\,\,
at\,\,\,\,\,\,\,\bar \as y \,\,\gg\,\,1\,\,.
\eeq
At small value of $y$, $F(\kappa)\,\,\rightarrow\,\,\bar \as \kappa$ and
we have:
\beq \label{108}
\kappa_{asymp}\,\,\rightarrow\,\,\kappa_{asymp} ( y = y_0 )
\,\,e^{\bar \as ( y - y_0)} \,\,.
\eeq
The solution is given in Fig.\ref{asy} for $\bar \as = 1/4$  in the whole region
 of $y$ for different nuclei in comparison with our calculations based
on the MF. We chose the value of $\kappa_{asymp} (y =y_0)$ from \eq{104}.
We claim this  solution is the asymptotic solution to \eq{103} and  will 
argue on 
this point a bit later.The calculations in Glauber approach for 
nucleon overshoot
 the  asymptotic solution
 at large values of $Q^2$  in HERA kinematic region ( at $x\,<\, 10^{-2}$ ). 
However, at small values of $Q^2$ the Glauber approach leads to stronger SC
than the asymptotic solution. For nuclei the SC incorporated in the
asymptotic solution turn out to be much stronger than 
the SC in the Glauber approach for any $Q^2 \,>\,1\,GeV^2$ at $x\,>\,10^{-2}$.
In this kinematic region  the solution of \eq{103} is drastically
different from the Glauber one.

A general conclusion for Fig.\ref{asy} is very simple:
the amount of  shadowing which was taken into account in the MF
 is not enough , at least for the gluon structure function in nuclei at
$x\,<\,10^{-2}$ and we have to solve \eq{103}
to obtain the correct behavior of the gluon structure function for nuclei.

Now, we would like to show that solution \eq{103} is the asymptotic solution
 of the new evolution equation.
 In order 
to check this let us try to find the solution to \eq{103} in the form
$\kappa \,=\,\kappa_{asymp} \,+\,\Delta \kappa$, 
anticipating
that $\Delta \kappa $ is small. If it is so, 
the following linear equation can be written for $\Delta \kappa$:
\beq \label{109}
\frac{\partial^2 \Delta \kappa( y, \xi  )}{\partial y\,\partial \xi}\,\,+\,\,
\frac{\partial \Delta \kappa( y, \xi}{\partial y}\,\,=\,\,\frac{d F(\kappa)}{
d \kappa}\,|_{\kappa = \kappa_{asymp}(y)}\,\,\Delta \kappa (y, \xi)\,\,.
\eeq
The general solution to \eq{109} one can find going to the Mellin transform
in respect to $\xi$, namely:
\beq \label{110}
\Delta \kappa ( y, \xi ) \,\,= \,\,\int_C \,\,\frac{ d \nu}{ 2 \,i\,\pi}
\Delta \kappa (y,\nu)\,\,e^{\nu (\,\xi\,-\,\xi_0\,)}\,\,,
\eeq
where contour $C$ is taken to the right of all singularities in  $\nu$ 
for function $\Delta \kappa (y,\nu)$.

For function  $\Delta \kappa (y,\nu)$ we have the equation:
\beq \label{111}
 ( \,\nu \,\,+\,\,1\,) \,\frac{ d \Delta \kappa (y,\nu)}{ d y}\,\,=
F'(y) \Delta \kappa (y,\nu)\,\,,
\eeq
where we denote $ F'(y)\,\,= \frac{d F(\kappa)}{
d \kappa}\,|_{\kappa = \kappa_{asymp}(y)}$.
Solving \eq{111} we obtain the answer:
\beq \label{112}
\Delta \kappa ( y, \xi ) \,\,= \,\,\int_C \,\,\frac{ d \nu}{ 2 \,i\,\pi}
\Delta \kappa (\nu)\,\,\exp[\,\nu (\xi\,-\xi_0) \,+\,\,\frac{1}{\nu + 1}\,\int^y_{y_0}
F'( y') d y'\,]\,\,=
\eeq
$$
\,\,=\,\,\int_C \,\,\frac{ d \nu}{ 2 \,i\,\pi}
\Delta \kappa (\nu)\,\,\exp[\,\nu(\, \xi\,-\,\xi_0\,)\,
\,+\,\,\frac{1}{\nu + 1}\,
\ln( F(\kappa_{asymp}(y))\,\,,
$$
where function  $ \Delta \kappa (\nu)$
should be find from the initial condition at $\xi = \xi_0$, namely
$\Delta \kappa (y,\xi = 0)$ = 0.
 
To satisfy the initial condition we will find  function  $
\Delta \kappa (\nu)$ from the equation:
\beq \label{113}
 \kappa_{asymp} ( y, \xi = \xi_0 ) \,\,= \,\,\int_C \,\,\frac{ d \nu}{ 2 \,i\,
\pi}
\Delta \kappa (\nu)\,\,\exp[\,\,\frac{1}{\nu + 1}\,\int^y_{y_0}
F'( y') d y'\,]\,\,.
\eeq 
%
%
%
In doing so we obtain $\Delta \kappa( y,\xi)$:
\beq \label{114}
\Delta \kappa ( y, \xi ) \,\,= \,\,\int_C \,\,\frac{ d \nu}{ 2 \,i\,
\pi}
\Delta \kappa (\nu)\,\,\exp[\,\,\frac{1}{\nu + 1}\,\int^y_{y_0}
F'( y') d y'\,]\,\,\{ \,\exp( \,\nu (\xi - \xi_0)\,)\,\,-\,\,1\,\}\,\,,
\eeq
which satisfies all our requirements. 
%
%
%
%

We claim that function:
\beq \label{115}
 \kappa ( y, \xi ) \,\,= \,\,\int_C \,\,\frac{ d \nu}{ 2 \,i\,
\pi}
\Delta \kappa (\nu)\,\,\exp[\,\,\frac{1}{\nu + 1}\,\int^y_{y_0}
F'( y') d y'\,]\,\,\exp( \,\nu (\xi - \xi_0)\,)\,\,
\eeq
is the approximate solution to the nonlinear equation (\ref{103}).

Let us check this point considering very large values of $\xi$ and $y$.
  In the region 
of large $y$,  $\int^y_{y_0}
F'( y') d y'\,\,\rightarrow \,\,\bar \as \ln (\bar \as y)$ and 
$ \Delta \kappa( \nu) \,\,\propto\,\,\frac{1}{\nu}$ which reproduces 
$ \kappa_{asymp} \,\,\rightarrow\,\,\bar \as y$ at large $y $ and $\xi = \xi_0$.
Making use of the saddle point approximation one can see that
\beq \label{116}
\Delta \kappa( y,\xi)\,\,\rightarrow\,\,e^{2 \sqrt{\xi \as \ln (bar \as y)}}
\,\,\ll\,\,\kappa_{asymp}\,\,.
\eeq
To give a solution able to describe the experimental data we have to adjust
the behavior of $\kappa_{asymp} $ at small  $y$ ( $ y \,<\,y_0 $ )  with
the available parameterization of the gluon structure function, in particular 
with the GRV parameterization which we have been using through this paper as
a standard one. We have not done this job in this paper and we are going to
publish the result of this  calculation elsewhere. However, to estimate
 the effect of the SC  which follow from \eq{103} we solve this equation using
the semiclassical approach. 
%
%
%
%
%

\subsubsection{Semiclassical Approach}

Here we solve eq.(\ref{103}) using the semiclassical approach, adjusted
to the solution of the nonlinear equation of eq.(\ref{103})-type in refs.
\cite{r1,Collins90,Bartels91}. For simplicity, we assume that $\as$ is 
fixed.

In the semiclasical approach we are looking for the solution of eq.
(\ref{103}) in the form
\bea
\kappa = e^S
\eea
where $S$ is a function with partial derivatives: $\frac{\partial S}{
\partial y} = \o $ and $\frac{\partial S}{\partial \xi} = \gamma $ 
which are smooth function of $y$ and $\xi$.
It means that
\bea
\frac{\partial^2 S}{\partial \xi \partial y} \ll \frac{\partial S}{
\partial y} \cdot \frac{\partial S}{\partial \xi} = \o \gamma
\label{118}
\eea

Using eq.(\ref{118}), one can easily rewrite eq.(\ref{103}) in the form
\bea
\frac{\partial S}{\partial y} \frac{\partial S}{\partial \xi} +
\frac{\partial S}{
\partial y} = e^{-S} F(e^{S}) \equiv \Phi(S)
\label{119}
\eea
or
\bea
\o (\gamma + 1) = \Phi (S)
\label{120}
\eea

We are going to use the method of characteristics( see, for example,
 ref.\cite{Sneddon}).
 For equation in the form
\bea
F(\xi, y, S, \gamma , \o ) = 0
\label{121}
\eea
we can introduce the set of characteristic lines $ (\xi(t),y(t), S(t), \o (t),
\gamma (t) ) $, functions of the variable $t$, which satisfy the following
equations
\bea
\frac{d \xi}{d t} = F_{\gamma} \, ; \,\, \frac{d y}{d t} = F_{\o} \,; \,
\, \frac{d S}{d t} = \gamma F_{\gamma} +\o F_{\o} \, ; \,\, \nonumber \\
\frac{d \gamma}{d t} =- ( F_{\xi} + \gamma F_{S})\, ; \,\,
\frac{d \o}{d t} =-( F_{y} + \o F_{S} )
\label{122}
\eea
where $F_{\xi} = \frac{\partial}{\partial \xi} F(\xi, y, S, \gamma , \o )$ 
etc. Eq.(\ref{122}) looks as follows for the case of eq(\ref{120})
\bea
\frac{d \xi}{d t} = \o  \, ; \,\, \frac{d y}{d t} = \gamma + 1 \,; \,
\, \frac{d S}{d t} =  \o (2 \gamma +1 ) \, ; \,\, \nonumber \\
\frac{d \gamma}{d t} = \Phi'_{S} \gamma \, ; \,\,
\frac{d \o}{d t} =\Phi'_{S} \o \,\, ,
\label{123}
\eea
where $\Phi'_{S} = \frac{\partial \Phi}{\partial S}$. For practical purpose
it is better to rewrite the set of equations(\ref{123}) in the form
\bea
\frac{d \xi}{d y} = \frac{\o}{\gamma +1}  \, ; \,\, 
\, \frac{d S}{d y} =  \o \frac{(2 \gamma +1 )}{\gamma +1} \, ; \,\, 
\frac{d \gamma}{d y} = \Phi'_{S} \frac{\gamma}{\gamma +1} \, .
\label{124}
\eea 
Using eq.(\ref{120}), eq.(\ref{124}) can be rewritten
\bea
\frac{d \xi}{d y} = \frac{\Phi(S)}{(\gamma +1)^2}  \, ; \,\,\,\, 
\, \frac{d S}{d y} =   \frac{2 \gamma + 1}{(\gamma +1)^2}\Phi (S) \, ; \,\,\,\, 
\frac{d \gamma}{d y} = \Phi'_{S} \frac{\gamma}{\gamma +1} \, .
\label{125}
\eea 
The initial condition for this set of equations we derive from eq.(\ref{104}),
namely
\bea
S_0 = ln \kappa_{in} (y_0, \xi_0) \nonumber \\
\gamma_0  = \left. \frac{\partial ln \kappa_{in} (y_0 , \xi )}
{\partial \xi} \right|_{\xi = \xi_0}
\label{129}
\eea

Let us discuss the main properties of the solution before numerical
calculations. The first observation is that $\Phi'_{S} < 0$ and $\Phi(S) > 0$
for all values of $S$. From the second equation of the set (\ref{125})
we see that $S$ decreases along all trajectories with $\gamma < -1/2$ and
increases for trajectories with $\gamma > -1/2$. Thus, it is useful to study
the qualitative behavior of the trajectories for two different
 regions of the initial condition.
 Namely, for $\gamma_0 < -1/2$ and for $\gamma_0 > -1/2$. 

From the third equation of the set (\ref{125}), we notice that
$\frac{d \gamma}{d y}\,>\,0$ for all $\gamma < 0$.
It means that $\gamma$ grows with $y$ starting from $\gamma_0$. However
for $\gamma > 0$, $\frac{d \gamma}{d y} < 0$ and $\gamma$ starts to fall 
down. In both cases, for $\gamma_0 > -1/2$, $S$ goes to infinity as $y$ grows,
and $\Phi'_{S}$ and the derivative $\frac{d \gamma}{d y}$ go to zero. Thus, we 
can conclude that 
$\gamma \rightarrow 0$ as $y \rightarrow \infty$.  


%
It is useful to study closer what is happening at small $\gamma$. 

From equations (\ref{125}), we can write
\bea
\frac{d S}{d \gamma} = \frac{2 \gamma +1}{\gamma (\gamma +1)}
\frac{\Phi (S)}{\Phi'_{S}(S)}
\eea
or
\bea
\frac{d ln \Phi}{d \gamma} = \frac{2 \gamma +1}{\gamma (\gamma +1)}
\label{126}
\eea
which has the solution
\bea
\Phi (S) = \Phi (S_0) \left( \frac{\gamma (\gamma +1)}
{\gamma_{0} (\gamma_{0} +1)} \right) \, .
\label{127}
\eea
For
small $|\gamma|$ the solution of eq.(\ref{127}) is correlated with large
values of $S$, since $\Phi (S)$ has maximum at $S=S_0$ and decreases at 
$S > S_0$. On the other hand, at large $S$, $\Phi'_{S} \rightarrow - \Phi (S)$.
Substituting this relation in the third equation of (\ref{125}), we obtain
%
\bea
\frac{d (\gamma /\gamma_0)}{d y} = -\frac{\Phi (S_0)}{\gamma_0 + 1} \frac{\gamma^2}{\gamma_0^2} \, ,
\label{131}
\eea
which  solution is
\bea
- \frac{1}{\gamma /\gamma_0} +1 = -\frac{\Phi (S_0)}{\gamma_0 +1} (y -y_0)
\label{132}
\eea
or
\bea
\frac{\gamma}{\gamma_0} = \frac{1}{ \frac{\Phi (S_0)}{\gamma_0 +1 } 
(y-y_0) +1} \, .
\label{133}
\eea
Therefore, we see that $\gamma$  approaches $\gamma = 0$ at large $y$ for
$\gamma_0 > -1/2$, either for $\gamma$ positive or negative.

In order to find the solution of whole set of equations (\ref{125}) at
large values of $y$ we use the fact that $\gamma \ll 1$. Indeed, neglecting
 $\gamma$ in comparison with $1$, eq's.(\ref{125}) are reduced to
\bea
S - S_0 = \xi - \xi_0 \nonumber \\
\frac{d S}{d y} = \Phi (S) \,\, .
\label{134}
\eea
Rewriting the second equation in terms of the function $\kappa = e^S$ we have
\bea
\frac{d \kappa}{d y} = F(\kappa) \, .
\label{135}
\eea
The solution of this equation is $\kappa = \kappa_{asymp}$. The first equation
gives the equation for trajectories and at $y \gg y_0$ we have
\bea
\xi - \xi_0 = ln \left(\frac{ \kappa_{asymp}}{\kappa} \right)
\rightarrow ln \bar \as
(y - y_0) + ln ln \bar\as (y -y_0)
\label{136}
\eea
These trajectories are the same as the trajectories of
 eq.(\ref{109}). The simplest way to see this is just to find 
the saddle point in the solution of eq.(\ref{112}).

Therefore the qualitative picture of the trajectories looks as follows.
At small values of $y - y_0$ we can start from initial 
condition in which $e^{S_0} \ll 1$. 
In this case $\Phi'_{S_0} \propto -e^{S_0}$, $ e^{S_0}
\ll 1$ and $\gamma$
remains close to $\gamma_0$ in the large interval of $y -y_0$. For these values
of $y-y_0$, $S$ grows as a function of $y$ and this grows leads $\Phi'_S$ to
approach zero.
This behavior reflects in the decrease of $\gamma$ versus $y - y_0$.
 Finally, at very large values of $y-y_0$ we approach the
 asymptotic solution for large region of $\xi$.

Now we will study the qualitative behavior of 
the trajectories of eq.(\ref{125})  for $\gamma < -1/2$. 
As we already know, $S$ decreases 
along all trajectories with $\gamma  < - 1/2$. As $S$ goes to negative values,
$\Phi(S)$ goes to $\frac{N_{C} \as}{\pi} $ and $\Phi'_{S}$ goes to zero from
negative values. It means that  $d \gamma / d y > 0$ but tends to zero as $y$
grows. Thus, $\gamma$ increases and tends to a constant. As $S$ decreases, the
value of $\kappa$ goes to zero. This behavior is enhanced for $\gamma_0$ 
closer to $-1$. We will show below that this solution approaches
the solution of the GLAP equation in DLA for $\gamma_0 < -1/2$


Let us recall
that in the  GLAP equation $\Phi'_{S} = 0$ and 
$\Phi (S) = \frac{N_{C} \as}{\pi} $. Thus, the set of  eq's(\ref{125}) can be
rewritten in the form:
\bea
\frac{d \xi}{d y} = \frac{N_{C} \as}{\pi   (\gamma_{0} + 1)^2}
\nonumber \\
\frac{d S}{d y} = \frac{ N_{C} \as}{\pi} \frac{2\gamma_{0} +1}
{(\gamma_{0} +1)^2}
\label{137}
\eea
where $\gamma_0$
is the initial value of the $\gamma$ which does not change along the 
trajectory. The solution of eq(\ref{137}) is
\bea
\xi - \xi_0 = \frac{N_C \as}{\pi (\gamma_0 + 1 )^2} ( y - y_0 )\,\,;
\nonumber \\
S - S_0 = \frac{ N_C \as }{\pi  (\gamma_0 +1 )^2} ( 2 \gamma_0 +1)
(y - y_0 ) \,\,;
\label{138}
\eea
or
\bea
\gamma_0 + 1 = \sqrt{\frac{N_C \as}{\pi} \frac{y - y_0}{\xi - \xi_0} }
\,\,;\nonumber \\
S - S_{0} = -( \xi - \xi_0) + 2 \sqrt{\frac{ N_C \as}
{\pi} (\xi - \xi_0)( y - y_0)}\,\,.
\label{139}
\eea
The above expression is the solution of DLA equation for $\kappa$. 
From eq.(\ref{138}) we see that $\kappa$ in DLA approximation goes to zero for
$\gamma_0 < -1/2$

In the region of the double log approximation we consider
\bea
\frac {N_c \as}{\pi} (\xi - \xi_0) (y - y_0)    \approx 1
\label{140}
\eea
while $\frac{N_C  \as}{\pi} (\xi - \xi_0) \ll 1 $ and 
$\frac{N_C \as}{\pi} (y - y_0) \ll 1$. Therefore
\bea
\gamma_0 + 1 \approx \frac{1}{\xi - \xi_0} \le 1
\label{141}
\eea
We set the  initial condition  $y = y_0 = 4.6$ ($ x_B = 10^{-2}$), 
where the shadowing correction is not big and the evolution starts from
$\gamma < 0$. In this case  $d \gamma / d y > 0$ and the value of 
$\gamma$ increases . At the same time $d S / d y < 0$ and $S$ decreases if
 $\gamma_0 < -1/2$. With the decrease of $S$, the value of $\Phi'_{S}$
becomes smaller and after short  evolution the trajectories of the nonlinear 
equation start to approach the trajectories of the GLAP equations. We face
this situation for any trajectory with $\gamma_0 $ close to -1. If the 
value of $\gamma_0$ is smaller than $- \,\frac{1}{2}$ but the value of
$S_0$ is sufficiently big, the decrease of $S$ due to evolution cannot
provide a small value for $\Phi'(S)$ and $\gamma$ increases until its value
becomes bigger than $ - \frac{1}{2}$ at some value of $y= y_c$. In this case
 for $y > y_c$   the trajectories behave as in the case with $\gamma_0 \,>
 \,-\,\frac{1}{2}$.
 For
$\gamma_0 > -1/2$, the picture changes crucially. In this case, $d S/ d y
> 0 $ , $d \gamma / d y > 0$  and both increase. Such trajectories go apart 
from the trajectories of the GLAP equation and nonlinear effects play more
and more important role with increasing $y$. These trajectories approach 
the asymptotic solution very quickly.

\begin{figure}[hptb]
\begin{center}
\begin{tabular}{ c c}
\psfig{file=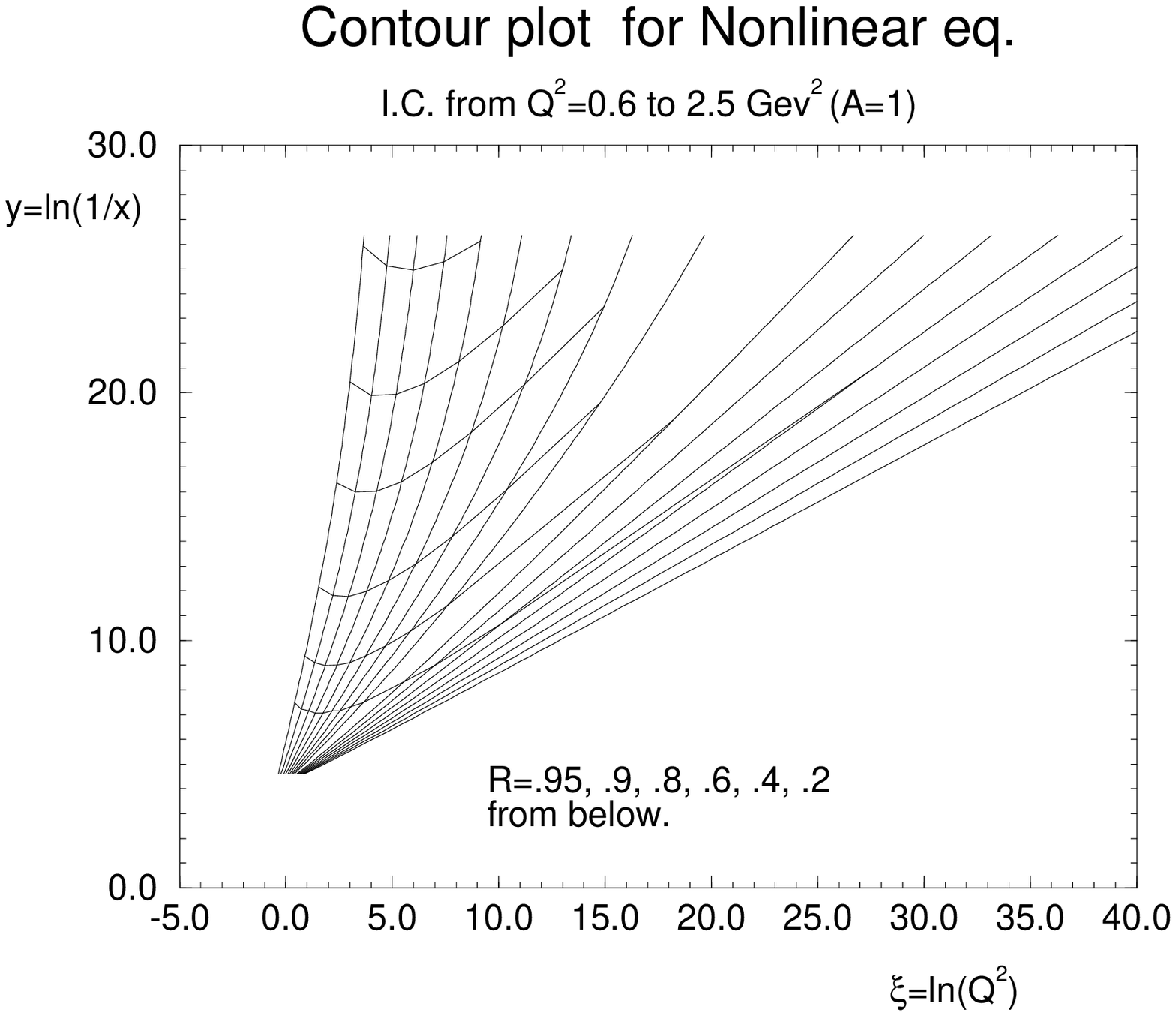,width=80mm} & \psfig{file=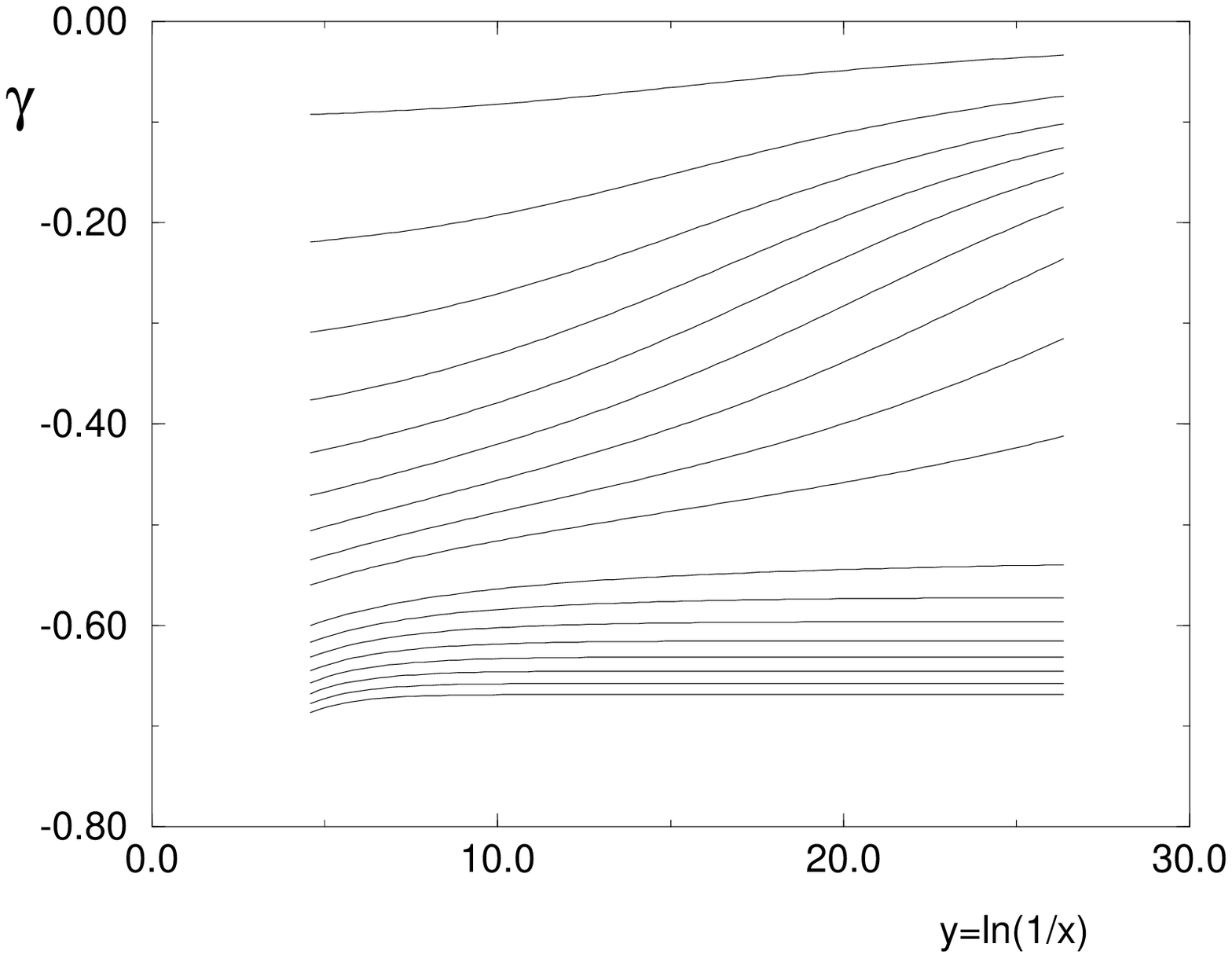,width=80mm}\\
\end{tabular}
\end{center}
\caption{ \em The trajectories and contour plot for the solution of 
the generalized
 evolution equation for N.\,\,\,$R\, = 
\,\frac{xG(x,Q^2)( generalized\,\,\, equation)}{x G(x,Q^2) (GLAP)}$.}
\label{scn}
\end{figure}

\begin{figure}[hptb]
\begin{center}
\begin{tabular}{ c c}
\psfig{file=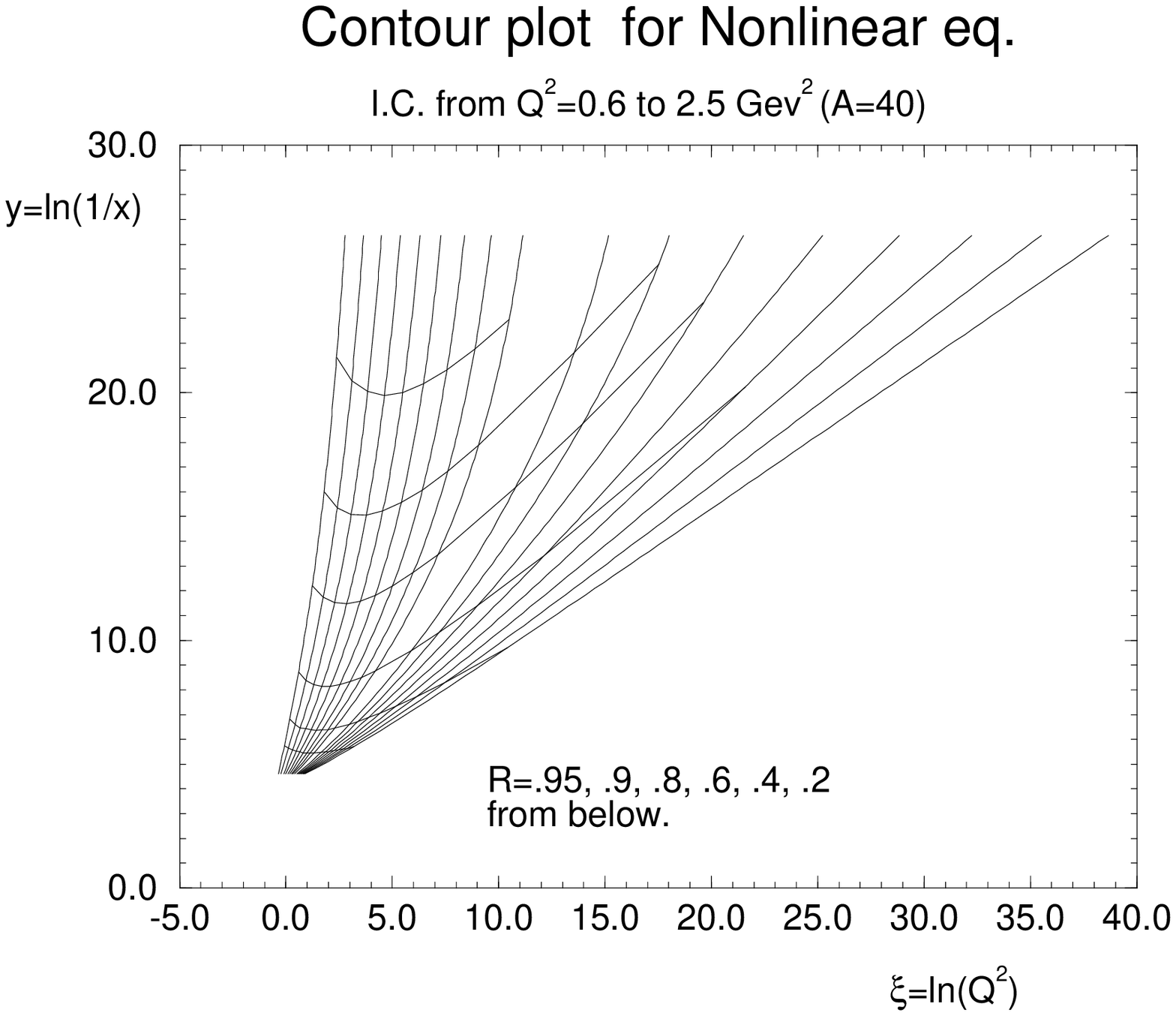,width=80mm} & \psfig{file=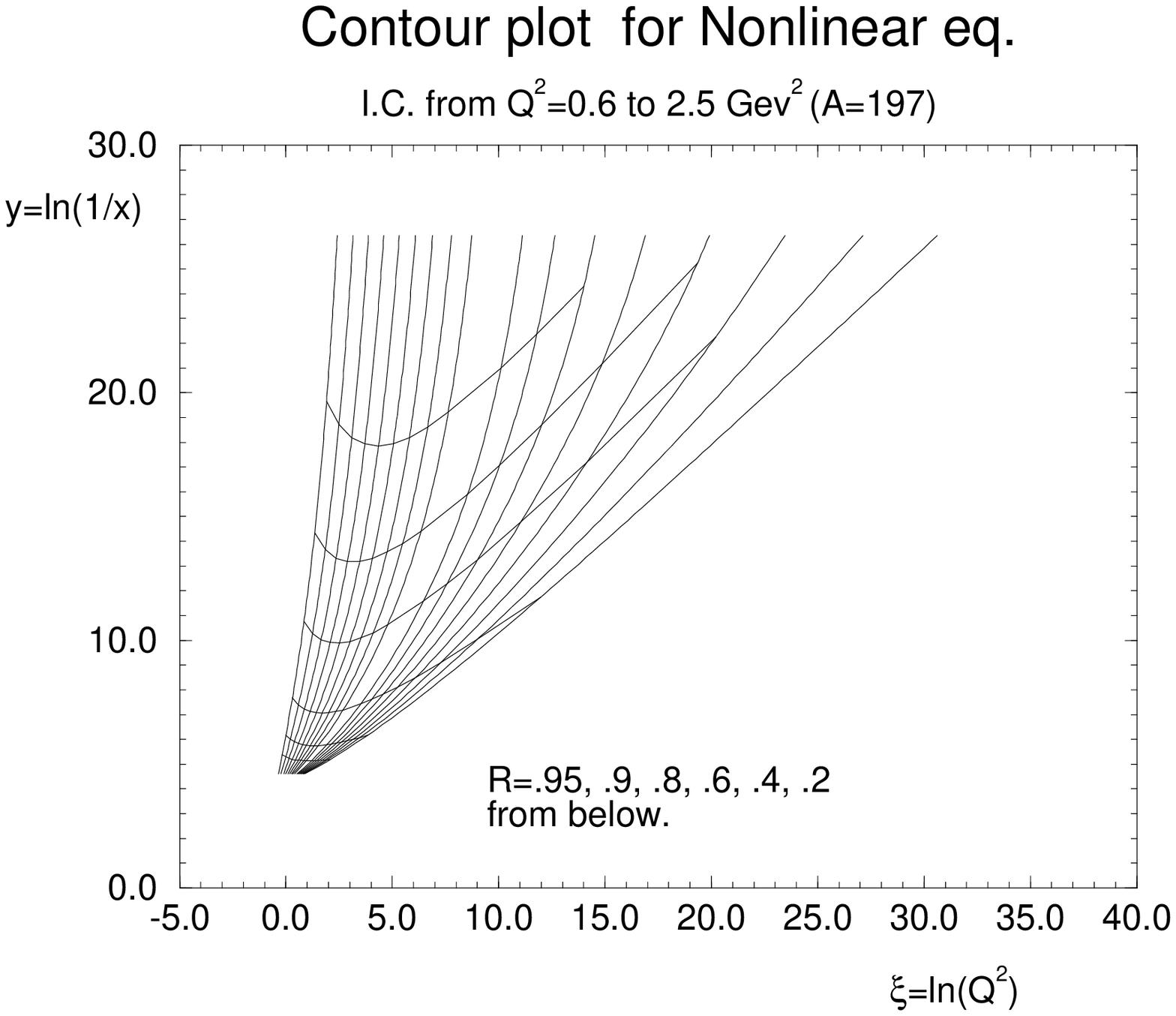,width=80mm}\\
\end{tabular}
\end{center}
\caption{\em The trajectories and  contour plot for the solution of the generalized
 evolution equation for Ca and Au.\,\,\,$R\, = 
\,\frac{xG(x,Q^2)( generalized \,\,\,equation)}{x G(x,Q^2) (GLAP)}$.}
\label{sca}
\end{figure}

For the numerical solution we use the 4th order Runge - Kutta method
to solve our set of equations with the initial distributions of  \eq{129}.
 The result of the solution is given in
Figs.\ref{scn} and \ref{sca}. 
In these figures we plot the bunch of the trajectories
with different initial conditions. For the nucleon ( Fig.\ref{scn} )
 we show also
the dependence of $ \gamma$ along these trajectories. One can notice 
that the trajectories behave in the way which we have discussed in our
qualitative analysis. It is interesting to notice that the trajectories,
which are different from the trajectories of the GLAP evolution equations,
 start  at $y = y_0 = 4.6$ with the values of $Q^2$ between $0.5 GeV^2$ and 
$2.5 GeV^2$ for a nucleon. It means that, guessing  which is the boundary
condition   at $Q^2 = Q^2_0 = 2.5 GeV^2$,  we can hope that the 
linear evolution equations ( the GLAP equations) will describe the evolution
of the deep inelastic structure function in the limited but sufficiently
wide range of $Q^2$. In other words, we can repeat the trick done in
subsection 4.3 deriving the GLR evolution equation for nucleus.

In Figs. \ref{scn} and \ref{sca} we plot 
also  the lines with definite value of the ratio
$R\, = \,\frac{xG(x,Q^2)( generalized\,\,\, equation)}{x G(x,Q^2) (GLAP)}$
 (horizontal lines). These lines give the way to estimate how big are the SC.
One can see that they are rather big.

\begin{figure}[hpbt]
\begin{tabular}{c c}
\psfig{file=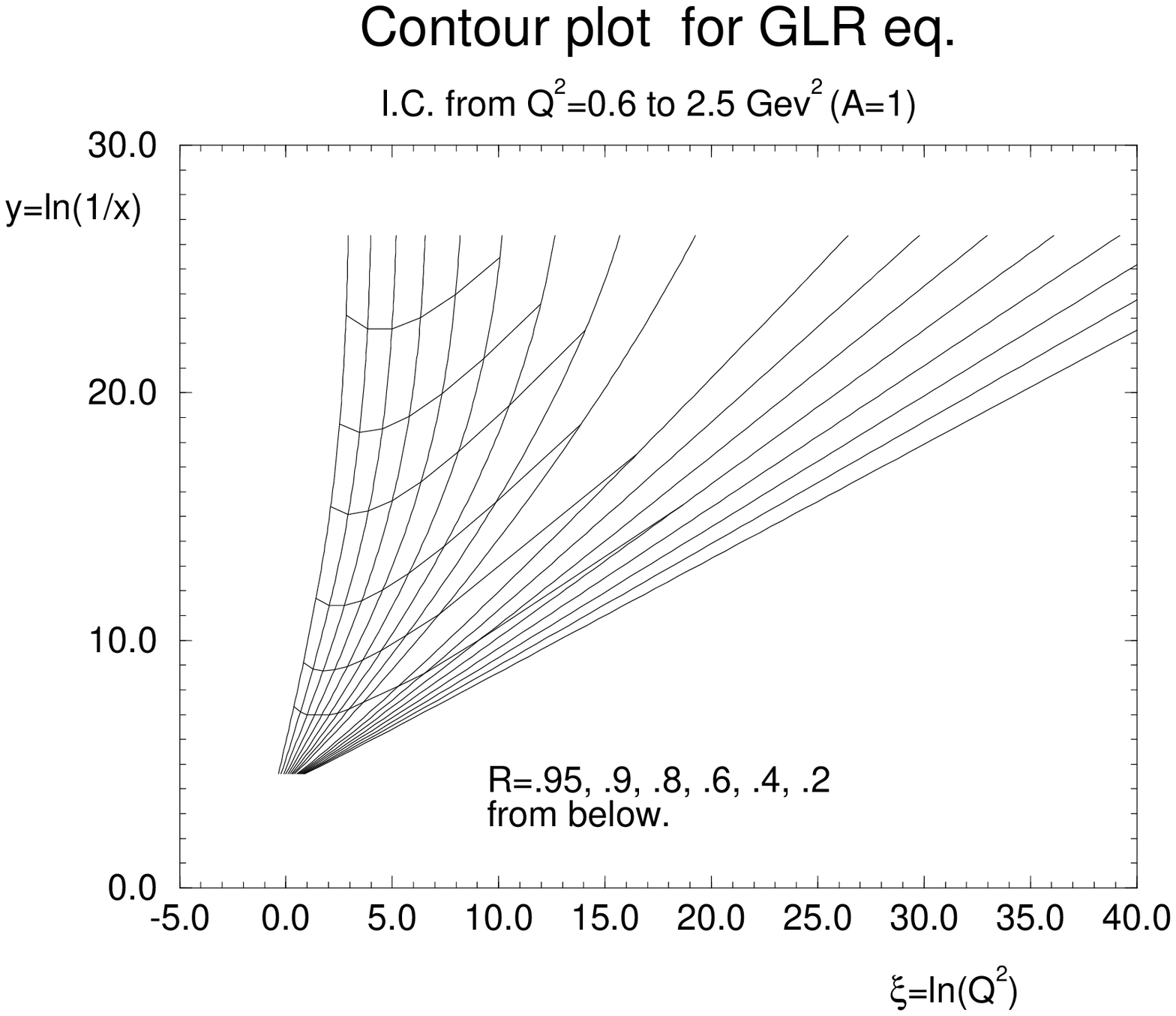,width=80mm} &\psfig{file=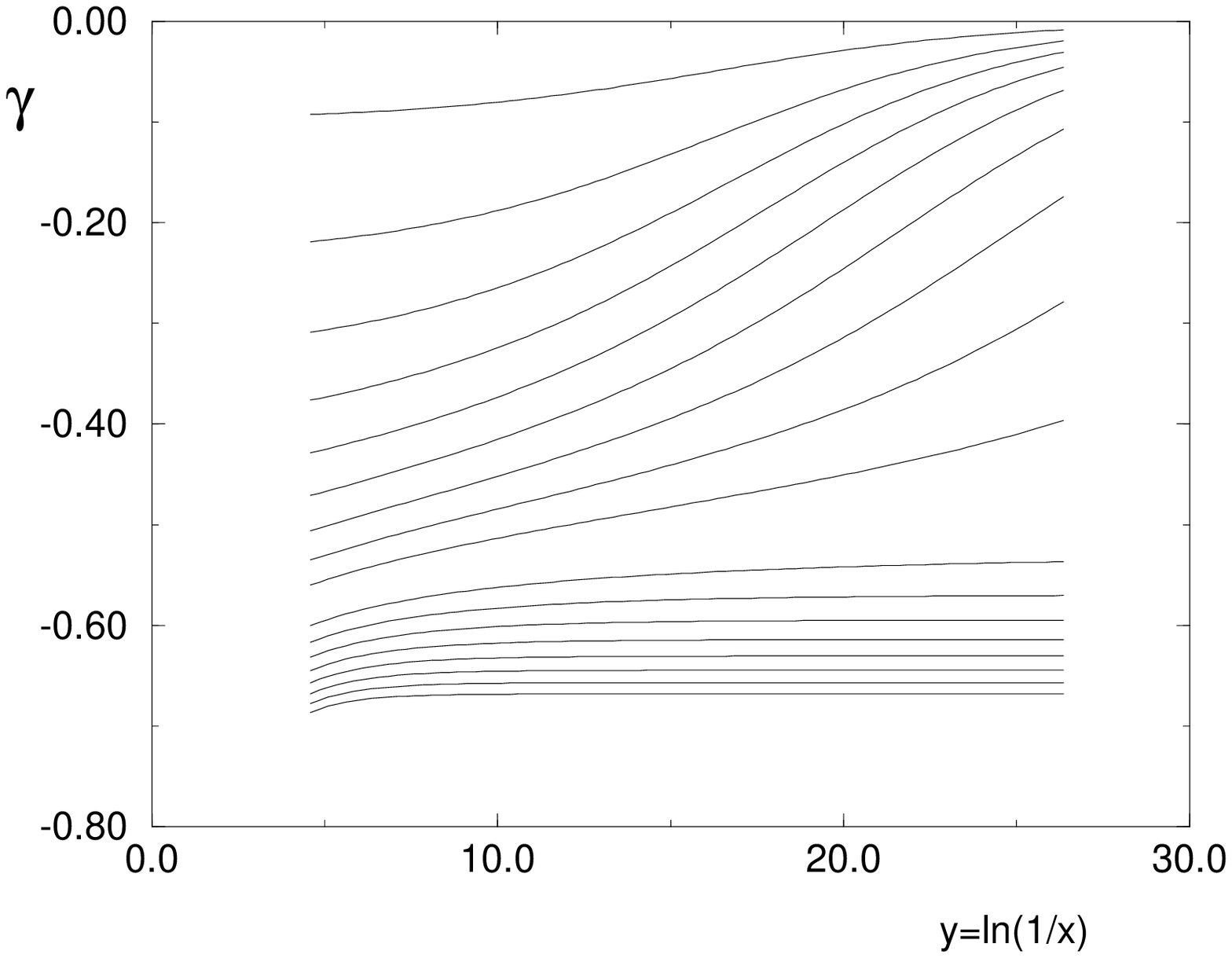,width=80mm} \\
\psfig{file=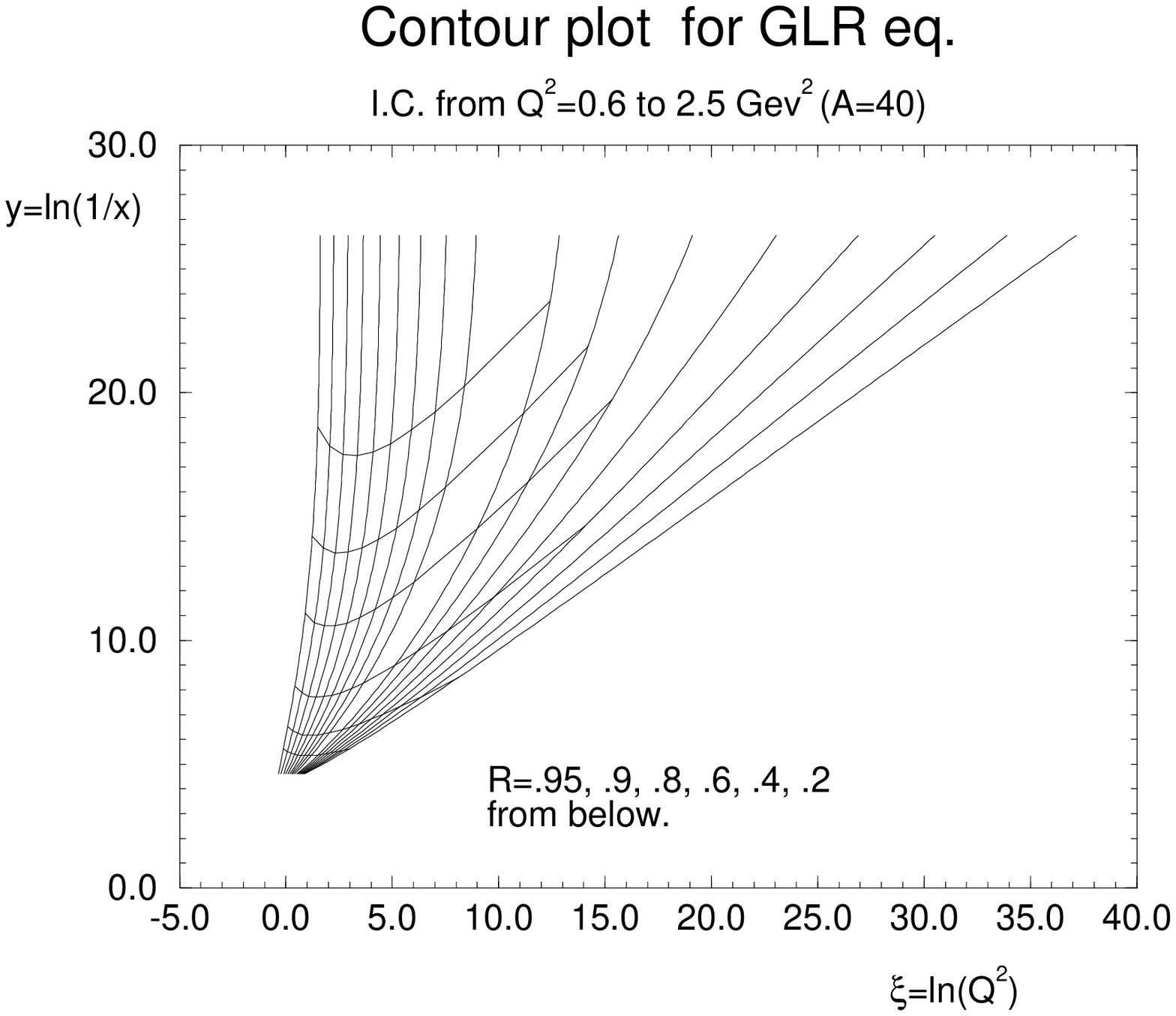,width=80mm} & \psfig{file=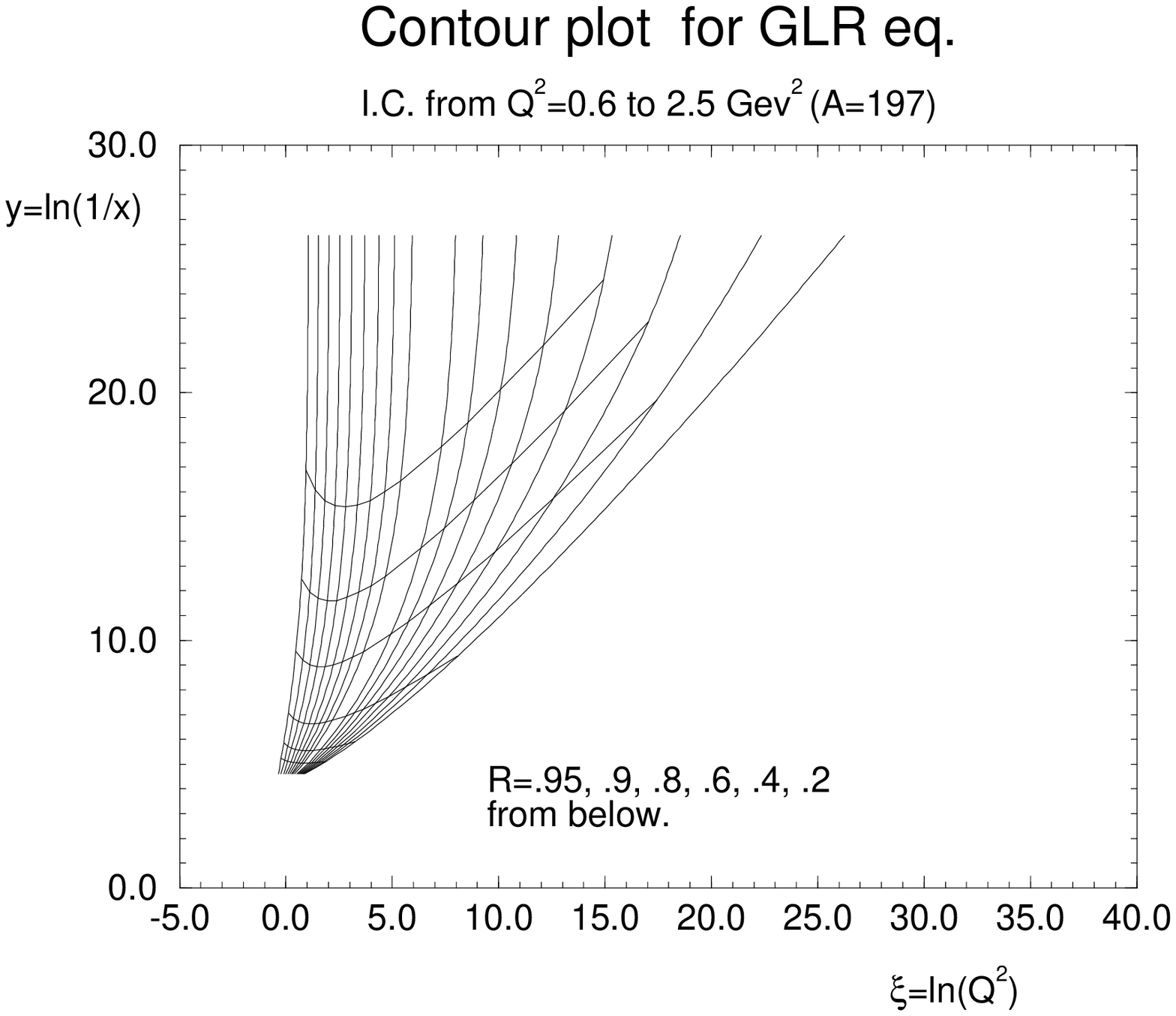,width=80mm}\\
\end{tabular}
\caption{\em The trajectories and contour plot for the solution of the GLR
 evolution equation for N,  Ca and Au.\,\,\,$R\, = 
\,\frac{xG(x,Q^2)( GLR )}{x G(x,Q^2) (GLAP)}$.}
\label{scglr}
\end{figure}

We have discussed only the solution with fixed coupling constant which we
put equal to $\as = 0.25$ in the numerical calculation. The problem how to
solve the equation with running coupling constant is still open.
The GLR equation is the limited case of the general one when we consider
only two first terms in the expansion of $F(\kappa)$ with respect to $\kappa$.
In Fig.\ref{scglr}
 we picture the trajectories and contour plot for the GLR equation.
One can see that the GLR equation gives stronger SC that the generalized
evolution equation. However, for nucleon the difference becomes sizable only
at very small values of $x$ out of the HERA kinematic region. 

We would like also to draw your attention to the fact that our asymptotic
solution turns out to be quite different from the GLR one. The GLR solution
in the region of very small $x$ leads to saturation of the gluon density
\cite{Collins90,Bartels91,BALE}. Saturation means that $\kappa$ tends
to a constant in the region of small $x$. The solutions of
\eq{103} approach the asymptotic solution  at $ x\,\rightarrow\,0$, which
 does not depend on $Q^2$, but exhibits sufficiently strong dependence
 of $\kappa$  on
$x$ ( see Fig.\ref{asy} ).

\section{Conclusions.}
\label{conclu}
1. The Glauber approach to the gluon structure function in a nucleus, suggested
 by Mueller in Ref.\cite{r7}, has been developed and studied in detail.
 Using the GRV
parameterization for the gluon structure function in a nucleon, we 
calculated the
value as well as energy and $Q^2$ dependence of the gluon structure function in 
a nucleus.It is shown that the shadowing corrections are important 
 in the region of small $x$ and changed
crucially the value and anomalous dimension of the nuclear structure function,
 unlike the nucleon one. The interesting observation is the fact that the
 average anomalous dimension for $Q^2$ bigger that $1 GeV^2$
does not exceed $\frac{1}{2}$ for nuclei. This fact makes the application of
 the GLAP evolution equations much more reliable in the case of nuclei than
for a nucleon.

2. The corrections to the Glauber approach have been investigated in a
 systematic way and it turned out that they are  essential and should be
taken into account. The  QCD technique has been developed to go beyond 
the Glauber approach and the new evolution equation for nuclear structure 
function has been derived in  perturbative QCD ( see \eq{103} ).

3. The new evolution equation was solved in the semiclassical approximation
and the main qualitative properties of the solution were discussed.
In particular, it was shown that this solution does not lead to the
 saturation of the gluon density but manifests a strong dependence of
 gluon density on $x$. We consider
 that this solution can provide  a selfconsistent interface  between  ``soft"
high energy phenomenology  and  ``hard" QCD physics. 
Surprisingly, the solution of the GLR equation gives a very accurate 
approximation to the solution of the new evolution equation in the
HERA kinematic region.

Several problems remain beyond  the scope of this  paper. First, we  described
the parton cascade in the GLAP evolution at low $x$ or in other words in
double log approximation of perturbative QCD (DLA). At first sight it is a
 controversial assumption since   the DLA works rather badly in the  accessible
region of $x$ and should be replaced by the BFKL dynamics at very small
values of $x$. We found the possible way out of this difficulty since the SC
change the value of the anomalous dimension in  such a 
way that it does not reach
the value of $\frac{1}{2}$ where the BFKL equation should be taken into
account. It means that the SC come first and the BFKL evolution will never
develop. We consider this result as a plausible explanation why the BFKL
evolution has not been seen yet in nucleon data at HERA. The experiments with
 nuclei will certainly shed light on this problem.

The second problem that we have not discussed in this paper is the corrections
 to the Glauber approach due to the selfinteraction of the partons belonging
to different branches of the parton cascade. We suppose to do this in further
publications using the general approach proposed in ref. \cite{r23}.

A certain limitation of all calculations in the paper is the fact that
we used only the GRV parameterization of the gluon structure function in the
nucleon. We would like only to recall that the goal of the paper is not to
provide a reliable prediction for an experiment but rather to study
the mechanism and amount  of the shadowing corrections. Therefore, the GRV 
parameterization was a tool for our theoretical experiment which is based on the
available experimental data and on the GLAP evolutions. Nevertheless, we are
 going to make full calculations using all parameterizations 
on the market in the
nearest future.

We also used the simplest assumption for the nucleon density in a nucleus,
namely,the 
 Gaussian one. The calculation with more general parameterization of the
nucleon density will come out soon.

We hope that our paper will convince the reader that the SC are essential for
the gluon density in a nucleus, and that the Glauber approach, 
in spite of the fact
 that it is widely used, is not enough both from the phenomenological and
 theoretical point of view. Fortunately, QCD gives us the first example
 how to treat the corrections to the Glauber approach theoretically.

{\bf Acknowledgements:} 
We are very grateful to the LAFEX-CBPF 
 and IF-UFRGS for kind
hospitality and use of their facilities during our work.
 MBGD thanks A. Capella and D.Schiff for enlightening discussions.
 Work partially
 financed by  CNPq, CAPES and FINEP, Brazil.

\end{document}